\pdfoutput=1
\documentclass[twocolumn, onecolappendix, iop]{openjournal}
\usepackage{natbib}
\usepackage{graphicx}
\usepackage{times}
\usepackage{rotate}
\usepackage{hyperref}
\hypersetup{colorlinks=true, linkcolor=blue, citecolor=blue, filecolor=blue, urlcolor=blue}
\usepackage{orcidlink}

\usepackage{multirow}
\usepackage{url} 
\usepackage{bm}
\usepackage[postdot,nostyles]{glossaries-extra}
\usepackage{glossary-long}
\setglossarystyle{long} 
\usepackage{amsmath,amssymb}
\usepackage[section]{placeins}
\usepackage{booktabs}
\usepackage{pifont}

\newcommand{\nout}{\ensuremath{n_\mathrm{out}}}

\newcommand{\Mpc}{\ensuremath{\mathrm{Mpc}}}

\newcommand{\expval}[1]{\ensuremath{\left< #1 \right>}}

\newcommand{\ds}{\ensuremath{\Delta\Sigma}}
\newcommand{\dsref}{\ensuremath{\ds_\mathrm{reference}}}

\newcommand{\ADS}{\ensuremath{A_{\Delta\Sigma}}}
\newcommand{\Sigmacrit}{\ensuremath{\Sigma_\mathrm{crit}}}
\newcommand{\rp}{\ensuremath{r_\mathrm{p}}}

\newcommand{\zl}{\ensuremath{z_\mathrm{l}}}
\newcommand{\zs}{\ensuremath{z_\mathrm{s}}}
\newcommand{\zf}{\ensuremath{z'}}
\newcommand{\asel}{a_\mathrm{sel}}
\newcommand{\aselhat}{\hat{a}_\mathrm{sel}}
\newcommand{\msel}{m_\mathrm{sel}}
\newcommand{\mselhat}{\hat{m}_\mathrm{sel}}

\newcommand{\gammat}{\ensuremath{\gamma_\mathrm{t}}}
\newcommand{\Dang}{\ensuremath{D_\mathrm{A}}}
\newcommand{\Dcom}{\ensuremath{D_\mathrm{C}}}

\newcommand{\dd}{\ensuremath{\mathrm{d}}}
\newcommand{\fbias}{\ensuremath{f_\mathrm{bias}}}

\setabbreviationstyle[acronym]{long-short-user}

\renewcommand*{\glsxtruserparen}[2]{
  \glsxtrfullsep{#2}%
  \glsxtrparen
   {#1\ifglshasfield{\glsxtruserfield}{#2}{;
     \expandafter\citealt\expandafter{\glscurrentfieldvalue}}{}}%
}

\glsdefpostdesc{acronym}{
 \ifglshasfield{\glsxtruserfield}{\glscurrententrylabel}%
 {~\expandafter\citep\expandafter{\glscurrentfieldvalue}}%
 {}%
}

\newacronym{ggl}{GGL}{Galaxy-Galaxy Lensing}
\newacronym{esd}{ESD}{excess surface-mass density}
\newacronym{lss}{LSS}{Large-Scale Structure}

\newacronym{bgs}{BGS}{Bright Galaxy Sample}
\newacronym{lrg}{LRG}{Luminous Red Galaxies}

\newacronym[
    user1={Kuijken:2015,Asgari:2021}
]{kids}{KiDS}{Kilo-Degree Survey}
\newacronym[
    user1={2015AJ....150..150F,Gatti:2021}
]{des}{DES}{Dark Energy Survey}
\newacronym[
  user1={2018PASJ...70S...8A,Aihara:2022}]
{hsc}{HSC}{Hyper Suprime-Cam Subaru Strategic Survey}
\newacronym{hscy1}{HSC-Y1}{first-year data of the HSC}
\newacronym{hscy3}{HSC-Y3}{third-year data of the HSC}
\newacronym{boss}{BOSS}{Baryon Oscillation Spectroscopic Survey}

\newacronym[
    user1={Mandelbaum:2013}
]{sdss}{SDSS}{Sloan Digital Sky Survey}
\newacronym{desi}{DESI}{Dark Energy Spectroscopic Instrument}
\newacronym{desiy1}{DESI DR1}{the first data release of DESI}
\newacronym{iip}{IIP}{individual-inverse-probability}
\newacronym{pip}{PIP}{pairwise-inverse-probability}

\defcitealias{Lange:2024}{L+24}
\defcitealias{Yuan:2024}{Y+24}
\defcitealias{Leauthaud:2022}{LWB I}

\makeglossaries

\begin{document}

\title{Lensing Without Borders: Measurements of galaxy-galaxy lensing and projected galaxy clustering in DESI DR1\vspace{-4em}}
\shorttitle{Lensing Without Borders: DESI DR1}

\author{\orcidlink{0000-0002-7273-4076}S.~Heydenreich,$^{1}$
\orcidlink{0000-0002-3677-3617}A.~Leauthaud,$^{1,2}$
\orcidlink{0000-0002-5423-5919}C.~Blake,$^{3}$
\orcidlink{0000-0002-8246-7792}Z.~Sun,$^{4}$
\orcidlink{0000-0002-2450-1366}J.~U.~Lange,$^{5}$
T.~Zhang,$^{6}$
M.~DeMartino,$^{1}$
\orcidlink{0000-0002-7522-9083}A.~J.~Ross,$^{7,8,9}$
J.~Aguilar,$^{10}$
\orcidlink{0000-0001-6098-7247}S.~Ahlen,$^{11}$
\orcidlink{0000-0001-9712-0006}D.~Bianchi,$^{12,13}$
D.~Brooks,$^{14}$
\orcidlink{0000-0001-7316-4573}F.~J.~Castander,$^{15,16}$
T.~Claybaugh,$^{10}$
\orcidlink{0000-0002-2169-0595}A.~Cuceu,$^{10}$
\orcidlink{0000-0002-1769-1640}A.~de la Macorra,$^{17}$
\orcidlink{0000-0002-0728-0960}J.~DeRose,$^{18}$
\orcidlink{0000-0002-4928-4003}Arjun~Dey,$^{19}$
\orcidlink{0000-0002-5665-7912}Biprateep~Dey,$^{20,6}$
P.~Doel,$^{14}$
N.~Emas,$^{3}$
\orcidlink{0000-0003-4992-7854}S.~Ferraro,$^{10,21}$
\orcidlink{0000-0002-3033-7312}A.~Font-Ribera,$^{22}$
\orcidlink{0000-0002-2890-3725}J.~E.~Forero-Romero,$^{23,24}$
\orcidlink{0000-0003-1481-4294}C.~Garcia-Quintero,$^{25}$
E.~Gaztañaga,$^{15,26,16}$
\orcidlink{0000-0003-3142-233X}S.~Gontcho A Gontcho,$^{10}$
G.~Gutierrez,$^{27}$
\orcidlink{0000-0002-2312-3121}B.~Hadzhiyska,$^{28,10,21}$
\orcidlink{0000-0002-6550-2023}K.~Honscheid,$^{7,29,9}$
\orcidlink{0000-0001-6558-0112}D.~Huterer,$^{30,31}$
\orcidlink{0000-0002-6024-466X}M.~Ishak,$^{32}$
N.~Jeffrey,$^{14}$
\orcidlink{0000-0001-8820-673X}S.~Joudaki,$^{33}$
\orcidlink{0000-0002-9253-053X}E.~Jullo,$^{34}$
\orcidlink{0000-0002-0000-2394}S.~Juneau,$^{19}$
\orcidlink{0000-0002-8828-5463}D.~Kirkby,$^{35}$
\orcidlink{0000-0003-3510-7134}T.~Kisner,$^{10}$
\orcidlink{0000-0001-6356-7424}A.~Kremin,$^{10}$
A.~Krolewski,$^{36,37,38}$
O.~Lahav,$^{14}$
\orcidlink{0000-0002-6731-9329}C.~Lamman,$^{25}$
\orcidlink{0000-0003-1838-8528}M.~Landriau,$^{10}$
\orcidlink{0000-0001-7178-8868}L.~Le~Guillou,$^{39}$
\orcidlink{0000-0003-4962-8934}M.~Manera,$^{40,22}$
\orcidlink{0000-0002-1125-7384}A.~Meisner,$^{19}$
R.~Miquel,$^{41,22}$
\orcidlink{0000-0001-9070-3102}S.~Nadathur,$^{26}$
\orcidlink{0000-0003-3188-784X}N.~Palanque-Delabrouille,$^{42,10}$
\orcidlink{0000-0002-0644-5727}W.~J.~Percival,$^{36,37,38}$
\orcidlink{0000-0002-2762-2024}A.~Porredon,$^{33,43,44,9}$
\orcidlink{0000-0001-7145-8674}F.~Prada,$^{45}$
\orcidlink{0000-0001-6979-0125}I.~P\'erez-R\`afols,$^{46}$
G.~Rossi,$^{47}$
\orcidlink{0000-0002-0394-0896}R.~Ruggeri,$^{48}$
\orcidlink{0000-0002-9646-8198}E.~Sanchez,$^{33}$
\orcidlink{0000-0002-0408-5633}C.~Saulder,$^{49}$
D.~Schlegel,$^{10}$
A.~Semenaite,$^{3}$
\orcidlink{0000-0002-3461-0320}J.~Silber,$^{10}$
D.~Sprayberry,$^{19}$
\orcidlink{0000-0003-1704-0781}G.~Tarl\'{e},$^{31}$
B.~A.~Weaver,$^{19}$
\orcidlink{0000-0002-5992-7586}S.~Yuan,$^{50}$
\orcidlink{0000-0002-7305-9578}P.~Zarrouk,$^{39}$
\orcidlink{0000-0001-5381-4372}R.~Zhou,$^{10}$
\orcidlink{0000-0002-6684-3997}H.~Zou,$^{51}$
}
\email{$^*$E-mail: sheydenr@ucsc.edu}
\shortauthors{S.~Heydenreich et al.}

\label{firstpage}

 
\begin{abstract} 
    We present \acrlong{ggl} measurements obtained by cross-correlating spectroscopically observed galaxies from the first data release of the \acrfull{desi} with source galaxies from the \acrlong{hsc}, the \acrlong{kids}, the \acrlong{sdss}, and the \acrlong{des}. Specifically, we measure the excess surface mass density $\ds$ and tangential shear $\gammat$ for the \acrlong{bgs} and \acrlong{lrg} measured within the first year of observations with \acrshort{desi}. To ensure robustness, we test the measurements for systematic biases, finding no significant trends related to the properties of the \acrshort{desi} lens galaxies. We identify a significant trend with the average redshift of source galaxies, however, this trend vanishes once we apply shifts to the \acrlong{hsc} redshift distributions that are also favored by their fiducial cosmology analysis. Additionally, we compare the observed scatter in the measurements with the theoretical covariance and find excess scatter, driven primarily by small-scale measurements of $r\leq 1 \, \Mpc/h$; measurements on larger scales are consistent at the $2\,\sigma$ level. We further present the projected clustering measurements $w_p$ of the galaxy samples in the \acrlong{desiy1}. These measurements, which are publicly available, serve as a foundation for forthcoming cosmological analyses.
\end{abstract}


\maketitle

\section{Introduction}
\label{sec:introduction}
The \gls{desi} is a state-of-the-art spectroscopic survey designed to map the large-scale structure of the Universe with unprecedented precision \citep{arxiv:1308.0847,arxiv:1611.00036,arxiv:1611.00037,arxiv:2205.10939,arxiv:2306.06309}. Using imaging data from the DESI Legacy Imaging Surveys \citep{arxiv:1702.03653,arxiv:1804.08657}, the \gls{desi} survey selects galaxy and quasar targets \citep{arxiv:2208.08518} and measures their spectra using a robotic fiber-fed spectrograph \citep{arxiv:2209.14482,2306.06310,2024AJ....168..245P}. Comprehensive survey validation was carried out during early observations, as described in \citet{arxiv:2306.06307}, and the Early Data Release (EDR) provides the first public spectroscopic dataset from \gls{desi} \citep{arxiv:2306.06308}. Cosmological constraints using measurements of baryonic acoustic oscillations and redshift space distortions from the \gls{desiy1} \citep{arxiv:2503.14745} have yielded exciting results, particularly pertaining to a potential for dynamic dark energy and tight constraints on neutrino masses \citep{arXiv:2404.03002,arXiv:2411.12022}, which have been improved upon by the recent results of \gls{desi}'s analysis of third-year data \citep{arXiv:2503.14738}.

In this work, we use spectroscopic galaxy samples from the \gls{desiy1} to measure \gls{ggl} signals. \gls{ggl}, the distortion of background source galaxy shapes due to the gravitational potential of foreground lens galaxies, provides a direct probe of the matter distribution around galaxies. We measure the excess surface mass density $\ds$ and tangential shear $\gammat$ for the \gls{desi} \gls{bgs} and \gls{lrg} samples. To achieve this, we cross-correlate DESI lens galaxies with source galaxy samples from three independent weak lensing surveys: the \gls{hsc}, the \gls{kids}, and the \gls{des}. Additionally, we include measurements from the \gls{sdss}, as the great overlap between \gls{sdss} and \gls{desi} allows us to rigorously test if the \gls{desiy1} lens sample is sufficiently homogeneous over the entire survey footprint, and not biased by, e.g., its varying completeness, the fact that its target selection is derived from different imaging surveys, or other local properties such as stellar density or extinction.

In recent years, analyses of cosmic shear surveys have provided precise measurements of the matter power spectrum and cosmological parameters \citep{arxiv:2007.15632, 2023OJAp....6E..36D, Amon:2022, 2023PhRvD.108l3518L, arxiv:2304.00704}. These measurements have significantly advanced our understanding of the Universe's large-scale structure and the nature of dark energy. However, these measurements report a value of the matter clustering parameter $S_8$ that is consistently lower than the one inferred by analyses of the cosmic microwave background \citep{2020A&A...641A...6P}. This discrepancy, known as the $S_8$ tension, has motivated the development of new cosmological probes and analyses to test the robustness of the results. In particular, a strong focus has been placed on potential biases arising in cosmic shear measurements, such as intrinsic alignments \citep{arxiv:1506.08730,arxiv:1811.09598,arxiv:1910.05994,2012JCAP...05..041B,
2015SSRv..193....1J,2015PhR...558....1T}, baryonic effects \citep{arxiv:2206.08591,arxiv:2403.20323,Bigwood:2024}, and systematics in the lensing measurements \citep{arxiv:2205.07892,arxiv:1710.03235}. One should note that the $S_8$ tension can be alleviated by allowing a more flexible prescription for the intrinsic alignment of galaxies \citep{arxiv:2407.04795}, and in the most recent analysis by \gls{kids} this discrepancy has been resolved primarily by a variety of updated calibration techniques \citep{2025arXiv250319441W}. On scales smaller than usually used for cosmology inference, weak gravitational lensing appears to give lower values for the matter clustering, even when directly compared to clustering measurements, which has become known as the `lensing is low' effect \citep{Leauthaud:2017,arxiv:2010.01143,2021MNRAS.502.2074L,2022MNRAS.509.1779L,2023MNRAS.520.5373L}, although some studies argue that this effect is caused by too simplistic assumptions in the galaxy-halo connection model \citep{arxiv:2211.01744,
2023MNRAS.525.3149C}.

This has spawned many efforts to compare lensing measurements between different surveys to identify potential causes of systematic biases \citep{2023MNRAS.520.5016L,2024JCAP...08..024G,2023OJAp....6E..36D,2024MNRAS.534.3305H,
2024MNRAS.528.2112S,2023JCAP...01..025G,
2018MNRAS.477.4285A}. For this, \gls{ggl} is a powerful probe \citep[see, e.g.][]{2005MNRAS.361.1287M}. It provides a direct measurement of the (physical) matter distribution around galaxies and thus serves as an ideal probe to compare lensing measurements at the data-level, without having to invoke a cosmological parameter inference. Several groups have already performed efforts to assess the consistency between different lensing surveys, such as \citet{2022MNRAS.509.2033L}. Most notably, \citet[][hereafter \citetalias{Leauthaud:2022}]{Leauthaud:2022} compare the amplitude of the excess surface density $\ds$ measured around galaxies of the \gls{boss} Data Release 12 \citep{Alam:2015,Reid:2016} using six different imaging surveys. They find a good agreement between the measurements, where differences between measurements are mostly consistent with the statistical scatter. They do find some potentially relevant trends of the lensing amplitude with the fraction of hemisphere coverage, and local stellar density, which is likely caused by inconsistencies in the \gls{boss} lens sample (although the local stellar density could also potentially impact the shear measurements of imaging surveys). Most notably, they also find a trend with the average redshift of sources, which is indicative of potential unidentified or underestimated biases in the lensing surveys. Overall, however, they conclude that the magnitude of the `lensing is low' effect can not be explained by any of their findings. A follow-up study by  \citet{2023MNRAS.518..477A} confirmed their finding and found consistent lensing signals between observations by the \gls{kids}, \gls{des}, and \gls{hsc} surveys.

In general, there is much debate about the current and future level of control over various lensing systematics, and the impact of these systematics on cosmological parameter inference (and, ultimately, whether they are the cause of the $S_8$ tension). For example, multiple analyses of data from \gls{hscy3} indicate that the redshift distributions inferred from the position of galaxies in color-magnitude space via multiple independent methods are biased significantly more than their respective methods indicate \citep{arxiv:2304.00701,2023PhRvD.108l3518L,arxiv:2304.00704}. Similarly, recent direct observations of baryon distributions indicate that baryon feedback is significantly stronger than previously assumed \citep{Hadzhiyska:2024}, and independent re-analyses of lensing data, utilizing different modeling strategies including more conservative baryon feedback or intrinsic alignment models, find no $S_8$ tension \citep{arXiv:2303.05537,arxiv:2407.04795}.

Therefore, a key focus of this work is the rigorous assessment of potential systematic errors. We perform this assessment by calculating the excess surface density $\ds$---a measure of the projected mass density---of the \gls{desiy1} galaxies. As $\ds$ is a physical property of the galaxy, it should not depend on properties of the source galaxies of different lensing surveys, their redshifts, or properties unrelated to the galaxies themselves, such as extinction or local stellar density. We test for consistency between lensing measurements from different source catalogs and tomographic redshift bins. We further evaluate potential biases related to the properties of the DESI lens galaxies. Furthermore, we empirically quantify a measure for systematic uncertainties by analyzing excess scatter between lensing measurements, and compare these with those estimated in \citet[][hereafter \citetalias{Lange:2024}]{Lange:2024}. Our analysis builds upon and refines the methodology of \citetalias{Leauthaud:2022} by leveraging the larger and deeper DESI dataset as well as state-of-the-art weak gravitational lensing surveys, which benefit both from a deeper and wider survey and a better control of systematics. Additionally, we perform more stringent cuts to our source and lens samples, which are motivated by independent analyses in \citetalias{Lange:2024} and \citet[][hereafter \citetalias{Yuan:2024}]{Yuan:2024}. This serves to minimize the impact of known systematics such as intrinsic alignments, which are a large potential uncertainty in the original \citetalias{Leauthaud:2022}. Finally, our \gls{ggl} measurements are performed on a unified pipeline that has been validated on simulations \citepalias{Lange:2024} and \gls{boss} data \citep[see e.g.][]{2023MNRAS.520.5373L}. Given that we are utilizing the \gls{desiy1} dataset, which is the largest spectroscopic dataset, combined with the most recent public releases of all major weak lensing surveys, this study represents the most powerful data-level comparison between \gls{ggl} measurements of different imaging surveys to date. We further demonstrate that the \gls{desiy1} dataset does not suffer from potential biases in the lens sample that \citetalias{Leauthaud:2022} found in their analysis of \gls{boss} data.

We further present the projected clustering measurements $w_p$ of the \gls{desiy1} \gls{bgs} and \gls{lrg} samples. These measurements provide a complementary probe of the large-scale structure and will be used as a basis of upcoming 3x2pt analyses, which utilize the combination of shear-shear (cosmic shear), position-shear (\gls{ggl}), and position-position (projected clustering) correlations to infer cosmological parameters.

The measurements presented in this paper provide critical inputs for upcoming cosmological analyses using DESI data. By directly constraining the matter distribution and galaxy-halo connection, these measurements will enhance our understanding of galaxy formation and evolution while providing robust tests of cosmological models. The results and data products are publicly available\footnote{\url{https://doi.org/10.5281/zenodo.22914838}} to enable further scientific studies.

This paper is structured as follows: In Section~\ref{sec:introduction}, we describe our efforts and put them in the context of existing work. Section~\ref{sec:DESI_data} describes the \gls{desiy1} data, including the galaxy samples and weights used to counter potential inhomogeneities. In Section~\ref{sec:lensing_data}, we discuss the data from lensing surveys used in this study. Section~\ref{sec:GGL_measurements} details the methodology for \gls{ggl} measurements. In Section~\ref{sec:amplitude_tests}, we outline the lensing amplitude tests conducted to ensure the fidelity of our measurements. We present the setup for our projected clustering measurements in Section~\ref{sec:clustering_measurements}. Section~\ref{sec:results} presents the results of our lensing and clustering measurements, including homogeneity tests and B-mode analysis. Finally, Section~\ref{sec:discussion} discusses the implications of our findings and potential causes of observed trends, and suggests directions for future research.

To calculate $\ds$, we need to fix a fiducial cosmology in order to calculate $\Sigmacrit$ and transform angular separations into projected comoving distances. The fiducial cosmology adopted for the analysis corresponds to the \texttt{base\_plikHM\_TTTEEE\_lowl\_lowE\_lensing} chain from the Planck 2018 release \citep{2020A&A...641A...6P}. This model serves as the baseline grid for the \textit{AbacusSummit} simulation suite (denoted as \texttt{c000}) used for covariance estimation and is \gls{desi}'s fiducial cosmology. Specifically, we assume a flat $\Lambda$CDM model defined by the matter density $\Omega_m = 0.31519$, the Hubble parameter $h = 0.6736$, the spectral index $n_s = 0.9649$, and the amplitude of fluctuations $\sigma_8 = 0.8079$. The physical baryon and cold dark matter densities are fixed to $\omega_b = 0.02237$ and $\omega_{cdm} = 0.1200$, respectively, with a single massive neutrino species summing to $\sum m_\nu = 0.06\,\mathrm{eV}$. We show in App.~\ref{app:alternaive_cosmologies} that our analysis is not substantially impacted by this choice.

\section{DESI Y1 data}
\label{sec:DESI_data}
    In this section, we introduce the \gls{desiy1} data, the galaxy samples, and the weights used to counter potential inhomogeneities in the data.
    \begin{figure*}
        \centering
        \includegraphics[width=\linewidth]{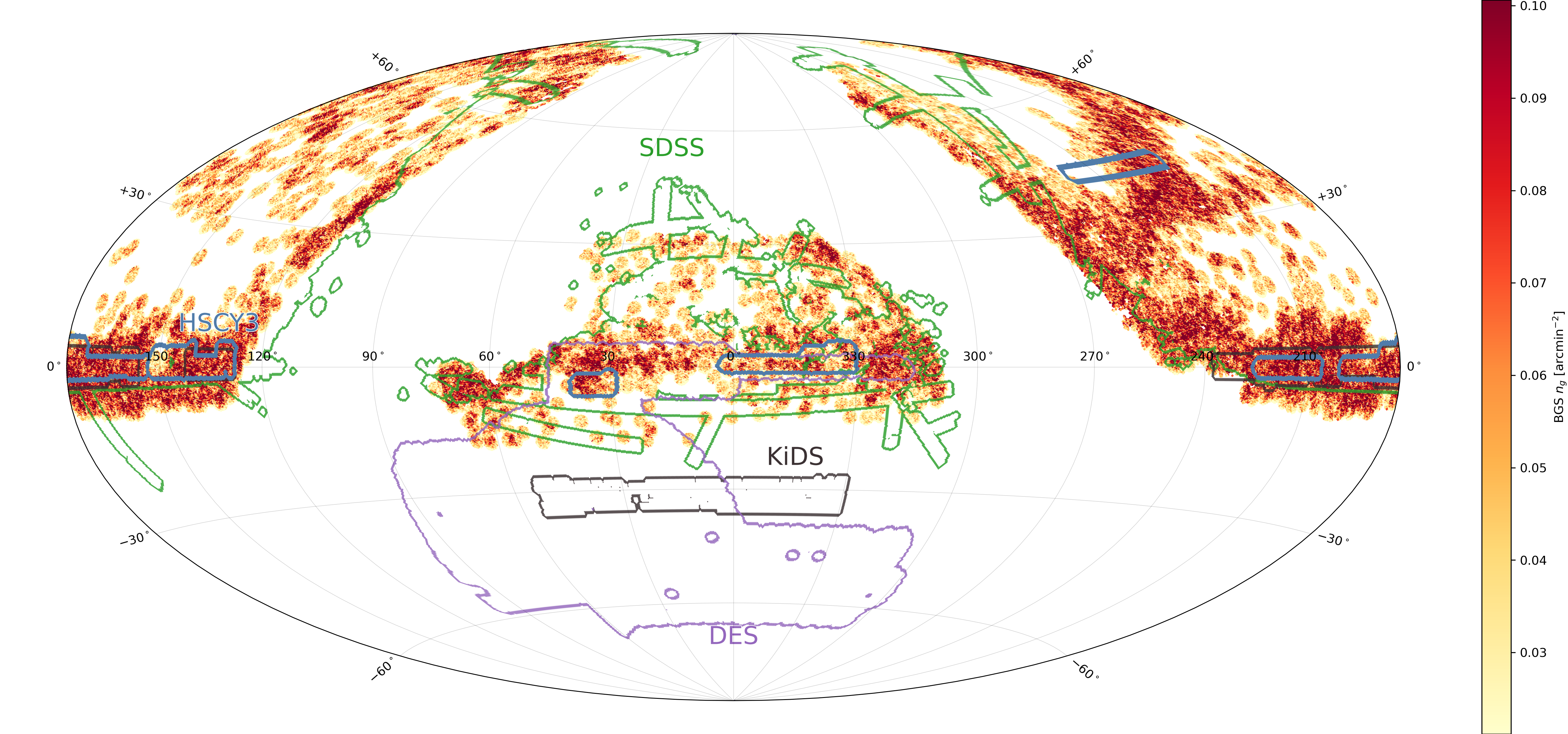}
        \caption{The observed number density of \gls{desiy1} \gls{bgs}, overlaid with the footprints of the four imaging surveys. One can see that \gls{desiy1} has very varying completeness over its footprint, in particular it is much more complete in the \gls{kids} and \gls{hsc} regions than in \gls{des}. We note that the footprint of the \gls{desiy1} \gls{lrg} looks very similar to the \gls{bgs}.}
        \label{fig:footprint_DESI_BGS_overlaps}
    \end{figure*}
    
\begin{table}
\centering
\begin{tabular}{|l|l|l|}
\hline
\textbf{\gls{desiy1} Tracer} & \textbf{Lensing Survey} & \textbf{Overlap Area [sq deg]} \\ \hline
\multirow{4}{*}{\gls{bgs}} & DES & 662 \\ \cline{2-3}
                           & HSCY3 & 453 \\ \cline{2-3}
                           & KiDS & 448 \\ \cline{2-3}
                           & SDSS & 5459 \\ \cline{2-3}
\hline
\multirow{4}{*}{\gls{lrg}} & DES & 793 \\ \cline{2-3}
                           & HSCY3 & 467 \\ \cline{2-3}
                           & KiDS & 460 \\ \cline{2-3}
                           & SDSS & 5383 \\ \cline{2-3}
\hline
\end{tabular}
\caption{Overlap areas of \gls{desiy1} tracers with various lensing surveys.}
\label{tab:intersection_areas}
\end{table}

    \subsection{DESI-Y1 samples}
    \label{sec:DESI_data:data_summary}
        We analyze the \gls{desiy1} \gls{bgs} \citep{arxiv:2010.11283,arxiv:2208.08512} and \gls{lrg} \citep{arxiv:2010.11282,arxiv:2208.08515} samples. The \gls{bgs} are an effectively flux-limited sample of bright galaxies at redshifts roughly $0<z\lesssim 0.4$. The \gls{lrg} sample has been selected to roughly have a constant comoving density between redshifts 0.4 and 1.1. We split the \gls{desiy1} \gls{bgs} into three bins with redshift boundaries $[0.1,0.2,0.3,0.4]$, the \gls{lrg} sample is also split into three bins with redshift boundaries $[0.4,0.6,0.8,1.1]$. These observed galaxies come with a set of randoms. Each set of randoms consists of points that cover the entire legacy survey imaging area at a density of $2500 / \mathrm{deg}^2$ and have been run through the same target selection algorithm as the galaxies \citep{arxiv:2208.08518}. Therefore, they trace the footprint of the \gls{desi} observations and can be used in both lensing measurements (to correct for the impact of large-scale density fluctuations in the \gls{lss} by subtraction of the lensing signal around random points) and clustering measurements (to correct for the impact of the survey window function). We assign redshifts to the randoms by drawing from the redshifts of the observed galaxies, weighted by their total weight $w_\mathrm{tot}$ (see Sect.~\ref{sec:DESI_data:weights}).
        
\begin{figure}
            \centering
            \includegraphics[width=\linewidth]{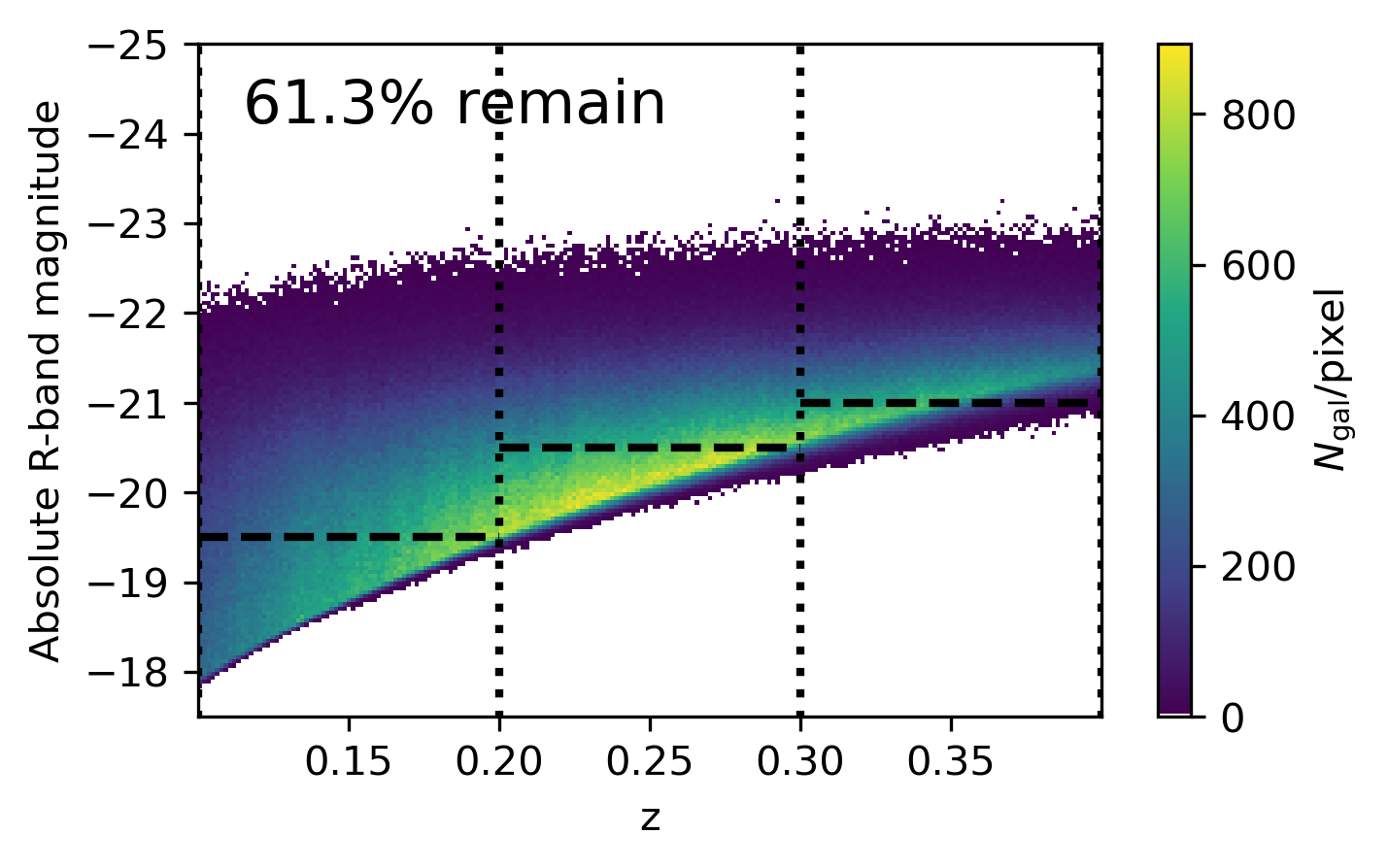}
            \caption{Absolute magnitudes of \gls{desiy1} \gls{bgs} as a function of redshift. The galaxies that pass the magnitude cut described in Sect.~\ref{sec:DESI_data:data_summary} are the ones above the dashed lines. In total, 61\% of all galaxies pass the magnitude cut.}
            \label{fig:absolute_magnitude_cut_bgs}
        \end{figure}

        Binning a galaxy sample in redshift bins of non-zero width runs the risk that the sample undergoes significant evolution within the redshift bin, in particular when that sample is predominantly flux-limited (compare Fig.~\ref{fig:absolute_magnitude_cut_bgs}, where the distribution of absolute magnitudes of observed galaxies evolves substantially with redshift). While this effect is not as relevant for the intra-survey comparison in this work, we follow the recommendation of \citetalias{Yuan:2024} and apply absolute R-band magnitude cuts of $M_\mathrm{R} < -19.5, -20.5, -21$ for the redshift bins of the \gls{bgs} sample $[0.1,0.2],[0.2,0.3]$ and $[0.3,0.4]$, respectively. This leads to approximately constant comoving density of the \gls{bgs} sample and minimizes impacts from redshift evolution of the galaxy sample within a single bin. The impact can be seen in Fig.~\ref{fig:absolute_magnitude_cut_bgs}. \citetalias{Yuan:2024} did not find significant redshift evolution for the \gls{lrg} sample, which has been designed to have a constant comoving density, therefore we do not employ additional magnitude cuts.
        
    \subsection{Weights for DESI-Y1 galaxies}
    \label{sec:DESI_data:weights}
        To mitigate the impact of systematic effects, \gls{desiy1} catalogues come equipped with different weights that are used to offset inhomogeneities in the sample. We will briefly describe the relevant weights here and refer to \citet{Ross:2024} for a more in-depth discussion.
        
        \subsubsection{Redshift failures}
        \label{sec:DESI_data:redshift_failures}
            \gls{desi} is incredibly successful at gathering redshifts for \gls{bgs} and \gls{lrg}, with redshift success rates of 98.9\% and 99.1\%, respectively \citep{2024arXiv241112020D}. However, a fraction of observations result in redshift failures. If uncorrected, these failures can bias the measurements, as they are correlated to intrinsic galaxy properties, such as their luminosity.
            \paragraph*{\texttt{WEIGHT\_ZFAIL}} is the inverse of the redshift failure rate for a given galaxy, based on its properties. This weight mitigates the impact of redshift failures on the analysis and will be denoted by $w_\mathrm{zfail}$.
    
        \subsubsection{Imaging systematics}
        \label{sec:DESI_data:imaging_systematics}
            The \gls{desi} target selection is performed on data from the legacy imaging surveys \citep{arxiv:1804.08657}, and the local stellar density and galactic extinction impact this target selection \citep{arxiv:2208.08512,arxiv:2208.08515}.
            \paragraph*{\texttt{WEIGHT\_SYS}} is the weight accounting for imaging systematics such as galactic extinction and stellar density, which potentially impact the observations. For our galaxy samples, they are computed via multi-linear interpolation of values for relevant imaging systematics on a healpix map of $\texttt{NSIDE}=256$. As we employ custom magnitude cuts for the \gls{bgs} sample, we re-compute the imaging systematics weights, meaning that our \gls{bgs} weights differ from the ones of other \gls{desiy1} analyses \citep[such as][]{2024arXiv241112021D,2024arXiv241112020D,2024arXiv240403000D}. The imaging systematics weights are denoted by $w_\mathrm{imsys}$.
            
        \subsubsection{Fiber incompleteness}
        \label{sec:DESI_data:fiber_incompleteness}
            DESI is unable to observe a complete set of galaxies due to a
            finite amount and extent of fibers. This effect is stronger in regions that have had fewer passes, and in regions that have a high number density of galaxies. \citetalias{Lange:2024} show that this effect strongly biases the lensing measurements, but can be completely removed by applying \gls{iip} weights to the lens galaxies, which we describe below.
            \paragraph*{\texttt{FRACZ\_TILELOCID}} is the inverse of the number of galaxies of the same tracer type at a given \texttt{TILELOCID} (i.e. the number of galaxies at a given position on the focal plane) and works as a fiber incompleteness weight, analogous to the close-pair weights in \gls{sdss}. We will denote our completeness \gls{iip} weights as $w_\mathrm{comp}=1/\texttt{FRACZ\_TILELOCID}$.
            \paragraph*{\texttt{BITWEIGHTS}} are created by running the fiber assignment software on potential targets with randomly permuted priorities between the targets a total of 128 times, yielding a total of 129 potential fiber assignments (including the observed one). For each galaxy we record in which of the 129 observations it got assigned a fiber. In a clustering measurement, we can quantify the likelihood for each pair of galaxies of being observed, and use its inverse as a weight \citep{Bianchi:2017,2018MNRAS.481.2338B}. These are usually referred to as \gls{pip} weights and are essential to calculate an unbiased estimate of the projected clustering on small scales. We describe the application of \gls{pip} weights in more detail in Sect.~\ref{sec:clustering_measurements:PIP_weights}.
            \paragraph*{\texttt{FRAC\_TLOBS\_TILES}} accounts for cases where a fiber was not assigned a specific target type (such as \gls{lrg}) because the associated fiber got assigned to measure either sky brightness or a higher-priority target type. We apply this weight to the randoms to ensure that respective regions on the sky are weighted equally between the data and the randoms. These weights are denoted with $w_\mathrm{tlobs}$.

        \subsubsection{Combined weights}
        \label{sec:DESI_data:combined_weights}
            Our total data weights $w_\mathrm{tot}$ and random weights $w_\mathrm{tot,r}$ are calculated as 
            \begin{align*}
            w_\mathrm{tot}={}&{}w_\mathrm{comp}w_\mathrm{zfail}w_\mathrm{imsys}\, ,\\
            w_\mathrm{tot,r}={}&{}w_\mathrm{tlobs}\, .
            \end{align*}
            We note that it would also be possible to assign no weights to the randoms, and instead divide the data weights by $w_\mathrm{tlobs}$. However, \citet{Ross:2024} have investigated this and shown that, while also unbiased, this choice of weights has worse noise properties.

        \begin{figure*}
            \centering
            \includegraphics[width=\linewidth]{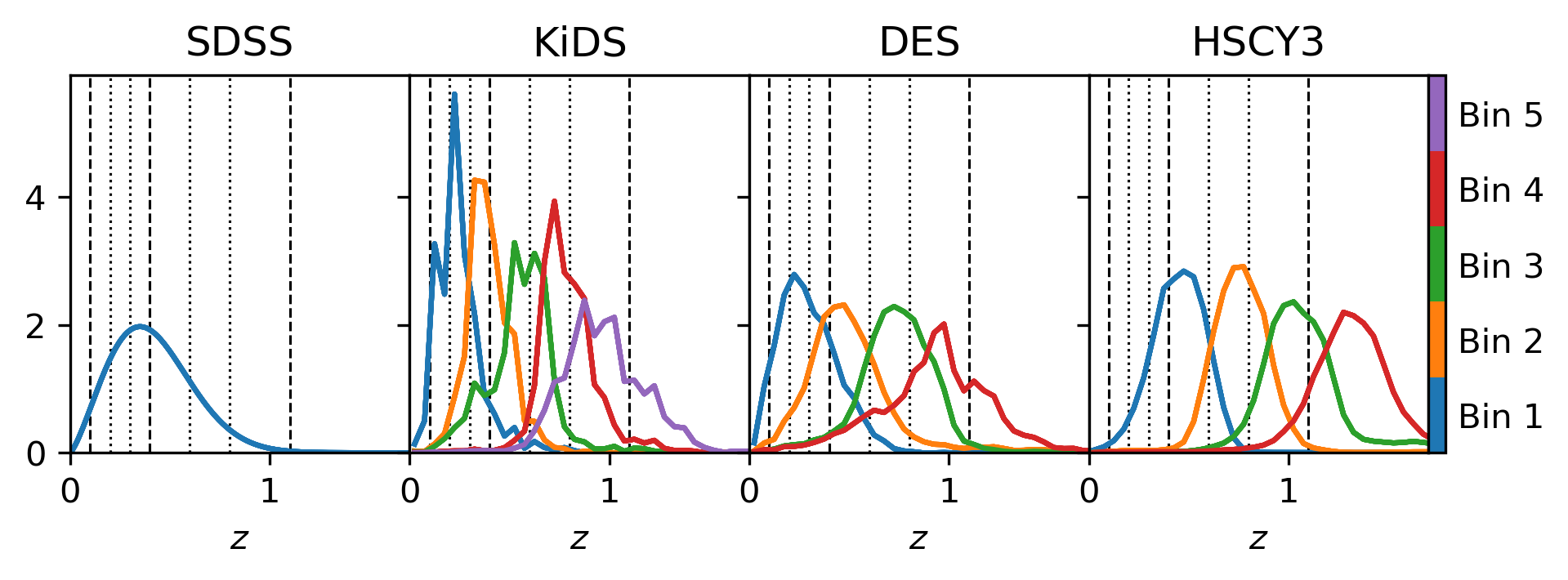}
            \caption{Photo-z distribution of source bins compared to \gls{desiy1} lens galaxies. The different panels show the source redshift distributions for the different imaging surveys, respectively. In every panel, the dashed lines denote the redshift boundaries of the \gls{bgs} and \gls{lrg} lens samples, where the \gls{bgs} extends from $0.1\leq z < 0.4$, and the \glspl{lrg} extend from $0.4\leq z < 1.1$. Each lens sample is subsequently split into three redshift bins, the boundaries of these bins are denoted by the dotted lines. To minimize contamination by lens-source overlaps, we perform stringent cuts on the allowed source-lens bin combinations. We note in Tab.~\ref{tab:source_lens_bins} which source bins are used for the different lens bin samples.}
            \label{fig:all_nofz_plot}
        \end{figure*}

        \begin{table}
            \centering
            \begin{tabular}{ll|ccc|ccc}
\hline
& & \multicolumn{3}{c|}{BGS BRIGHT} & \multicolumn{3}{c}{LRG} \\
& & Bin 1 & Bin 2 & Bin 3 & Bin 1 & Bin 2 & Bin 3 \\
\hline
\multirow{1}{*}{SDSS} & Bin 1 & c & c & c & x & x & x \\
\hline
\multirow{5}{*}{KiDS} & Bin 1 & x & x & x & x & x & x \\
  & Bin 2 & lc & x & x & x & x & x \\
  & Bin 3 & lc & lc & x & x & x & x \\
  & Bin 4 & c & c & c & c & x & x \\
  & Bin 5 & c & c & c & c & x & x \\
\hline
\multirow{4}{*}{DES} & Bin 1 & lc & x & x & x & x & x \\
  & Bin 2 & lc & lc & lc & x & x & x \\
  & Bin 3 & c & c & c & lc & x & x \\
  & Bin 4 & c & c & c & c & x & x \\
\hline
\multirow{4}{*}{HSCY3} & Bin 1 & lc & lc & x & x & x & x \\
  & Bin 2 & c & c & c & lc & x & x \\
  & Bin 3 & c & c & c & c & c & x \\
  & Bin 4 & c & c & c & c & c & lc \\
\hline
\end{tabular}

            \caption{Usage of source bins for different lens samples. `c': used in case of conservative source-lens cuts, as in this analysis. `lc': can be used in less conservative case, not used in this analysis. `x': not used.}
            \label{tab:source_lens_bins}
        \end{table}

\section{Data from lensing surveys}
\label{sec:lensing_data}
    We cross-correlate spectroscopically observed galaxies from \gls{desiy1} with source galaxies from imaging surveys \gls{hsc}, \gls{kids}, \gls{des} and the \gls{sdss}. In Tab.~\ref{tab:intersection_areas} we list the overlap areas between the \gls{desiy1} galaxy samples and the respective lensing surveys. This is also visualized in Fig.~\ref{fig:footprint_DESI_BGS_overlaps}.

    \subsection{HSC}
    \label{sec:lensing_data:HSC}
        The Hyper Suprime-Cam Subaru Strategic Survey (HSC) is a deep, wide-field imaging survey conducted with the Hyper Suprime-Cam instrument on the 8.2-meter Subaru Telescope in Hawaii. The HSC-SSP program provides high-resolution imaging with excellent depth and a wide field of view, making it a premier dataset for weak gravitational lensing studies. In this work, we use source galaxy shapes and photometric redshift data from the HSC Year 3 dataset (HSC-Y3). The photometric redshifts in HSC are computed using three different methods, the template-fitting based code \texttt{Mizuki}, the Direct Empirical Photometric Code (DEmPz), and a similar, neural-network based estimator (DNNz). The photometric redshifts are rigorously calibrated using external spectroscopic samples \citep{Tanaka:2018, Nishizawa:2020,arxiv:2211.16516}, and validated with clustering redshifts. We utilize the public redshift distributions, which were obtained by combining the DNNz estimates with clustering redshifts, for our analysis. We exclude the region within the HSC-Y3 footprint that shows an abnormal amount of B-modes \citep{arxiv:2304.00703}. The survey’s depth and quality allow for precise \gls{ggl} measurements, particularly at higher redshifts. For detailed information, see \citet{Bosch:2018}, \citet{Aihara:2022}, and \citet{arxiv:2304.00703}.

    \subsection{DES}
    \label{sec:lensing_data:DES}
        The Dark Energy Survey (DES) conducted a wide-area imaging survey covering approximately 5000 square degrees of the southern sky in five optical bands ($grizY$). The survey utilized the 570-megapixel Dark Energy Camera (DECam) on the 4-meter Blanco Telescope at the Cerro Tololo Inter-American Observatory. In this analysis, we use the DES Year 3 (DES-Y3) data, which provides a high-quality source galaxy shear catalog based on the {\sc METACALIBRATION} algorithm for shear estimation. The DES-Y3 catalog includes photometric redshift distributions calibrated with external spectroscopic samples and source-injection simulations \citep{2021MNRAS.505.4249M}. With its large area and high galaxy density, DES-Y3 enables robust \gls{ggl} measurements across a wide range of scales. For more details, refer to \citet{Amon:2022}, \citet{Gatti:2021}, and \citet{Sevilla-Noarbe:2021}. We note that we are using the catalog of \gls{des} source galaxies with the updated assignment to source tomographic bins, as pointed out in \citet{2024arXiv241022272M}.

    \subsection{KiDS}
        \label{sec:lensing_data:KiDS}
            The Kilo-Degree Survey (KiDS) is a deep imaging survey designed to study the large-scale structure of the Universe and weak gravitational lensing. KiDS covers approximately 1350 square degrees in four optical bands ($ugri$), observed with the OmegaCAM wide-field imager on the VLT Survey Telescope (VST) at the European Southern Observatory (ESO) Paranal Observatory. In addition to optical bands, the fourth data release, KiDS-1000, incorporates infrared data from the VISTA Kilo-Degree Infrared Galaxy (VIKING) survey. This inclusion of infrared data significantly enhances the accuracy of photometric redshifts by reducing degeneracies, particularly at high redshifts.
    
            A key component of KiDS-1000 is the ``KiDS Gold sample'', a subset of approximately 21 million galaxies selected for their well-calibrated photometric redshift distributions. This sample achieves an effective number density of 6.17 galaxies per square arcminute, providing high-quality data essential for weak lensing studies. The selection process ensures precise redshift measurements, which are critical for accurate excess surface mass density and tangential shear calculations.
    
            The photometric redshifts are computed using the Bayesian Photometric Redshift (BPZ) code and validated against spectroscopic samples \citep{2020A&A...637A.100W,Hildebrandt:2020}. KiDS-1000 also includes an improved weak lensing shear catalog derived from the {\sc lensfit} algorithm, validated through extensive image simulations. The survey’s accurate shape and photometric redshift measurements make it a powerful dataset for \gls{ggl} studies. For further details, see \citet{Hildebrandt:2020}, \citet{Giblin:2021}, and \citet{Kuijken:2019}.
    \subsection{SDSS}
    \label{sec:lensing_data:SDSS}
        The \gls{sdss} is a pioneering wide-area imaging and spectroscopic survey that has significantly advanced studies of large-scale structure and weak gravitational lensing. It provides imaging data in five optical bands ($ugriz$) and has obtained spectra for millions of galaxies, stars, and quasars. The SDSS lensing sample \citep{Mandelbaum:2013} utilizes shape measurements from the SDSS imaging dataset, which, while not as deep as more recent surveys, benefits from exceptionally large sky coverage and high overlap with the DESI footprint (see Fig.~\ref{fig:footprint_DESI_BGS_overlaps}). However, as shown in Fig.~\ref{fig:all_nofz_plot}, the SDSS sample predominantly overlaps with the redshift range of the DESI \gls{bgs} and is at lower redshifts than the \gls{lrg}.

        Despite its strengths, the SDSS lensing sample faces limitations in higher-redshift applications due to its relatively shallow depth. Moreover, as discussed in Sect.~\ref{sec:amplitude_tests:analysis_considerations_zsource}, intrinsic alignments of source galaxies introduce significant contamination in the lensing signal when using SDSS data, especially for higher-redshift lens bins. For these reasons, we exclude SDSS measurements in our comparisons between lensing surveys. Nevertheless, the large overlap of SDSS with the DESI footprint provides a valuable opportunity to test the homogeneity of the DESI BGS sample, as described in Sect.~\ref{sec:amplitude_tests:analysis_considerations_homogeneity}.
        
\section{Galaxy-galaxy lensing measurements}
\label{sec:GGL_measurements}
    In this section we describe how we perform the \gls{ggl} measurements for the different lensing surveys.
    \subsection{Theoretical background}
    \label{sec:GGL_measurements:theory}
        Massive galaxies and their dark matter halos constitute overdensities in the \gls{lss} of the Universe. These overdensities exert a tangential shear $\gamma_\mathrm{t}$ on the images of background source galaxies. The tangential ellipticities of source galaxies $\epsilon_\mathrm{t}$ can be used as an unbiased estimator of the reduced tangential shear
        \begin{equation}
            g_\mathrm{t} = \gamma_\mathrm{t}/(1-\kappa)\;,
            \label{eq:reduced_shear_defn}
        \end{equation} where the convergence $\kappa$ is defined as the projected density along the line of sight $\Sigma$ divided by the critical surface mass density $\Sigmacrit$, where
        \begin{equation}
            \Sigmacrit(\zl,\zs) = \begin{cases}
                \frac{c^2}{4\pi G} \frac{1}{(1 + \zl)^2} \frac{\Dang(\zs)}{\Dang(\zl) \Dang(\zl, \zs)}\quad & \zl < \zs \\
                \infty & \zl \geq \zs
            \end{cases} \; .
            \label{eq:sigmacrit_defn}
        \end{equation}
        Here, $\zl$ and $\zs$ are the redshift of lens and source galaxy, respectively, and $\Dang(z)$ is the angular diameter distance to redshift $z$.
        
        Given a population of lens and source galaxies, we can estimate the average tangential shear via 
        \begin{equation}
            \hat{\gamma}_\mathrm{t} (\theta) \approx \hat{g}_\mathrm{t}(\theta) = \frac{\sum_{ls} w_{\rm tot} w_\mathrm{ls} \epsilon_\mathrm{t}}{\sum_{ls} w_{\rm tot} w_{\rm ls}} \, ,
            \label{eq:gammat_raw_estimator}
        \end{equation}
        where the sum encompasses all galaxy pairs with angular separation $\theta$, $w_\mathrm{tot}$ are the systematic weights for lens galaxies (as detailed in Sect.~\ref{sec:DESI_data:weights}), $w_\mathrm{ls}$ are lens-source weights, and $\hat{\gamma}_\mathrm{t}$ and $\hat{g}_\mathrm{t}$ are the estimators for $\gammat$ and $g_\mathrm{t}$, respectively. For measurements of $\gammat$, usually the lens-source weights are equal to the weight of the source galaxy $w_\mathrm{s}$. 
        
        In practice, the measured ellipticities $\epsilon_\mathrm{t}$ are corrected for various calibration factors, including multiplicative shear biases, shear responsivity, and other survey-specific corrections, to ensure accurate and unbiased estimates of the shear. These corrections are described in detail in Sect.~\ref{sec:GGL_measurements:correction_choices_for_lensing_surveys}. For now, we proceed with the general formulation, assuming that these factors have been accounted for.
        
        Assuming circular symmetry\footnote{While a single galaxy does not obey circular symmetry, an ensemble of lens galaxies does.}, the average tangential shear translates to the projected density contrast $\ds$ via
        \begin{equation}
            \ds(\theta) = \bar{\Sigma}(<\theta) - \Sigma(\theta) = \gammat(\theta)\,\Sigmacrit(\zl,\zs) \; ,
            \label{eq:ds_defn}
        \end{equation}
        where $\bar{\Sigma}(<\theta)$ describes the projected density $\Sigma$ averaged over the disk of radius $\theta$, whereas $\Sigma(\theta)$ is the average value of $\Sigma$ on a circle of radius $\theta$.
        
        We assume that the galaxy ellipticities are an unbiased estimator of the shear $\gamma$. This is not exactly true, as the ellipticities trace the reduced shear field $g=\frac{\gamma}{1-\kappa}$. As $\expval{\kappa}=0$ and usually $\kappa\ll 1$ holds, this is a good approximation for cosmic shear analyses. However, in \gls{ggl}, the convergence $\kappa$ is systematically larger than 0 due to the \gls{esd} of the lens galaxy. This effect can lead to an artificial boost of the \gls{ggl} signal by up to 5\% on the small scales, potentially biasing cosmological constraints by up to $\gtrsim 0.3\sigma$ \citepalias{Lange:2024}, if these small scales are used in the analysis.\footnote{We note that this is not the case for most cosmological parameter inferences, as small scales are usually disregarded to avoid contamination by hard-to-quantify systematic effects.} For a comparison between imaging surveys, this is less of a concern. The convergence $\kappa$ is defined as the surface mass density $\Sigma$ divided by the critical surface mass density $\Sigmacrit$. The former is a pure physical property that depends only on the lens galaxy, so it should be the same for all three lensing surveys. While $\Sigmacrit$ depends on the source redshift distribution, this dependency is fairly small. The reduced shear approximation thus impacts all \gls{ggl} measurements in a very similar way, the difference between different source redshift bins of \gls{hsc} is generally at the sub-percent level for all source-lens combinations we utilize in this analysis; aside from a few percent-level differences on the smallest scales \citepalias[compare Fig.~12 of][]{Lange:2024}.

    \subsection{The lensing estimator and its correction terms}
    \label{sec:GGL_measurements:estimator}
        We use the public and well-tested code dsigma\footnote{\url{https://github.com/johannesulf/dsigma}} \citep{2022ascl.soft04006L} to measure the \gls{ggl} signal (both \gls{esd} $\ds$ and tangential shear $\gamma_\mathrm{t}$).

        Given a catalogue of lens and source galaxies, we estimate the raw average tangential shear $\gamma_{\mathrm{t,raw}}$ via Eq.~\eqref{eq:gammat_raw_estimator}. To estimate the raw \gls{esd}, we multiply the tangential shear by the critical surface mass density $\Sigmacrit$:
        \begin{equation}
            \ds_\mathrm{raw}(\rp) = \frac{\sum_\mathrm{ls}w_\mathrm{tot}w_\mathrm{ls}\epsilon_\mathrm{t}\Sigmacrit{}_\mathrm{,ls}}{\sum_\mathrm{ls}w_\mathrm{ls}} \; ,
            \label{eq:estimator_raw_esd}
        \end{equation}
        where $\Sigmacrit{}_\mathrm{,ls}=\Sigmacrit(\zl,\zs)$ is the critical surface mass density for the lens-source galaxy pair, and $r_{\rm p}$ the comoving distance from the lens galaxy projected perpendicular to the line-of-sight direction. One also usually sets \mbox{$w_\mathrm{ls} = \Sigmacrit^{-2}(\zl,\zs)\,w_\mathrm{s}$}, as this minimizes the impact of shape noise.
        
        We again emphasize that $\epsilon_\mathrm{t}$ is in practice not an unbiased estimator of the tangential shear and requires correction through calibration factors discussed below.

        \subsubsection{Random subtraction} 
        \label{sec:GGL_measurements:random_subtraction}
            While the \gls{desiy1} footprint spans a substantial fraction of the sky, the overlap areas with the weak lensing surveys are comparatively small, and \gls{ggl} measurements are thus contaminated by imprints of large-scale density fluctuations in the \gls{lss}, that are comparable in size to the survey window function. To address this, we calculate $\ds_\mathrm{random}$ and $\gammat{}_\mathrm{,random}$, the \gls{ggl} signal around the randoms (as described in Sect.~\ref{sec:GGL_measurements:estimator}) via Eqs.~\eqref{eq:gammat_raw_estimator} and \eqref{eq:estimator_raw_esd}, respectively, where we replace galaxy positions by random positions and $w_\mathrm{tot}$ with $w_\mathrm{tot,r}$. We then define:
            \begin{align}
                \ds(\rp) = {}&{} \ds_\mathrm{raw}(\rp) - \ds_\mathrm{random}(\rp) \notag\\
                \gammat(\theta) = {}&{} \gammat{}_\mathrm{,raw}(\theta) - \gammat{}_\mathrm{,random}(\theta) \; .
            \end{align}
            This effort reduces the sample variance of the measurements by removing contamination from the large-scale fluctuations, it also thereby decreases the covariance of \gls{ggl} measurements \citep{Singh:2017}. Last but not least, a random subtraction also removes additive shear measurement and joint lens-source selection biases\footnote{If lens and source galaxies are observed with the same imaging survey (such as in \gls{desi} x \gls{des}), this introduces correlated selection biases that are removed via random subtraction. But even if source galaxies are observed with a different imaging survey, some correlated selection biases (such as extinction and stellar contamination) remain.}.
        
        \subsubsection{Boost factor correction} 
            \label{sec:lensing_measurements:lensing_estimator:boost_factor}
            \gls{ggl} measurements on small scales are impacted by the fact that, due to source-lens clustering, an excess of source galaxies is expected around the lens galaxy. This means that the expected redshift distribution $n(z)$ of source galaxies around a lens does not equal the average $n(z)$ of source galaxies that has been determined by the lensing survey collaborations. The excess of source galaxies with redshift similar to the lens leads to an under-estimation of the true lensing signal, as these galaxies exhibit little to no lensing. One can correct the measurements for this dilution by a boost factor $b$ to $\ds = b\,\ds_\mathrm{raw}$ and $\gammat = b\,\gammat{}_\mathrm{,raw}$, where $b$ is determined via
            \begin{equation}
                b(r_p) = \frac{\sum_r w_{\rm sys}}{\sum_l w_{\rm sys}} \frac{\sum_{ls} w_{\rm sys} w_{ls}}{\sum_{rs} w_{\rm sys} w_{rs}} \, .
                \label{eq:boost_factor_correction}
            \end{equation}
            Here, $w_{rs}$ are the equivalent lens-source weights for the randoms and the sum extends over all lens-source pairs whose projected distance is within the specified $r_p$ bin.
            
            Unfortunately, this measurement is biased by several unknown factors, such as blending \citep{Simet:2015}, contamination of background galaxies \citep{Leauthaud:2017, Everett:2022}, and source magnification, where it is unknown whether these biases are negligible or not. The first two of these effects depend on the telescope, instrument, and data reduction pipeline. This means that a boost factor correction potentially introduces a more complex, and less homogeneous, bias in the measurements, complicating a comparison between different lensing surveys (which is the focus of this work). Especially for source-lens bin combinations with little to no overlap, the boost factor can become noisy and introduce unphysical corrections (compare Fig.~\ref{fig:boost_factor} in App.~\ref{app:boost_factor}). Therefore, we are not employing any boost factor corrections. \citetalias{Lange:2024} show that this boost factor induces percent-level changes on the datavector, but stay well below a $\Delta\chi^2=0.1$ significance level for all scales above $1\,h^{-1}\Mpc$. More importantly, they show that the difference of the boost factors between the lensing surveys is negligible. We compare our estimated boost factors to the ones from \citetalias{Lange:2024} in App.~\ref{app:boost_factor} and observe a reasonable agreement.

            We note that the reasoning for our scenario (comparison between lensing surveys at the data-level, and very conservative exclusion of source-lens bins with any substantial overlap) does not necessarily extend to uses of these measurements in cosmological analyses where the \gls{ggl} signal is compared to models. We thus also provide measurements that include boost factor corrections in our published dataset.

        \subsubsection{Photometric redshift dilution} 
            \label{sec:GGL_measurements:correction_choices_for_lensing_surveys:photometric_redshift_dilution}
            When measuring the \gls{esd} via the estimator in Eq.~\eqref{eq:estimator_raw_esd}, we can estimate the critical surface mass density for each individual lens-source pair, in case individual photometric redshift estimates are available. However, the combination of all individual photometric redshift estimates is not equal to the redshift distribution of a tomographic bin. For example, Fig.~\ref{fig:all_nofz_plot} shows the redshift distributions of \gls{kids}, where the source tomographic bins are determined by best estimates for photometric redshifts falling within the bin boundaries $[0.1,0.3,0.5,0.7,0.9,1.2]$, but one can clearly see that the combined redshift distribution of the respective tomographic bins extends beyond these boundaries. To correct for this effect, we use a calibration catalogue that has both photometric redshift estimates $z$ and high-quality (usually spectroscopic) redshifts $\hat{z}$, associated with calibration weights $w_\mathrm{c}$. This calibration catalogue is the same as the one employed for photometric redshift calibrations, which implies that this catalogue is representative of the whole lensing catalogue for all relevant galaxy properties (after accounting for the weights $w_\mathrm{c}$). We calculate a multiplicative correction factor $\fbias$ via
            \begin{align}
                    \Sigmacrit{}_\mathrm{,ls} &= \Sigmacrit (\zl, \zs) \times \frac{\sum_c w_c w_{ls}}{\sum_c w_c w_{ls} \Sigmacrit(\zl, \zs) / \Sigmacrit (\zl, \hat{z}_\mathrm{s})} \notag\\
                    &= \Sigmacrit (\zl, \zs) \times \fbias (\zl)\, .
                \label{eq:f_bias}
            \end{align}
            However, in weak lensing surveys, source galaxies usually do not have spectroscopic redshifts, and while some weak lensing surveys provide individual photometric redshift estimates for galaxies, the primary (and best-validated) data product is a redshift distribution for a tomographic sample of galaxies, in which case $\Sigmacrit(\zl,\zs)$ is not defined for individual source galaxies. One can thus define the effective critical surface mass density for a population of source galaxies with normalized effective redshift distribution $n_\mathrm{eff}(z)\, \dd z$ via
            \begin{equation}
                \Sigmacrit{}_\mathrm{,ls} (\zl) = \left[ \int_0^\infty \dd \zs\, n_\mathrm{eff}(\zs) \Sigmacrit^{-1} (\zl, \zs) \right]^{-1} \, ,
                \label{eq:sigma_crit_eff}
            \end{equation}
            and use this quantity for all source galaxies of a tomographic bin in Eq.~\ref{eq:estimator_raw_esd}. While this prevents the use of shape-noise minimizing weights \mbox{$w_\mathrm{ls} = \Sigmacrit^{-2}(\zl,\zs)\,w_\mathrm{s}$}, it circumvents the need for an $f_\mathrm{bias}$ calculation via the spectroscopic calibration catalogue.

            As we believe the tomographic redshift distribution to be the best-validated data products for \gls{kids}, \gls{des}, and \gls{hsc}, we use the photometric redshift distributions detailed in Fig.~\ref{fig:all_nofz_plot} to calculate the source-lens weight $w_\mathrm{ls}$ (meaning effectively $\fbias=1$ holds); for \gls{sdss} (whose photometric redshift calibration has been performed much earlier and is thus not as robust as the ones from the Stage-III weak lensing surveys) we use individual best-estimates for the photometric redshift and correct the measurement by $f_\mathrm{bias}$, as discussed in Sect.~\ref{sec:GGL_measurements:correction_choices_for_lensing_surveys:sdss}.

            We note that both methods (point estimates and effective critical surface density) have been used in the literature. The former estimator has been derived to maximize S/N in the shot-noise limited regime and can yield a slightly higher S/N in certain applications, while effectively having a lower S/N at low lens redshifts \citep{2018MNRAS.478.4277S}. However, the point estimates are inherently biased, which the photometric dilution correction is supposed to correct. Previous studies \citepalias[in particular][]{Leauthaud:2022} have not found evidence for systematic differences between these two approaches. As photometric redshift calibrations have become more and more complex, we believe that using average photo-z distributions and an effective critical surface density is the more conservative choice.
        
        \subsubsection{Lens magnification bias}
            Gravitational lensing by the \gls{lss} not only leads to a distortion of galaxy images, but also to a (de)magnification $\mu$ of their flux. This effect is predominantly relevant for lens galaxies: In a region of high magnification $\mu$, which is usually behind an overdensity in the \gls{lss}, the number density of lens galaxies decreases with $1/\mu$, but at the same time each lens galaxy is magnified by $\mu$, so its individual probability of being observed increases. This means that their probability of being observed is correlated with the \gls{lss} between galaxy and observer, which in turn exerts a shear on background galaxies. This correlation can be quantified via \citep[see e.g.][]{2020A&A...638A..96U}
            \begin{align}
                \ds_{\rm lm} (\rp) = {}&{} 2\,\Sigmacrit(\overline{\zl}, \overline{\zs}) (\alpha_\mathrm{l} - 1) \gamma_{\rm LSS} (\rp / \Dcom (\overline{z_l}), \overline{z_l}, \overline{z_s}) \notag\\
                \Delta \gammat = {}&{} 2 (\alpha_l - 1) \gamma_{\rm LSS} (\theta, \overline{z_l}, \overline{z_s}) \, ,
                \label{eq:lens_magnification}
            \end{align}
            where $\overline{\zl}$ and $\overline{\zs}$ are the effective lens and source redshifts, defined as the mean redshifts of all lens-source pairs weighted by $w_\mathrm{tot}w_{ls}$, and $\Dcom$ is the comoving angular diameter distance. Assuming a fraction $f_\mathrm{l}$ of lens galaxies passes the selection cut, then $\alpha_\mathrm{l}$ parametrizes the behavior of $f_\mathrm{l}$ to magnification effects via 
            \begin{equation}
              \alpha_\mathrm{l} = \left. \frac{\dd \ln f_\mathrm{l}}{\dd \mu} \right|_{\mu = 1} \, .
              \label{eq:alpha_defn}
            \end{equation}
            The quantity $\gamma_\mathrm{LSS}$ parametrizes the average lensing by the \gls{lss} and can be calculated via \citep{Lange:2024}
            \begin{align}
                \gamma_{\rm LSS} (\theta, z_l, z_s) = &\frac{9 H_0^3 \Omega_{\rm m, 0}}{4 c^3} \int\limits_0^\infty \dd \ell J_2(\ell \theta) \ell \int\limits_0^{\zl} \dd \zf \notag\\
                &\frac{(1 + \zf)^2}{2 \pi} \frac{H_0}{H(\zf)} \frac{\Dang(\zf, \zl) \Dang(\zf, \zs)}{\Dang(\zl) \Dang(\zs)}\notag\\
                &P_\mathrm{m}\left( \frac{\ell + 1/2}{(1 + \zf) \Dang(\zf)}, \zf \right) \, ,
            \end{align}
            where $\Omega_{\mathrm{m},0}$ is the matter density parameter, $J_2$ the second order Bessel function of first kind, $P_\mathrm{m}$ the non-linear matter power spectrum and $H_0$ is the Hubble parameter. We calculate the parameter $\alpha_\mathrm{l}$ for both the \gls{bgs} and the \gls{lrg} sample using the methodology developed by \citet{Wenzl:2024} and apply this correction to our measurements. The values can be found in Tab.~\ref{tab:magnification_bias_table}, and a more rigorous description of our measurements is in App.~\ref{app:magnification_bias}.
            
            A less studied effect is the source magnification bias \citep[see][]{Unruh:2020}, however, \citetalias{Lange:2024} demonstrate that this effect is negligible for current-generation surveys. The lens magnification bias is more dominant at higher redshifts and can reach up to 16\% for the second LRG bin.
            
            All measurements we present, if not otherwise stated, have been corrected for lens magnification bias. The measurements we publish on \url{https://doi.org/10.5281/zenodo.22914838}, however, do not include this correction. This is because, while the determined $\alpha_\mathrm{l}$ value is independent of the assumed cosmology, the quantity $\gamma_\mathrm{LSS}$ is cosmology-dependent. While this dependency is relatively weak (within reasonable choices for cosmological parameters), it is advantageous to forward-model a cosmology-dependent magnification bias that also marginalizes over the uncertainty of the estimated $\alpha_\mathrm{l}$-values.
            
            \begin{table}
                \centering
                reg\begin{tabular}{l|c|c}
 & BGS BRIGHT & LRG\\
\hline
Bin 1  & $0.94 \pm 0.01\,{\color{blue} \pm 0.00 }$ & $2.54 \pm 0.02\,{\color{blue} \pm 0.03 }$\\
Bin 2  & $1.62 \pm 0.01\,{\color{blue} \pm 0.00 }$ & $2.49 \pm 0.01\,{\color{blue} \pm 0.12 }$\\
Bin 3  & $2.19 \pm 0.02\,{\color{blue} \pm 0.01 }$ & $2.58 \pm 0.01\,{\color{blue} \pm 0.48 }$\\
\end{tabular}

                \caption{Coefficients $\alpha_\mathrm{l}$ measured for the magnification bias. We quote the uncertainties of the numerical derivative to compute $\alpha_\mathrm{l}$ (black), and the systematic uncertainty caused by the assumption that all galaxies follow a deVaucouleurs profile (blue). We discuss this in detail in App.~\ref{app:magnification_bias}.}
                \label{tab:magnification_bias_table}
            \end{table}
            \label{sec:methods:lensing_measurements:magnification_bias}
        
        \subsubsection{Multiplicative shear bias correction}
            Measuring the shape of tiny galaxy images is no trivial task, and the ellipticity measurements are usually biased. This is corrected by multiplying our measurements with a best estimate for the multiplicative shear measurement bias $\mathcal{M}$ to
            \begin{equation}
                \ds = \bar{\mathcal{M}}\left[\ds_\mathrm{raw}-\ds_\mathrm{random}\right]\; .
            \end{equation}
            The quantity $\mathcal{M}$ is determined differently for each weak lensing survey, and we will describe the individual methods below.
            
        \subsubsection{Combined corrections} 
            Combining all corrections, we arrive at
            \begin{align}
                \ds(\rp) = {}&{} \bar{\mathcal{M}}\left[b\fbias \ds_\mathrm{raw}(\rp)-\ds_\mathrm{random}(\rp)\right] -\Delta\Sigma_\mathrm{lm}(\rp) \notag\\
                \gammat(\theta) = {}&{}\bar{\mathcal{M}}\left[b \gammat{}_\mathrm{,raw}(\theta)-\gammat{}_\mathrm{,random}(\theta)\right]-\gammat{}_{,lm}(\theta) \; ,
            \end{align}
            where $\fbias\equiv 1$ holds when the effective critical surface mass density is used for Eq.~\eqref{eq:estimator_raw_esd}, and $b\equiv 1$ holds as we do not employ boost factor corrections. We note that the correction for the magnification bias is not applied in the published measurements, as it is usually preferred to forward model its impact in a cosmological analysis.

    \subsection{Correction choices for lensing surveys}
    \label{sec:GGL_measurements:correction_choices_for_lensing_surveys}

        Both the multiplicative shear bias $\mathcal{M}$ and the photometric redshift dilution $f_\mathrm{bias}$ depend on properties of the imaging survey that provides the source galaxies. Therefore, our methods to calculate these correction terms differ for each lensing survey.
        
        \subsubsection{HSC} 
            The galaxy shape measurements for the HSC third-year weak lensing analysis (HSC-Y3) were conducted using the pipeline detailed in \cite{Bosch:2018}, with galaxy ellipticities computed via the {\sc HSM} approach \citep{Hirata:2003}. Due to the specific definition of ellipticity used, the mean tangential ellipticity is related to the intrinsic per-component ellipticity dispersion $e_{\mathrm{rms}}$ as $\langle e_t \rangle = 2 (1 - e_{\mathrm{rms}}^2) \gammat$. Residual multiplicative shear bias $m$ was calibrated in \cite{arxiv:2107.00136} such that $\langle \hat{e} \rangle = 2 (1 - e_{\mathrm{rms}}^2) (1 + m) \langle e \rangle$, by first characterizing the shape dispersion $e_{\mathrm{rms}}$ as a function of signal-to-noise $S/N$ and resolution $R_2$, and then estimating $m$ in discrete bins of $S/N$ and $R_2$, interpolating to obtain per-object values.
            
            From the above, the shear calibration factor $\mathcal{M}$ for the HSC weak lensing catalog is computed via:
            \begin{equation}
            \mathcal{M}_{\mathrm{HSC}} (\rp) = \frac{1}{2 \mathcal{R} (\rp) \left[ 1 + \overline{m} (\rp) \right]} \, ,
            \label{eq:mbias_HSC}
            \end{equation}
            where the shear responsivity $\mathcal{R} (\rp)$ is:
            \begin{equation}
            \mathcal{R} (\rp) = 1 - \frac{\sum_\mathrm{ls} w_\mathrm{tot} w_\mathrm{ls} e_{\mathrm{rms}}^2}{\sum_\mathrm{ls} w_\mathrm{tot} w_\mathrm{ls}} \; ,
            \end{equation}
            and the weighted mean multiplicative bias $\overline{m} (\rp)$ is given by:
            \begin{equation}
            \overline{m} (\rp) = \frac{\sum_\mathrm{ls} w_\mathrm{tot} w_\mathrm{ls} m}{\sum_\mathrm{ls} w_\mathrm{tot} w_\mathrm{ls}} \,.
            \label{eq:HSC_m}
            \end{equation}
            \gls{hsc} provides individual redshift probability distributions $p(z)$ for each source galaxy. The best approach would arguably be to calculate an effective critical surface density (Eq.~\ref{eq:sigma_crit_eff}) for each lens-source pair, but that would be computationally extremely demanding. We instead opt to calculate the effective critical surface density for the fiducial redshift distributions of the cosmology analyses from \gls{hsc}. In the absence of strong correlations between (weighted) shapes and redshift, we expect this to be unbiased.

            \gls{hsc} also employs a selection bias correction to mitigate the impact of a selection criterion that correlates with the true lensing shear or the PSF anisotropy. The additive and multiplicative selection bias ($\asel$ and $\msel$, respectively) are calibrated from image simulations \citep[compare Eqs.~(18) and (19) in][]{arxiv:2304.00703} and quantify the impact of the size and magnitude limit on the shear estimation. They are mitigated via \citep[see][]{arxiv:2304.00703,2023PhRvD.108l3518L}
            \begin{align}
                \ds = {}&{} \frac{1}{1+\mselhat}\left(\ds_\mathrm{raw}-\aselhat\ds^\mathrm{psf}\right)   
                \label{eq:hscy3_selection_bias} \\
                \ds^\mathrm{psf} = {}&{} \frac{\sum_{ls} w_{ls} e_{t,ls}^\mathrm{psf}}{\sum_{ls} w_{ls}} \; ,
            \end{align}
            where $e^\mathrm{psf}$ is the PSF ellipticity at the position of the source galaxy.
        
        \subsubsection{KiDS} 
            The KiDS-1000 data set, which represents the fourth data release of the Kilo-Degree Survey (KiDS), employs a self-calibrating variant of the {\sc lensfit} algorithm \citep{Miller:2007, Miller:2013, FenechConti:2017} to measure galaxy ellipticities. The residual multiplicative shear biases $m$ for this data set were evaluated for each tomographic redshift bin \citep{Hildebrandt:2020, Giblin:2021}, considering source galaxies in aggregate. Consequently, the shear calibration factor for KiDS-1000 takes the form:
            \begin{equation}
            \mathcal{M}_{\mathrm{KiDS}} (\rp) = \frac{1}{1 + \overline{m} (\rp)} \, ,
            \end{equation}
            with the weighted mean multiplicative bias $\overline{m}$ calculated in the same manner as for HSC (i.e., via Eq.~\ref{eq:HSC_m}).
            
            We use the public photometric redshift distributions to estimate an average effective critical surface mass density via Eq.~\eqref{eq:sigma_crit_eff} and set $f_\mathrm{bias}\equiv 1$.

        \subsubsection{DES} 
            The DES Year 3 (DES Y3) analysis measures source galaxy ellipticities in the $riz$ bands using the {\sc NGMIX} code \citep{Sheldon:2015}. The {\sc METACALIBRATION} algorithm \citep{Huff:2017, Sheldon:2017} is employed to estimate the shear response matrix $\mathbf{R}$, defined as
            \begin{equation}
            R_{ij} = \frac{\partial{\hat{e}}_i}{\partial e_j},
            \end{equation}
            by artificially perturbing the intrinsic ellipticities of galaxies in the survey data. Additional biases stem from selection effects \citep{Gatti:2021} and blending \citep{MacCrann:2022}. For DES, the shear calibration factor is given by
            \begin{align}
            \mathcal{M}_{\mathrm{DES}}^{-1} = {}&{} (1+\overline{m}_s)(1+\overline{R}_{\mathrm{t}} + \overline{R}_\mathrm{sel}) \notag\\
             = {}&{} \left(1+ \frac{\sum_\mathrm{ls} w_\mathrm{tot} w_\mathrm{ls} R_{\mathrm{t}}}{\sum_\mathrm{ls} w_\mathrm{tot} w_\mathrm{ls}} + \frac{\sum_\mathrm{ls} w_\mathrm{tot} w_\mathrm{ls} R_{\mathrm{sel}}}{\sum_\mathrm{ls} w_\mathrm{tot} w_\mathrm{ls}} \right) \notag\\
             {}&{}\times \left(1+\frac{\sum_\mathrm{ls} w_\mathrm{tot} w_\mathrm{ls} m_s}{\sum_\mathrm{ls} w_\mathrm{tot} w_\mathrm{ls}}\right),
            \end{align}
            where $R_{\mathrm{t}}$ is the projection of the response matrix $\mathbf{R}$ onto the vector connecting the lens and source, given by
            \begin{equation}
            \begin{split}
            R_{\mathrm{t}} =& R_{11} \cos^2 (2 \phi) + R_{22} \sin^2 (2 \phi)\\&+ (R_{12} + R_{21}) \sin (2 \phi) \cos (2 \phi),
            \end{split}
            \end{equation}
            with $\phi$ denoting the polar angle of the source in the source coordinate system. We use $R_\mathrm{sel}$ to denote the \textsc{METACALIBRATION} selection response and $m_s$ to describe the multiplicative shear bias induced by blending \citep{MacCrann:2022}. Unlike HSC and KiDS, the DES shear response factors are computed for each galaxy individually. However, as the individual estimates for $\mathbf{R}$ and $R_t$ can be quite noisy, and $\langle x \rangle^{-1} \neq \langle x^{-1} \rangle$ in general, it is crucial to calculate ensemble-averaged estimates for $R_{\mathrm{t}}$ as correction factors for the raw $\gammat$ and $\ds$ values, rather than correcting each source galaxy using its individually estimated $R_{\mathrm{t}}$.
            
            As for \gls{kids}, we use the public redshift distribution of \gls{des} to calculate an effective critical surface mass density and set $f_\mathrm{bias}\equiv 1$.
        
        \subsubsection{SDSS}
        \label{sec:GGL_measurements:correction_choices_for_lensing_surveys:sdss}
            The shape measurements of \gls{sdss} need to be corrected for the shear responsivity and multiplicative shear bias in the same way as \gls{hsc}. The only difference is that the shear responsivity $\mathcal{R}$ and the multiplicative shear bias $\bar{m}$ are computed globally for the entire catalogue to $1+\bar{m}=1.04\pm 0.02$ and $\mathcal{R}=0.87$ \citep{Mandelbaum:2012,Mandelbaum:2018}, and we apply these corrections to all \gls{sdss} measurements.
            
            In contrast to the other three lensing surveys, we use the individual best-fit photometric redshifts to calculate the critical surface mass density $\Sigmacrit$. To calculate the $f_\mathrm{bias}$-correction, we have received a table (R.~Mandelbaum, priv.~comm.) detailing $f_\mathrm{bias}$ as a function of lens redshift. We then calculate the mean $f_\mathrm{bias}$ for each sample of lens galaxies from that.

    \subsection{Measurement setup}
    \label{sec:GGL_measurements:measurement_setup}
        Before performing the actual measurements, we cut both the \gls{desiy1} and the lensing catalogues to their respective overlaps. We then measure the gravitational lensing signal $\ds$ in 15 radial bins between $0.08\,h^{-1}\Mpc$ and $80\,h^{-1}\Mpc$ and employ subsequent scale cuts (described below and visualized in Fig.~\ref{fig:deltasigma_datavector}) to minimize the impact of systematic biases. We perform two types of measurements: For the source redshift tests (and the subsequent cosmological analyses) we perform a \emph{tomographic} measurement where we measure the lensing signal for each individual combination between \gls{desiy1} lens bins and KiDS/HSC/DES source bins. 
        For the lens homogeneity tests (described in Sect.~\ref{sec:amplitude_tests:outliers_in_lens_homogeneity}), we employ a \emph{non-tomographic} measurement setup, where we combine all source galaxies of the allowed tomographic bins for a single measurement for each lens bin.
        
        In both cases, we only allow combinations of source-lens bins that do not have significant overlap in redshift, such as not to contaminate our measurements by systematic effects. The combinations of allowed source-lens bin combinations are detailed in Fig.~\ref{fig:all_nofz_plot}. We disregard all source-lens combinations where more than 10\% of sources overlap with the redshift range of the lens galaxies, which is denoted in Fig.~\ref{fig:all_nofz_plot} as the `optimistic source-lens cut'. As we do not model any systematic effects, we further narrow our source-lens combinations such that the data from \citetalias{Lange:2024} predicts a contamination of all lens amplitudes $\ADS$ by less than 3\%, corresponding to the `conservative source-lens cut' in Fig.~\ref{fig:all_nofz_plot}.
        
        Considering the geometry of the lensing surveys, we place further restrictions on the radial ranges we use for our analysis. To minimize the impact from blending between source and lens galaxies, we disregard all radial bins where the lens-source pairs are on average less than $0.\!'5$ apart. We place our maximum scale at the range where the ratio of measured source-lens pairs is more than half of what would be measured in an infinitely large survey with the same galaxy density, as above these scales the estimation of the covariance becomes unreliable. For a detailed discussion and the numerical values of these cut-offs we refer to Sect.~\ref{sec:GGL_measurements:covariance}. In this work, where we compare lensing amplitudes between different surveys, we only use points that satisfy the most conservative cut-off (given by \gls{hsc}) to ensure consistency between the measurements.
        
        In our analysis, we further separate the measurements into `small scales' ($\rp\leq 1\,h^{-1}\Mpc$), `large scales' ($\rp > 1 \,h^{-1}\Mpc$) and `all scales'. While the measurement on small scales has the smallest statistical uncertainties, it is contaminated in particular by the boost factor (compare Sect.~\ref{sec:GGL_measurements:estimator}), which can not be removed, although we minimize its impact by following a strict criterion for lens-source separation. Discrepancies between lensing surveys that only occur on small scales could be attributed to this.
        
        To calculate $\ds$, we need to fix a fiducial cosmology in order to calculate $\Sigmacrit$ and transform angular separations into projected comoving distances. Our fiducial parameters are the best-fit results from the Planck-mission \citep{2020A&A...641A...6P}. We test the impact of this choice by comparing with measurements that use alternative cosmologies in App.~\ref{app:alternaive_cosmologies}.

    \subsection{Covariance Computations}
    \label{sec:GGL_measurements:covariance}

        Unless otherwise specified, we use the analytical covariances for our GGL measurements as described and validated in \citetalias{Yuan:2024}, following methods previously presented by \cite{Singh:2017, Shirasaki:2018, Dvornik:2018, 2020A&A...642A.158B}. We will give a brief summary of their findings and encourage the reader to consult \citet{Yuan:2024} for a more in-depth description and the validation of statements that will be made here. These Gaussian covariances include sample variance, shape and shot noise and mixed contributions. In particular, they include correlations between different lensing surveys, correlated shape noise, and corrections taking into account the exact number of galaxy pairs in each bin, given the survey footprint.
        We excluded non-Gaussian contributions to the covariance, given that these are negligible compared to the other terms across all scales for current survey parameters.
        On small scales (below several $h^{-1}$ Mpc, depending on lens and source configurations), usually non-Gaussian terms would substantially contribute to the sample covariance. However, on these scales the covariances are shape noise dominated, so the Gaussian approximation is still sufficiently accurate \citep[see][for details]{Joachimi:2021,Yuan:2024}. We generated covariances for a full data vector including lens and source samples and separation bins.  We also included the cross-covariances between different weak lensing surveys, where relevant for a joint analysis in which surveys overlap and share common sources.  When computing the analytical covariance we assumed a non-linear matter power spectrum generated in the fiducial DESI cosmology, the linear bias factors determined for early DESI data by \cite{Prada:2023}, the effective source number densities and shape noise provided by the weak lensing surveys \citep{Hikage:2019, Giblin:2021, Amon:2022}, and effective lens densities computed from each DESI dataset within the lensing survey overlap area.
        
        \citetalias{Yuan:2024} demonstrated that the geometrical effects of the survey window function have a significant effect on the GGL covariance on scales approaching the width of the survey footprint, where the number of accessible source-lens pairs is reduced by the survey boundaries.  This effect is much more significant for the B-mode measurements (of the ``cross'' component $\Delta \Sigma_\times$) than for E-mode measurements ($\Delta \Sigma$).  As described in Appendix B of \citetalias{Yuan:2024}, we used an ensemble of lognormal simulations to calibrate these effects, and scaled our analytical covariances by this correction factor.  Specifically, the analytical prediction of the error is increased by $10\%$ ($30\%$) for E-mode (B-mode) measurements at angular separations for which $50\%$ of source-lens pairs are ``lost'' due to the boundary effect.  For the overlaps between the (KiDS-1000, DES-Y3, HSC-Y3) and DESI-Y1 datasets, this separation is $(3.39^\circ, 4.29^\circ, 2.18^\circ)$, respectively. After these corrections, \citetalias{Yuan:2024} showed that the analytical covariance determination agreed with the error estimate using an ensemble of mock catalogues generated with the Buzzard DESI-Lensing simulations \citep{DeRose:2019,arXiv:2105.13547,2024arXiv241212548B,Lange:2024} within around $10\%$ across scales, redshifts and surveys. We also compared the diagonal of the analytical covariance to jackknife estimates of the data, also achieving an agreement within $10\%$. A more rigorous covariance validation is beyond the scope of this paper, as data jackknife covariance have very noisy off-diagonal terms, and comparisons of Fisher information are hampered by the bias introduced in inverting sample covariance matrices \citep[see e.g.][]{Hartlap:2007,Sellentin:2015}.
        
        Analytical covariance estimates become significantly less accurate for survey footprints which are non-contiguous, i.e., possess significant structure on the scales of interest \citep{Kilbinger:2004, Sato:2011, Shirasaki:2019, Friedrich:2021, Joachimi:2021, Yuan:2024}.  Hence, for our measurement comparison tests in which the survey footprint is sub-divided in a complex and non-contiguous manner by an observational or galaxy selection property, we utilised a jack-knife covariance instead of an analytical covariance. We will note these cases below, when they occur.

\section{Lensing amplitude tests}
\label{sec:amplitude_tests}
    In this work, we perform two tests to investigate the fidelity of our measurements and the underlying data. We call our first test the \emph{lens homogeneity test}, the second test is the \emph{source redshift test}. We describe how we perform these tests below.

    \subsection{Fitting a lensing amplitude}
    \label{sec:amplitude_tests:lensing_amplitude}
        To identify discrepancies in the measurements of the excess surface density from various surveys, one could directly compare the measurements of $\ds(\rp)$. However, this is not an optimal approach as this type of comparison lacks interpretability and is not necessarily very robust.
        A more robust method involves utilizing a \emph{matched filter}, which relies on understanding the shape and covariance of $\ds(r)$. Fitting a model's amplitude to the data is equivalent to finding an optimal linear combination of the data vector, resulting in a single value of interest. A $\chi^2$-test can then be conducted using this value.

        For a data vector $d$ with covariance matrix $\mathsf{C}$, a linear combination $A=w^{\mathsf{T}}d=\sum_i w_i d_i$ can be defined using a weight vector $w$. Assuming Gaussian statistics, the variance of this combination is $\sigma_{\rm A}^2=w^\mathsf{T}\mathsf{C}w=\sum_{ij}w_i w_j C_{ij}$. When the noiseless signal's true shape is known as $t$, the optimal linear compression is achieved by the matched filter amplitude $A$ with weights $w\propto\mathsf{C}^{-1}t$.
        In our analysis, $d$ represents the $\Delta\Sigma$ measurements, and we will call the respective lensing amplitude $\ADS$. Most major known lensing systematics, such as shear or redshift calibration errors constitute multiplicative errors in the lensing signal to first order, implying that the difference in the measurements has a radial shape similar to $\Delta\Sigma$ itself, and the matched filter technique is ideally suited to detect these discrepancies. We note that the sensitivity of the test depends on this assumption (i.e. we may be unable to detect differences between data vectors that follow a significantly different radial profile), but its validity does not, meaning that a measured difference between lensing amplitudes necessarily indicates a difference in the data vectors. We explain this argument below and mathematically prove it in App.~\ref{sec:app:matched_filter_template}.
        
        This method requires a reference model ($\dsref$) for the data vector. While the model doesn't need to be perfect, it should reasonably match the observed lensing signal's shape. To achieve this, we use galaxy catalogs derived from AbacusSummit simulations, which were calibrated using a halo occupation distribution (HOD) fit to three-dimensional clustering measurements akin to the method presented in \citet{2024MNRAS.532..903S}. These simulations provide the reference $\ds$ datavectors for the \gls{bgs} and \gls{lrg} galaxy samples.

        Assuming that the measurements are uncorrelated, the covariance we use for the data compression is $\sum_j\mathsf{C}_j$, where $j$ iterates over all measurements of $\ds$ of a single lens bin (i.e.~all measurements that are shown in a single panel of Figs.~\ref{fig:deltasigma_amplitude_ntile_split}, \ref{fig:source_redshift_slope}, etc.). We use a single covariance to construct the weights for all measurements, as using a matched filter amplitude with specific covariance for each survey ($\bm{w}_j = \mathsf{C}_j^{-1}\dsref$), could lead to non-zero amplitude differences for consistently calibrated surveys with different covariance matrix structures if there is a scale-dependent offset between $\dsref$ and the correct model. For measurement $j$ and radial range (small, large, and all radii), we determine a matched amplitude $A_j = \frac{w^{\mathsf{T}}\Delta\Sigma_j}{w^{\mathsf{T}}\dsref}$ and its uncertainty $\sigma_j^2=\frac{w^\mathsf{T} \mathsf{C}_j w}{(w^{\mathsf{T}}\dsref)^2}$, where $w = \left(\sum_j \mathsf{C}_j\right)^{-1}\dsref$. We report $A-\overline{A}$, where $\overline{A}=\left(\sum_j A_j/\sigma_j^2\right) / \left(\sum_j 1/\sigma_j^2\right)$, which is insensitive to amplitude differences between the lensing signal and the prediction from the AbacusSummit simulations.
        
        The validity of our tests does not depend on the model's shape, but the sensitivity does, as we detail in App.~\ref{sec:app:matched_filter_template}. If the lensing measurements show a difference that has a significantly different shape than the $\dsref$-profile, the conversion to lensing amplitudes $\ADS$ might still yield the same result.

    \subsection{Lens homogeneity test}
    \label{sec:amplitude_tests:analysis_considerations_homogeneity}
        When comparing the signal between different lensing surveys, it is crucial to ensure that potential differences between the measurements are not due to residual biases in the lens galaxies observed by \gls{desi}. To investigate whether we correctly account for these biases, we perform our \emph{lens homogeneity test} by splitting our lens galaxies in the overlap regions into four quantiles, determined by the value of a quantity that might bias the \gls{desiy1} targets, and comparing the measurements between these quantiles. We identify the \gls{desiy1} completeness, the local stellar density, and the seeing and depth of the legacy imaging survey as potential systematics.
        
        The potentially largest impact stems from the fact that \gls{desi} has a finite amount of fibers that have a finite thickness. This means that it is impossible to observe all galaxies in the field of view, and that one can not observe galaxies that are close together due to fibre collisions. This effect is partially mitigated by the fact that \gls{desi} passes over the same area multiple times, but it can not be neglected. The fiber incompleteness weights were designed to account for this effect \citepalias[compare][]{Lange:2024}. To test whether we correctly mitigate fiber incompleteness, we split our sample by the number of passes that the \gls{desi} field of view has performed on that part of the sky (encoded by the \texttt{NTILE} flag). The impact can be seen in Fig.~\ref{fig:deltasigma_amplitude_ntile_split}.
        
        The local stellar density has been suggested to impact the measurements of spectroscopic surveys \citep{2017MNRAS.464.1168R}. Furthermore, the seeing and depth of the imaging survey that is used for target selection has the potential to impact the subsequent selected lens galaxies. To test our imaging systematics weight, we split our sample by local stellar density (\texttt{STARDENS}), as well as the seeing and depth of the corresponding imaging survey (\texttt{PSFSIZE} and \texttt{PSFDEPTH\_Z}). We expect any trends or outliers in these tests to be caused by an inhomogeneity in the \gls{desiy1} data.

        After unblinding, independent efforts detected trends in some properties of the \gls{lrg} sample with $B-V$ extinction, captured by the \texttt{EBV} flag \citep{arxiv:2405.16299}. We thus decided to add this to our lens homogeneity tests. This decision was made post-unblinding.
        
    \subsection{Source redshift test}
    \label{sec:amplitude_tests:analysis_considerations_zsource}
        To test the validity of the lensing data, we investigate the lensing signal as a function of effective source redshift. This not only tests the consistency of the measurements between the different lensing surveys, it also flags any redshift-dependent systematic biases (such as photometric redshift calibration). For this, intrinsic alignments are by far the most concerning systematic effect in this work. \citetalias{Lange:2024} show that, for a given lens bin, the strength of the IA effect depends on the source redshift bin, as a larger overlap between source and lens bin yields a stronger suppression of the lensing signal. This means that intrinsic alignments have the potential to induce a slope of the lensing amplitude $\ADS$ as a function of source redshift, which is one of the main effects that we are investigating. We address this issue by performing stringent cuts on source-lens bin pairs (detailed in Tab.~\ref{tab:source_lens_bins} and described in Sect.~\ref{sec:GGL_measurements:measurement_setup}) to minimize the redshift-dependent impact of systematic biases investigated in \citetalias{Lange:2024}.

    \subsection{Identifying trends and outliers}
    \label{sec:amplitude_tests:trends_and_outliers}
        To investigate potential systematics, we want to be sensitive to both consistent trends and individual outliers. We therefore employ two different tests to quantify both.
        To test for the presence of trends, we fit a linear function $f(x) = c_0 + c_1x$ to the lensing amplitudes $\ADS(x)$, where $x$ is an arbitrary tracer for a potential systematic effect (such as the effective source redshift, the local stellar density, or others). Any significant deviation from zero slope shows a consistent trend, that points to a systematic bias in the data.
        
        Our second key objective is to utilize the measured discrepancy between the lensing amplitudes $\ADS$ as a comprehensive and data-driven assessment of systematic uncertainties, denoted as $\sigma_{\rm sys}$. The calculation of $\sigma_{\rm sys}$ proceeds as follows: 
        We assume that each lensing amplitude is sampled from a Gaussian distribution with variance $\sigma^2=\sigma_{\rm stat}^2+\sigma_{\rm sys}^2$, where $\sigma_{\rm stat}$ represents the statistical uncertainty on the amplitude for each survey. To our lensing amplitudes $\ADS{}_{,i}$ and statistical uncertainties $\sigma_{{\rm stat},i}$ we then fit a model with a single lensing amplitude, $\ADS$, and a systematic scatter $\sigma_{\rm sys}$, as free parameters, where we define the log-likelihood as
        \begin{equation}
            \log\mathcal{L} = -\frac{1}{2}\sum_{i} \left[\frac{(\ADS{}_{,i} - \ADS)^2}{\sigma_{{\rm stat},i}^2+\sigma_{\rm sys}^2} + \log\left( \sigma_{{\rm stat},i}^2+\sigma_{\rm sys}^2\right) \right] \; ,
        \end{equation}
        where the sum extends over all measurements for one lens bin. We use a uniform prior between -1 and 1 for the lensing amplitude and a uniform prior between 0 and 2 for the systematic uncertainty.

    \subsection{Quantifying outlier probabilities}
    \label{sec:amplitude_tests:outlier_probabilities}
        The goal of this work is to identify the presence of potential systematics purely from the available data. We achieve this by counting the number of outliers, which we define as deviations above a certain threshold of the slope $c_1$ of the linear function fit to the lensing amplitudes.
        
        In this work, we primarily analyze the amplitude of the lensing signal $A_{\ds}$ in two sets of measurements, the lens homogeneity tests (Sect.~\ref{sec:amplitude_tests:analysis_considerations_homogeneity}), and the dependence on effective source redshift (Sect.~\ref{sec:amplitude_tests:analysis_considerations_zsource}). In both these measurements, we analyze 5 different lens bins on three scales (only small scales, only large scales, and all scales). If all $N_\mathrm{meas}$ measurements were uncorrelated, we would be able to estimate the probability that the number of outliers $\nout$ is equal to or larger than $N_\mathrm{out}$ via the binomial distribution
        \begin{equation}
            P(\nout \geq N_\mathrm{out}) = 1-\sum_{k=0}^{N_\mathrm{out}-1} \binom{N_\mathrm{meas}}{k} p^k (1-p)^{N_\mathrm{meas}-k} \; ,
            \label{eq:binomial_outlier_fraction}
        \end{equation}
        where $p$ is the probability for an individual outlier to occur (i.e., for a normal distribution with a $3\sigma$ limit for an outlier, $p\approx 0.003$).
        However, all these measurements are correlated to some degree, so estimating the significance of outliers becomes a nontrivial task. For both the lens homogeneity tests (Sect.~\ref{sec:amplitude_tests:analysis_considerations_homogeneity}) and the investigations on the source redshift (Sect.~\ref{sec:amplitude_tests:analysis_considerations_zsource}), we therefore generate $100\,000$ random realizations from a multivariate normal distribution with our reference datavector and estimated covariance.
        
        We compute an accurate analytic covariance for the source redshift tests, as detailed in Sect.~\ref{sec:GGL_measurements:covariance}. For the lens homogeneity tests, due to the small and patchy nature of the individual regions, we employ a Jackknife method for covariance estimation.       We note that we exclude correlations between different potential systematics in the lens homogeneity tests as well as correlations between measurements in the lens homogeneity tests and the source redshift tests. We thus treat the outliers in these analyses independently, justified by the distinct nature of the tests: the lens homogeneity tests assess systematic biases in the lens sample from \gls{desiy1}, while the source redshift tests evaluate systematics in the lensing surveys (\gls{hsc}, \gls{des}, and \gls{kids}).
        
        We are unable to validate our Jackknife covariances for the lens homogeneity tests to the standards required for cosmological analysis; thus, interpretations of these tests should be approached with caution. We enhance our confidence in these results by comparing clustering signals across different \texttt{NTILE} values, which serve as a critical indicator of potential \gls{desiy1} incompleteness effects.
        
        Based on Fig.~\ref{fig:probability_of_finding_outliers_for_different_sigmas}, we determine that a $3\sigma$ limit is suitable to classify the presence of an outlier. For the lens homogeneity tests, our randoms indicate a $\lesssim 10\%$ chance that such an outlier occurs by chance, while it is $\lesssim 2\%$ for the source redshift investigations. We note that the former is likely an upper limit: Despite our best efforts, we find it impossible to account for all potential cross-correlations between the lens bins, source surveys, and sub-samples based on potential systematic biases.

    \subsection{Blinding}
    \label{sec:amplitude_tests:blinding}

        We apply a constant offset and a slope in $\log(R)$ to all our $\ds$ measurements. To mask a potential slope of the lensing amplitude as a function of effective source redshift (which was tentatively found in \citetalias{Leauthaud:2022}), we also offset the measurements as a function of effective source redshift. In total, our blinding function $f_\mathrm{blind}(R,z)$, which we multiply to our $\ds$ measurement, is defined as
        \begin{align}
                f_\mathrm{blind}(R,z) {}&{}= \frac{1}{4}\left[(\beta-\alpha)\log(R/[0.15\,\mathrm{Mpc}/h])+5\alpha-\beta\right]\notag\\
                {}&{}\qquad\times\left[1+\nu(\expval{z_\mathrm{source}}-0.8)\right] \; ,
                \label{eq:DS_blinding_redshift}
        \end{align}
        where $\alpha,\beta$ are randomly drawn from a uniform distribution $\mathcal{U}[0.8,1.2]$, and $\nu$ is drawn from a normal distribution with zero mean and standard deviation 0.75, $\nu\in\mathcal{N}(0,0.75)$. We keep the same blinding function for the different source tomographic bins of the lensing surveys, but draw separate random numbers for each combination between lensing survey and lens bin.
        
        This procedure allows us to perform lens homogeneity comparisons on blinded data\footnote{During the blinded analysis, we noticed that this is not entirely correct. The lens homogeneity comparisons are based on splitting the lens sample into four quantiles ordered by the value of a potential systematic (for example, depth of the corresponding imaging data). While the blinding does not change the lensing amplitude $A_\mathrm{lens}$ between these four quantiles, it can still lead to a perceived offset when more than one lensing survey contributes to the analysis. For example, the \gls{desi} overlap with \gls{des} has a substantially deeper imaging, as the corresponding DECam observations were incorporated into the \gls{desi} legacy imaging survey. If the \gls{des} measurements are blinded high, we see a positive slope of the lensing amplitude with respect to imaging depth. We decided to inspect these cases by eye pre-unblinding, ensuring that potential slopes with respect to systematic offsets are just caused by this, and can not be seen in any of the single lensing surveys.}, while masking a potential slope of the lensing signal as a function of source redshift, as was discovered in \citetalias{Leauthaud:2022}.

    \subsection{Unblinding}
    \label{sec:amplitude_tests:unblinding}

        We decide on the following unblinding criteria:
        \begin{itemize}
            \item The \textbf{unblinded} cross components $\ds_\times = \gamma_\times\Sigma_\mathrm{crit}$ and $\gamma_\times$ are consistent with zero (i.e.~there are no significant B-modes in the data)
            \item The \textbf{unblinded} $\ds$ measured around random positions is consistent with zero on scales $\lesssim 1\mathrm{Mpc}/h$ (see Sect.~\ref{sec:GGL_measurements:random_subtraction})
            \item The jackknife covariance matrix is consistent with the theory covariance matrix on scales $\lesssim 5\mathrm{Mpc}/h$
            \item The scale cuts and redshift bins have been fixed
            \item We have performed the analysis with $10^6$ random noise realisations of analytic predictions (see Sect.~\ref{sec:amplitude_tests:outlier_probabilities}). Based on these tests, we determine that a 3 sigma cut results in probability of finding one deviation caused by a noise fluctuation of about 3\% for the 15 measurements in the source redshift tests, and 10\% for the 60 measurements in the lens homogeneity tests. We have determined tentative plans how to proceed if outliers show up in the unblinded data (detailed in Sect.~\ref{sec:amplitude_tests:post-unblinding plans}). 
            \item \textbf{Upon unblinding, should further analyses be deemed necessary, these must be explicitly highlighted to guard against the look-elsewhere effect.}
        \end{itemize}

    \subsection{Plan for post-unblinding analysis}
    \label{sec:amplitude_tests:post-unblinding plans}
        To avoid over-emphasizing the statistical significance of outliers, it is crucial to develop a plan to deal with potential outliers before unblinding. Here we describe plans for various post-unblinding scenarios.

        \subsubsection{Outliers in the lens homogeneity tests}
        \label{sec:amplitude_tests:outliers_in_lens_homogeneity}
            The lens homogeneity tests consist of a total of 60 measurements, meaning that there is a non-negligible chance to find a random outlier ($\sim 10\%$ for a $3\sigma$ outlier). We therefore decide to investigate $3\sigma$ outliers, but we will mainly be concerned about $4\sigma$ outliers, for which there is a $\sim 1\%$ chance that one would randomly occur.
            
            In this work, we have adapted the choices from the \gls{desi} key projects \citep[see][]{Ross:2024} for imaging systematics weights $w_\mathrm{imsys}$ and fiber incompleteness weights $w_\mathrm{comp}$. If we notice a trend in the measurements split by \texttt{NTILE}, this is a strong sign that the fiber incompleteness weights are not robust, and we will see if the problem persists with the alternative. If a trend shows up in measurements that were split by a different quantity (such as imaging depth), this is a strong sign that the imaging systematics weights are not robust, and we will see if the trend persists with a different choice of these weights (as detailed in Sect.~\ref{sec:DESI_data:weights}).
            
            We stress that, due to the uncertain nature in our covariance calculations affecting both the outlier determinations (Sect.~\ref{sec:amplitude_tests:outlier_probabilities}) and the actual error bars (Sect.~\ref{sec:amplitude_tests:lensing_amplitude}), results from these tests should be interpreted with some caution.

        \subsubsection{Outliers in the source redshift tests}
        \label{sec:amplitude_tests:outliers_zsource_test}
            The presence of an outlier in the source redshift tests is a far more intriguing possibility. First and foremost, our error analysis for this test is far more robust, as we obtained reliable analytic cross-covariances that account for the correlation between different source surveys. The resulting outlier detection is thus far more robust and we can make confident statements about its significance.
            If we detect an outlier, we will assume that it is driven by an erroneous source redshift calibration, as we have taken great care to robustly minimize the impact of other potential biases, such as intrinsic alignments or magnification bias. We therefore calculate the shift in mean redshift, $\Delta \bar{z}$, that would be required to explain this outlier and show this in Fig.~\ref{fig:delta_z}, by calculating which linear offset of the source $n(z)$ would lead to a $\Sigmacrit$ that yields a lensing amplitude of unity. 
            The interpretation of this outlier depends on where it occurs. In case we find an outlier, we will investigate whether it is driven by an extreme redshift bin that we already suspect as a potential error source. These can be bins with a small lens-source separation, leading to potential contamination by systematics (as discussed in Sect.~\ref{sec:amplitude_tests:analysis_considerations_zsource}), or a high source redshift, where the calibration of photometric redshifts is most prone to errors. In case we do find significant outliers, we will inflate our covariance by the determined $\sigma_\mathrm{sys}$ value (see Sect.~\ref{sec:amplitude_tests:trends_and_outliers}). In any case, any subsequent cosmological analyses should show that their posteriors are not driven by the outlier found in this work.

\section{Clustering measurements}
\label{sec:clustering_measurements}
    \subsection{Definition of \texorpdfstring{$w_p(r_p)$}{wp(rp)}}
        In addition to \gls{ggl} measurements, we present here the projected two-point galaxy correlation function, $w_p(r_p)$. We begin with the three-dimensional two-point correlation function (2PCF), $\xi(\mathbf{r})$, which measures the excess probability over random of finding a pair of galaxies separated by a vector $\mathbf{r}$. Due to the isotropy of the universe, we can decompose $\mathbf{r}$ into components perpendicular and parallel to the line of sight (LoS), denoted as $r_p$ and $r_\pi$, respectively. This yields the two-dimensional correlation function $\xi(r_p, r_\pi)$. To estimate $\xi(r_p, r_\pi)$ from data, we use the Landy-Szalay estimator \citep{Landy:1993}:
        \begin{equation}
            \xi(r_p, r_\pi) = \frac{DD - 2DR + RR}{RR},
        \label{eq:xi_def}
        \end{equation}
        where $DD$, $DR$, and $RR$ are the normalized counts of data-data, data-random, and random-random galaxy pairs in each bin of $(r_p, r_\pi)$. To simplify the analysis and mitigate the complexities arising from redshift-space distortions—such as the “finger-of-god” effect—we project $\xi(r_p, r_\pi)$ along the line of sight to obtain the projected correlation function $w_p(r_p)$:
        \begin{equation}
        w_p(r_p) = 2 \int_0^{r_{\pi,\text{max}}} \xi(r_p, r_\pi) \, \mathrm{d}r_\pi.
        \label{eq:wp_def}
        \end{equation}
        This integration effectively averages out redshift-space distortions, making $w_p(r_p)$ more straightforward to model. Although $w_p(r_p)$ contains less information than $\xi(r_p, r_\pi)$ because it integrates over the line-of-sight direction of the correlation function, it offers practical advantages. The reduction in dimensionality simplifies the covariance matrix, and the integration over line-of-sight separations impacted by redshift-space distortions makes theoretical predictions more tractable.
        
        In combined-probe analyses, $w_p(r_p)$ is often used alongside weak lensing measurements like $\Delta\Sigma$ to break parameter degeneracies associated with galaxy bias. By jointly analyzing these statistics, we can achieve tighter constraints on cosmological parameters and gain a deeper understanding of the galaxy-dark matter connection.
        
    \subsection{Measurement with PIP Weights and Angular Upweighting}
        
        As discussed in \citet{2024arXiv241112020D}, correcting for observational effects such as fiber assignment incompleteness is crucial for accurate clustering measurements. The fiducial approach in most DESI 2024 cosmological analyses is to remove angular scales within a DESI fiber patrol radius, specifically angles less than $\theta < 0.05^\circ$ \citep{KP3s5-Pinon}. The fiducial Data Release 1 (DR1) Large-Scale Structure (LSS) catalogs are optimized for this method. Instead, we choose to correct for fiber assignment incompleteness at all scales using the combination of \gls{pip} weights and angular upweighting \citep{Bianchi:2017}. This requires LSS catalogs constructed differently, released as version \texttt{v1.5pip}. Below, we detail the methodology for measuring $w_p(r_p)$ using PIP weights and angular upweighting.
        
        \subsubsection{PIP Weights}
        \label{sec:clustering_measurements:PIP_weights}
        We obtain a total of 129 realizations of the fiber assignment of the DESI DR1 data: one from the real data and 128 using the \texttt{altmtl} method described in \cite{KP3s7-Lasker}. Each target in the full catalogs has a bit array $w^{(b)}$ indicating whether it was assigned in each of the 128 realizations. The total probability of assignment is stored as:
        \begin{equation}
            p_{\rm obs} = \frac{1 + c}{1 + 128},
        \end{equation}
        where $c = \mathrm{popcnt}(w^{(b)})$ counts the number of realizations in which the target is assigned. The completeness weights for the data are then determined as the individual inverse probability (IIP):
        \begin{equation}
            w_{\rm comp} = \frac{1}{p_{\rm obs}}.
        \end{equation}
        We note that these completeness weights $w_{\rm comp}$ are equivalent to the purely data-driven \texttt{FRACZ\_TILELOCID} and \texttt{FRAC\_TLOBS\_TILES}-based completeness weights described in Sect.~\ref{sec:DESI_data:fiber_incompleteness}.
        When calculating pair weights for the 2PCF, each pair $(i, j)$ is assigned a PIP weight:
        \begin{equation}\label{eq:w_pip}
            w^{\rm PIP}_{ij} = w_{\rm comp,i} w_{\rm comp,j}  \frac{w^{\rm eff}_{ij}}{g_{128}(c_i, c_j)},
        \end{equation}
        where
        \begin{equation}
            w^\mathrm{eff}_{ij} = \frac{1 + 128}{1 + \mathrm{popcnt}(w^{(b)}_i \, \& \, w^{(b)}_j)},
        \end{equation}
        and $g_{128}(c_i, c_j)$ represents the expectation value of $w^{\rm PIP}_{ij}$ in the limit of independent probabilities (see \cite{KP3s6-Bianchi} for details). This formulation enhances robustness against scale-dependent artifacts arising from low probabilities and a limited number of realizations. The weights for the randoms are not multiplied by $w_{\rm tlobs}$ (in contrast to the weights for the \gls{ggl} measurements described in Sect.~\ref{sec:DESI_data:fiber_incompleteness}), as the \textsc{altmtl}-based completeness weights account for any source of fiber assignment incompleteness.
        
        \subsubsection{Angular Upweighting}
        
        Due to much of the DESI DR1 area being covered by only one or two tiles, many pairs have zero probability of being observed and are excluded from PIP weights. To address this, angular upweighting is employed. This method uses the angular separation of pairs in the full LSS catalog (using all data, referred to as “parent”) and the selection of observed targets (referred to as “fibered”). To apply angular upweighting consistently, we include weights that account for the variation of $w_{\rm tot}/w_{\rm comp}$ and $w_{\rm FKP}$ with $n_{\rm tile}$, the number of overlapping tiles \citep[where $w_{\rm FKP}$ defines redshift-dependent weights to optimally extract the power spectrum,][]{1994ApJ...426...23F}. The mean values of these quantities are determined as a function of $n_{\rm tile}$ and added as new columns to the full LSS catalog.\footnote{These columns are \texttt{WEIGHT\_NTILE} and \texttt{WEIGHT\_FKP\_NTILE}.} Angular upweights are then computed as:
        \begin{equation}
            w_{\rm angular}(\theta) = \frac{DD_{\rm parent}(\theta)}{DD_{\rm fibered}(\theta)},
        \end{equation}
        where $\theta$ is the angular separation. These weights are pre-computed as a function of angular separation with logarithmic binning and are interpolated during pair counting. While PIP weights and angular upweighting correct for fiber assignment incompleteness, they introduce increased variance in the completeness weights $w_{\rm comp}$. This variance leads to larger uncertainties in clustering measurements compared to the fiducial catalogs.\footnote{
        The IIP method shows approximately 60\% higher uncertainty due to the increased variance in $w_{\rm comp}$. When applying PIP weights and angular upweighting, the results at large scales are nearly identical, but the primary difference is in how $w_{\rm comp}$ is obtained. There is more variance in the $w_{\rm comp}$ obtained from the IIP than from our method for the \gls{ggl} measurements.}
        
        In summary, by properly accounting for observational incompleteness using PIP weights and angular upweighting, we can obtain unbiased clustering measurements across all scales. This is essential for combined-probe analyses that aim to constrain cosmological parameters using small scales and understand the galaxy-dark matter connection.

    \subsection{Clustering homogeneity tests}
        While we do not perform elaborate amplitude tests with the projected clustering measurements, we do verify our correct application of the pairwise weights, which are designed to offset fiber-incompleteness of \gls{desi}, by splitting the survey area into four quantiles ordered by the respective \texttt{NTILE}-value. We then verify that we can not observe substantial differences between clustering measurements of different \texttt{NTILE}-values on small scales. This is motivated by the fact that any fiber-incompleteness effects are expected to strongly correlate with \texttt{NTILE}, as fiber incompleteness is substantially more impactful when \gls{desi} has had fewer passes on the same area of the sky.
        
\section{Results}
\label{sec:results}
    \subsection{$\ds$ measurements and homogeneity tests}
    \label{sec:results:ds_measure}
\begin{figure*}
            \centering
            \includegraphics[width=0.9\linewidth]{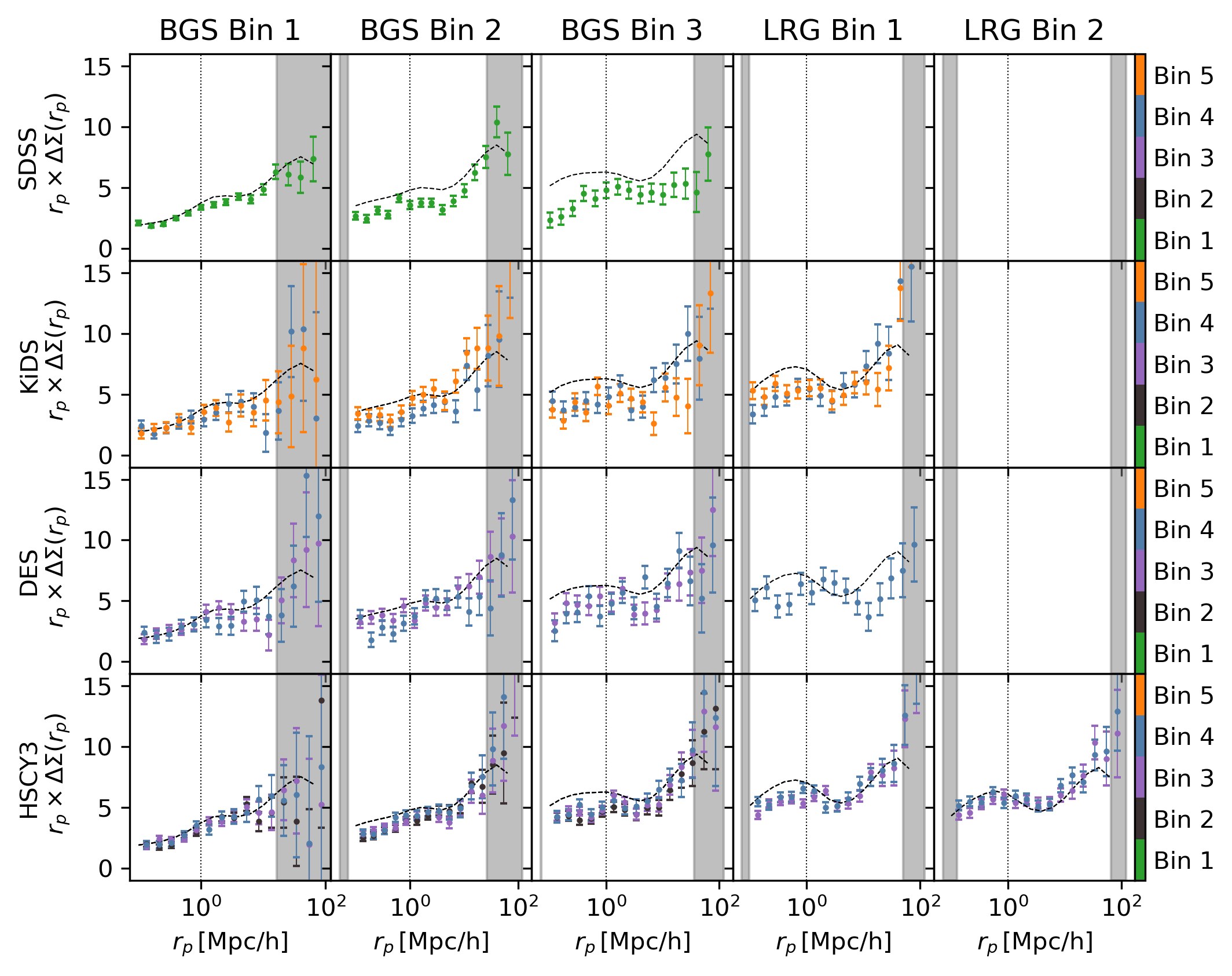}
            \caption{Lensing signal for the different source samples. The grey shaded regions are excluded due to our scale cuts outlined in Sect.~\ref{sec:GGL_measurements:measurement_setup}. The dotted line corresponds to the boundary between small and large scales. The dashed lines corresponds to the reference datavector extracted from the AbacusSummit simulations.}
            \label{fig:deltasigma_datavector}
        \end{figure*}

        We measure the $\ds$ signal as a function of comoving projected separation $\rp$. The results can be seen in Fig.~\ref{fig:deltasigma_datavector}. A by-eye inspection reveals that the measurements appear to be reasonably consistent. Further, our reference datavector captures the true $\ds$ signal reasonably well. While this is an encouraging sign, we discuss the results of our quantitative analysis in the next subsections. 

        We perform the lens homogeneity tests as outlined in Sect.~\ref{sec:amplitude_tests:analysis_considerations_homogeneity}, testing the dependence of the lensing amplitude on properties that might introduce systematic biases into DESI observations. We show an example of this test, here for the \texttt{NTILE} flag, in Fig.~\ref{fig:deltasigma_amplitude_ntile_split}. We do not detect a significant trend for any of the potential systematics. As none of the tests flagged anything concerning, we omit plots for the other potential systematics, but we do provide the best-fit values and uncertainties for all measured slopes in Tab.~\ref{tab:slopes_systematics} in the Appendix.

    \subsection{B-modes}
    \label{sec:results:bmodes}
        We measure the B-modes of the $\ds$ signal, $\ds_\times$, for each combination of lens bin and source survey. We also calculate the $\chi^2$ deviation from the expected zero signal by combining the measurement and the analytic B-mode covariance of the depicted source bins and scale cuts. We further calculate a statistical probability to exceed the measured $\chi^2$ value, assuming the measurements follow a normal distribution. The results can be seen in Fig.~\ref{fig:deltasigma_bmodes}. In the first \gls{bgs} bin, the \gls{hsc} sources show a $p$-value that would traditionally be considered to be of moderate significance ($0.02 < p < 0.05$). However, assuming the 13 B-mode calculations are independent, the chance for finding at least one moderate outlier is 48\%, and the chance to find at least one outlier with $p\leq 0.025$ is 28\%, implying that the B-mode detection in itself is not significant whatsoever. We note that \citet{2023PhRvD.108l3518L} report a B-mode detection on large angular scales in their cosmic shear measurements, which for us would be most visible in the first \gls{bgs} bin, although it is questionable how much cosmic shear B-modes relate to B-modes in \gls{ggl}. All things considered, we claim no significant detection of B-modes in our measurements.

    \subsection{Source redshift tests}
    \label{sec:results:zsource_trend}
        \begin{figure*}
            \centering
            \includegraphics[width=0.9\linewidth]{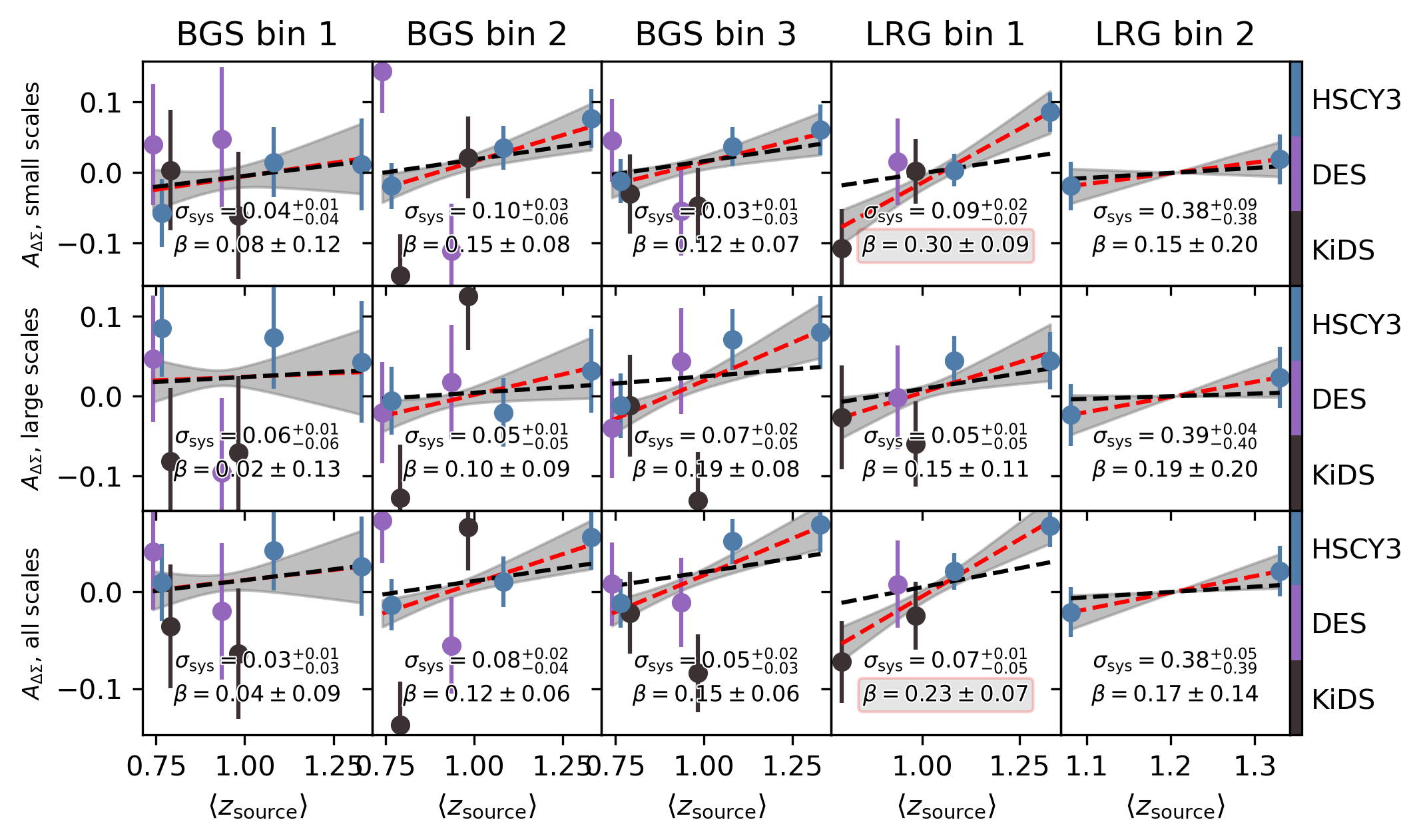}
            \caption{Lensing amplitude as a function of source redshift. The red line represents the best-fit slope $\beta$; the grey band denotes its 1$\sigma$ uncertainty. The black line represents the slope we find when feeding the datavectors contaminated by all uncorrected lensing systematics (intrinsic alignments, boost factor, source magnification, and the reduced shear approximation) from the analysis in \citet{Lange:2024} through our pipeline.}
            \label{fig:source_redshift_slope}
        \end{figure*}
                
        Fig.~\ref{fig:source_redshift_slope} displays the lensing amplitude $\ADS$ as a function of source redshift. We detect a statistically significant trend with source redshift for the first LRG bin. Most other bins also show positive trends with source redshift, albeit at lower levels of significance. We also show the slope that we find when feeding a datavector from \citetalias{Lange:2024} that is contaminated by all uncorrected lensing systematics (intrinsic alignments, boost factor, source magnification, and the reduced shear approximation) through our pipeline. 
        As photometric redshifts are one of the main uncertainties in current cosmic shear surveys, we compute the shift in mean redshift that would bring all lensing amplitudes in agreement in Fig.~\ref{fig:delta_z}: As $\Delta\Sigma = \gammat\Sigmacrit$, we calculate the ratio $\Sigmacrit(\bar{\zl},z_\mathrm{s,shifted})/\Sigmacrit(\bar{\zl},\zs)$ and multiply our lensing amplitude by that value\footnote{We note that, technically, one has to compute this ratio for each individual lens. We have verified that the difference between those two methods is negligible compared to the measurement uncertainties.}. We then fit the shift $\Delta z$ for each source bin such that they all equal the average of all measurements for that lens bin. This is analogous to similar (albeit more sophisticated) efforts to test photo-z distributions with shear ratio tests \citep{2025arXiv251005539E}, and complementary to efforts to constrain redshift distribution of imaging surveys via clustering redshifts \citep{2025arXiv251215963R}. We find that there is a preference to shift the higher redshift bins to even higher mean redshifts in order to achieve a uniform lens amplitude $\ADS$ in Fig.~\ref{fig:delta_z}.
        
        For the first LRG bin (at all scales), we find a preference for the 4th \gls{hscy3} bin to be shifted up by $\Delta z = 0.115^{+0.062}_{-0.057}$, the 3rd \gls{hscy3} redshift bin to remain as-is ($\Delta z = 0.000^{+0.024}_{-0.023}$), and the other lensing surveys to be at lower (DES Bin 4: $-0.010$, \gls{kids} Bin 4: $-0.041$, \gls{kids} Bin 5: $-0.039$) mean redshift. We note that this is just a relative adjustment of redshifts to bring the lens amplitude to the same value and should not be confused with an indication for absolute redshift shifts. Assuming the other redshifts (from \gls{des} and \gls{kids}) to be correctly calibrated, this indicates a preference for the 3rd and 4th \gls{hscy3} bins to be at higher redshift by $\Delta z_3\sim 0.03$ and $\Delta z_4\sim 0.13$. As outlined in \citet{2023PhRvD.108l3518L}, redshift distributions of the \gls{hsc} source sample may require adjustments, with reported shifts of $\Delta z_3=0.115$ and $\Delta z_4=0.192$ to higher redshifts. These shifts align with our findings, suggesting they could explain the observed source redshift trends. Implementing these shifts results in a lensing signal trend consistent with mock predictions, as shown in Fig.~\ref{fig:source_redshift_slope_hsc_shifts}. We will explore the implications of this in detail in Sect.~\ref{sec:discussion:potential_causes_of_redshift_trends}.

    \subsection{Outlier rates}
    \label{sec:results:outlier_rates}        
        \begin{figure}
            \includegraphics[width=\linewidth]{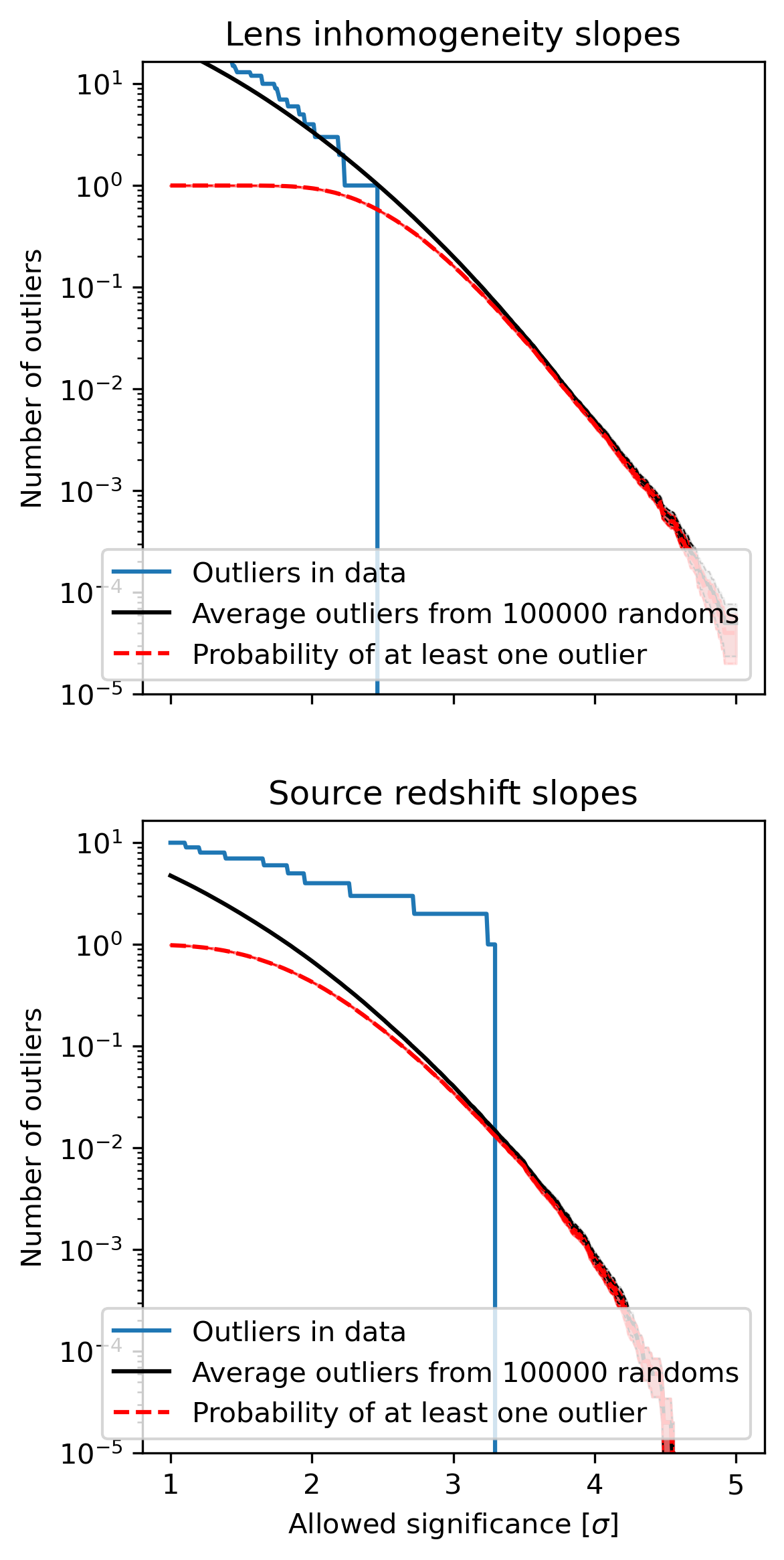}
            \caption{Estimating significance of $\ADS$ slopes. Top panel: The number of outliers from the 108 measurements of degeneracies with potential systematics. We plot the number of outliers in the data (blue), the average number of outliers in the randoms (black), and the probability of having at least one outlier in the randoms (red).
            Bottom panel: Same as the top panel, but we investigate the number of outliers from the 15 measurements of source redshift slopes.}
            \label{fig:probability_of_finding_outliers_for_different_sigmas}
        \end{figure}

        We compare the number of outliers in the data, as a function of significance level, to the number of outliers in the randoms that were used to estimate the uncertainty in the fitted slopes (see Sect.~\ref{sec:amplitude_tests:outlier_probabilities}) in Fig.~\ref{fig:probability_of_finding_outliers_for_different_sigmas}. As one can see, the number of outliers in the lens homogeneity tests closely follows the number one would expect from the random tests. This shows that our finding of no significant trends in the lens homogeneity tests does not depend on our choice of significance level -- for all levels we see very good agreement between the data and the randoms.
        
        The number of outliers in the source redshift tests, however, is consistently much larger than the number expected from the random tests. This implies that our report of a redshift-dependent systematic in the imaging surveys again does not depend on our chosen significance level -- at all levels, the data show a clear excess of outliers with respect to the randoms. The only exception is at significances $\gtrsim 3.2\sigma$, where we do not measure any outliers in the data anymore. We also see that the chance of finding an outlier of the magnitude we found in the first \gls{lrg} bin is less than 2\%, indicating that this finding is statistically significant even in the context of multiple simultaneous outlier searches.

    \subsection{Estimating an amplitude systematic}
    \label{sec:results:sigma_sys}
\begin{figure*}
            \centering
            \includegraphics[width=0.9\linewidth]{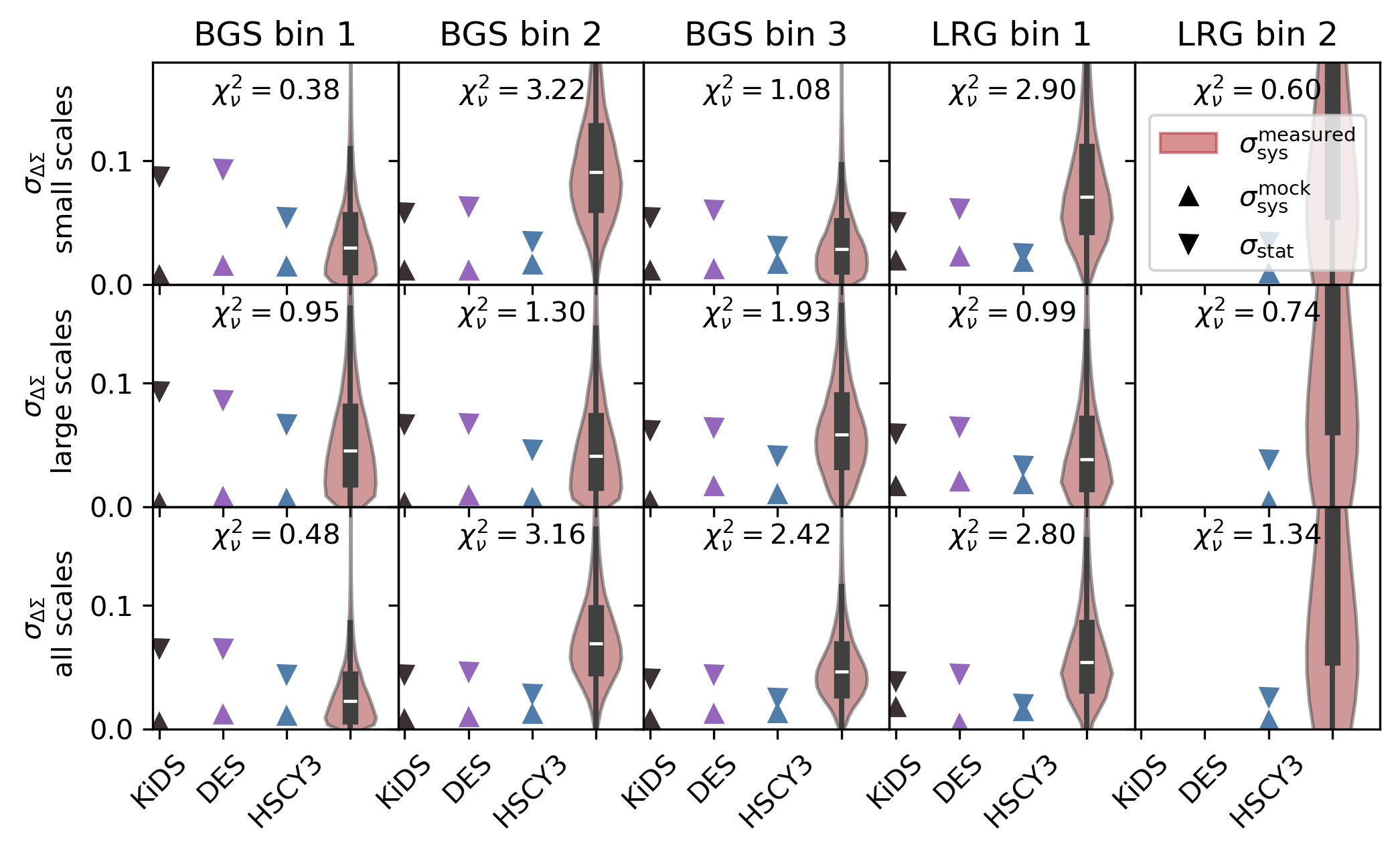}
            \caption{Estimate of the systematic uncertainty between different lensing surveys, compared to the systematic uncertainties of the lensing surveys themselves. The downwards-facing triangles denote the statistical uncertainties of the measurements determined from our covariance estimation. The red dots represent the systematic uncertainty we estimate from our method detailed in Sect.~\ref{sec:amplitude_tests:trends_and_outliers}, with 1- and 2-$\sigma$ confidence intervals in red and blue, respectively.  The upwards-facing triangles denote the systematic uncertainties of the lensing surveys, which were estimated from the measurements in \citetalias{Lange:2024} and only include impacts of intrinsic alignments, source magnification, the boost factor, and the reduced shear approximation, and therefore under-estimate the true systematic uncertainty of the measurement. The best-fit value for $\sigma_\mathrm{sys}$ for the second \gls{lrg} bin equals $0.38$ for all three scales, but it is fully consistent with zero as it is derived from only two data points and has an extremely large uncertainty.}
            \label{fig:sigma_sys}
        \end{figure*}

        We compare the scatter between amplitudes in the source redshift tests with the one we would expect from the estimated covariance in Fig.~\ref{fig:sigma_sys}. We note that the posteriors for the excess scatter between amplitudes, $\sigma_\mathrm{sys}$, is highly non-Gaussian, so we always cite 1- and 2-$\sigma$ confidence intervals, with the latter being denoted in blue. We can see that the scatter between points is larger than the estimated covariance (indicated by a $\sigma_\mathrm{sys}>0$) for the 2nd and 3rd \gls{bgs} bin, with a $\gtrsim 2\sigma$ evidence for a non-zero additional amplitude systematic. 
        
        For the 2nd \gls{bgs} bin at all radial scales, we get $\sigma_\mathrm{sys}=0.077_{-0.045}^{+0.022}\color{blue}{}_{-0.070}^{+0.084}$, driven by a measurement of $\sigma_\mathrm{sys}=0.102_{-0.059}^{+0.029}\color{blue}{}_{-0.092}^{+0.112}$ on small scales. Comparing with Fig.~\ref{fig:source_redshift_slope}, we can see that this is driven by outliers of the 4th \gls{kids} and 3rd and 4th \gls{des} source tomographic bins.
        
        The measurement of the 3rd \gls{bgs} bin at all scales, $\sigma_\mathrm{sys}=0.052_{-0.032}^{+0.016}\color{blue}{}_{-0.050}^{+0.060}$, is also non-zero at the 2-$\sigma$ level, but although the large scales $\left(\sigma_\mathrm{sys}=0.066_{-0.048}^{+0.023}{\color{blue}{}_{-0.066}^{+0.085}}\right)$ show a stronger preference for a non-zero $\sigma_\mathrm{sys}$ than the small scales, they are still consistent with zero at the 2-$\sigma$ level. Comparing again with Fig.~\ref{fig:source_redshift_slope}, we can see that this is driven by a single outlier in the 5th \gls{kids} source tomographic bin, as well as by a non-zero trend in source redshift, inflating the measured scatter between amplitudes.
        
        For the 1st \gls{lrg} bin, a non-zero amplitude systematic of $\sigma_\mathrm{sys}=0.068_{-0.051}^{+0.014}\color{blue}{}_{-0.068}^{+0.099}$ is also preferred, albeit at a lower statistical significance. This appears to be driven by the small-scale measurements $\left(\sigma_\mathrm{sys}=0.089_{-0.066}^{+0.017}{\color{blue}{}_{-0.089}^{+0.130}}\right)$. We note that the source redshift trend certainly inflates the scatter between lensing amplitudes, and the presence of a non-zero $\sigma_\mathrm{sys}$ is partly caused by that.
        
    \subsection{Clustering measurements}
        \begin{figure*}
            \centering
            \includegraphics[width=\linewidth]{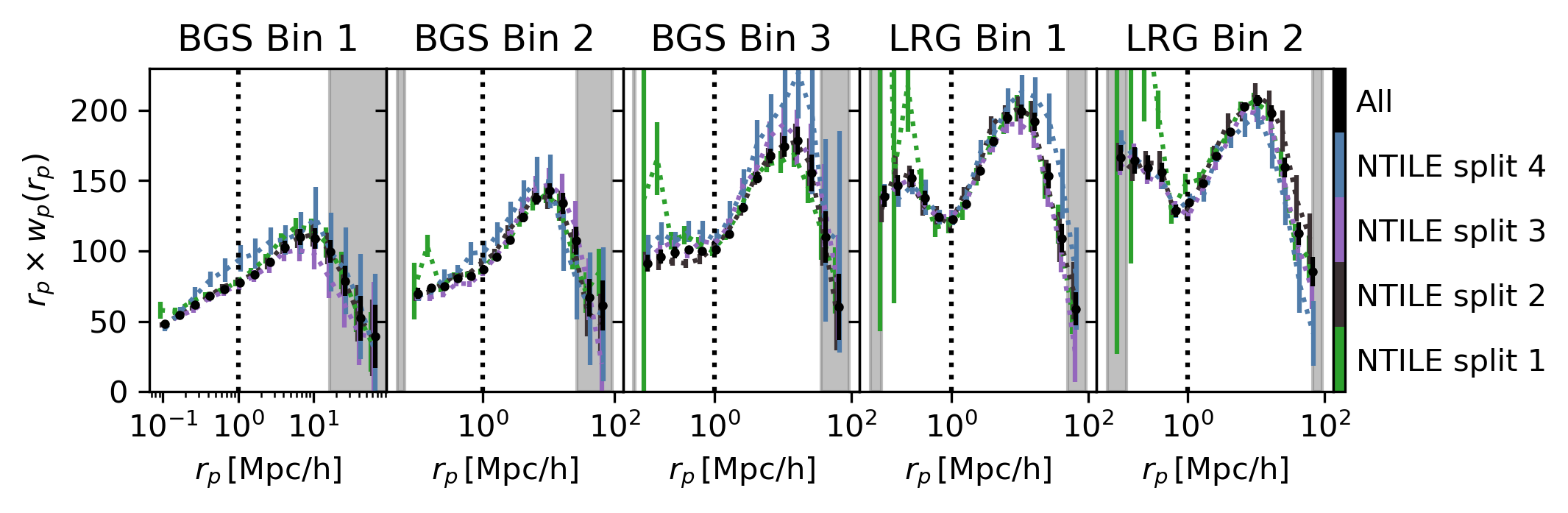}
            \caption{The projected clustering measurements $w_p$ for the \gls{desiy1} \gls{bgs} and \gls{lrg}. One can see that, apart from some noise upticks at small scales in the regions with small NTILE values, the projected clustering is broadly consistent between different \texttt{NTILE} values.}
            \label{fig:wp_including_ntile}
        \end{figure*}
        We show the clustering measurements, including the ones split by \texttt{NTILE} values, in Fig.~\ref{fig:wp_including_ntile}. We note that we do observe a higher value of $w_p$ for the highest \texttt{NTILE} value in some lens bins of the \gls{bgs} sample. We found the statistically most significant outlier to be the 1st \gls{bgs} bin. This discrepancy is on large scales, which are unaffected by both PIP weights and angular upweighting. We investigate the statistical significance of the outliers in the \gls{bgs} bins: Firstly, we estimate Jackknife covariances for all \texttt{NTILE} values using 64 regions. We then combine the measurements for the first three \texttt{NTILE} values and their covariance into a single measurement and covariance \citep[according to the procedure detailed in App.~C of ][]{2020A&A...642A.158B}. We then calculate the deviation of the 4th \texttt{NTILE} measurement from the combination of the other three. We note that, as all covariances are estimated from a finite set of data, the inverse covariance is biased and a naive $\chi^2$ evaluation would yield substantially biased $p$-values. As detailed in App.~C of \citet{2024MNRAS.534.3305H}, we instead perform the analogous test using Hotelling's $T^2$ distribution. We could not find an entirely satisfactory answer for the number of independent datavectors $N_\mathrm{sims}$ that is used to estimate the covariance matrix, which is an important input for the $T^2$ distribution. While a single covariance was calculated from 64 datavectors, the covariance matrix we use to calculate the $T^2$ statistic is a combination of four covariances, each calculated from 64 datavectors. We therefore estimate $p$-values for both $N_\mathrm{sims}=64$ and $N_\mathrm{sims}=256$, assuming that the truth lies somewhere in between. We get $p=0.58$ and $p=0.38$, respectively, while a naive $\chi^2$ test would yield $p=0.31$. All three values are highly consistent with the hypothesis that the outlier is a random fluctuation, but their difference highlights the importance of debiasing the inverse covariance matrix. We also repeated this test with all three lens bins, as well as all possible combinations of the three lens bins (setting the cross-covariance between different lens bins to zero), all of which yield lower significances of outliers. The observation that an outlier that appears to be substantial by-eye has such low significance can be attributed to the fact that the area with $\texttt{NTILE}=4$ covers just $234\,\text{deg}^2$, meaning a strong presence of sample variance and a high correlation between measurements at large scales. We also see some artifacts in the $\texttt{NTILE}=1$ region at small scales, which can be attributed to noise fluctuations in the weighting normalization. As we do not observe an impact of these fluctuations on the full sample, we believe that these are not a concern for the analysis.
\section{Discussion}
\label{sec:discussion}
    We have conducted \gls{ggl} measurements cross-correlating positions of \gls{desiy1} lenses with shapes of \gls{hsc}, \gls{kids}, \gls{des}, and \gls{sdss} background sources. We have not found evidence for any trends with respect to potential systematics that might bias the lens galaxies observed by \gls{desi}. We have found ample evidence for an effect that increases the measured $\ds$ signal as a function of source redshift. This effect is partly expected from known systematics, as can be seen in the black lines of Fig.~\ref{fig:source_redshift_slope}, but our measurement far surpasses this expectation. We have analyzed several potential causes for this effect, that we outline below.
    \subsection{Post-unblinding: Potential causes of redshift trends}
    \label{sec:discussion:potential_causes_of_redshift_trends}
        \subsubsection{Under-estimation of IA contamination in mocks}
            The mocks analyzed in \citet{Lange:2024} predict a slope of $\beta = 0.05$ induced by intrinsic alignments for the first LRG bin on all scales. One potential explanation of the strong trend visible here is that the true IA signal is stronger than the one predicted in the mocks. In order to test this, we repeat our analysis while including all lens-source combinations where the average source redshift $\langle z_\mathrm{source}\rangle$ is larger than the average lens redshift $\langle z_\mathrm{lens}\rangle$, and show this in Fig.~\ref{fig:source_redshift_slope_allbins}. This sample is contaminated much stronger by intrinsic alignments due to a substantially larger source-lens overlap (with a $\sim 20\%$ impact on the lensing amplitude), so if IA were under-represented in the mocks we would expect to see a stronger slope in the data. The measured slope, however, is in agreement with the mocks. This means that the data suggest an accurate level of systematics in the mocks, and that the present slope is unlikely to be caused by an under-representation of IA in the mocks.
            
        \subsubsection{Shifts in mean redshift}
\begin{figure*}
            \includegraphics[width=0.9\linewidth]{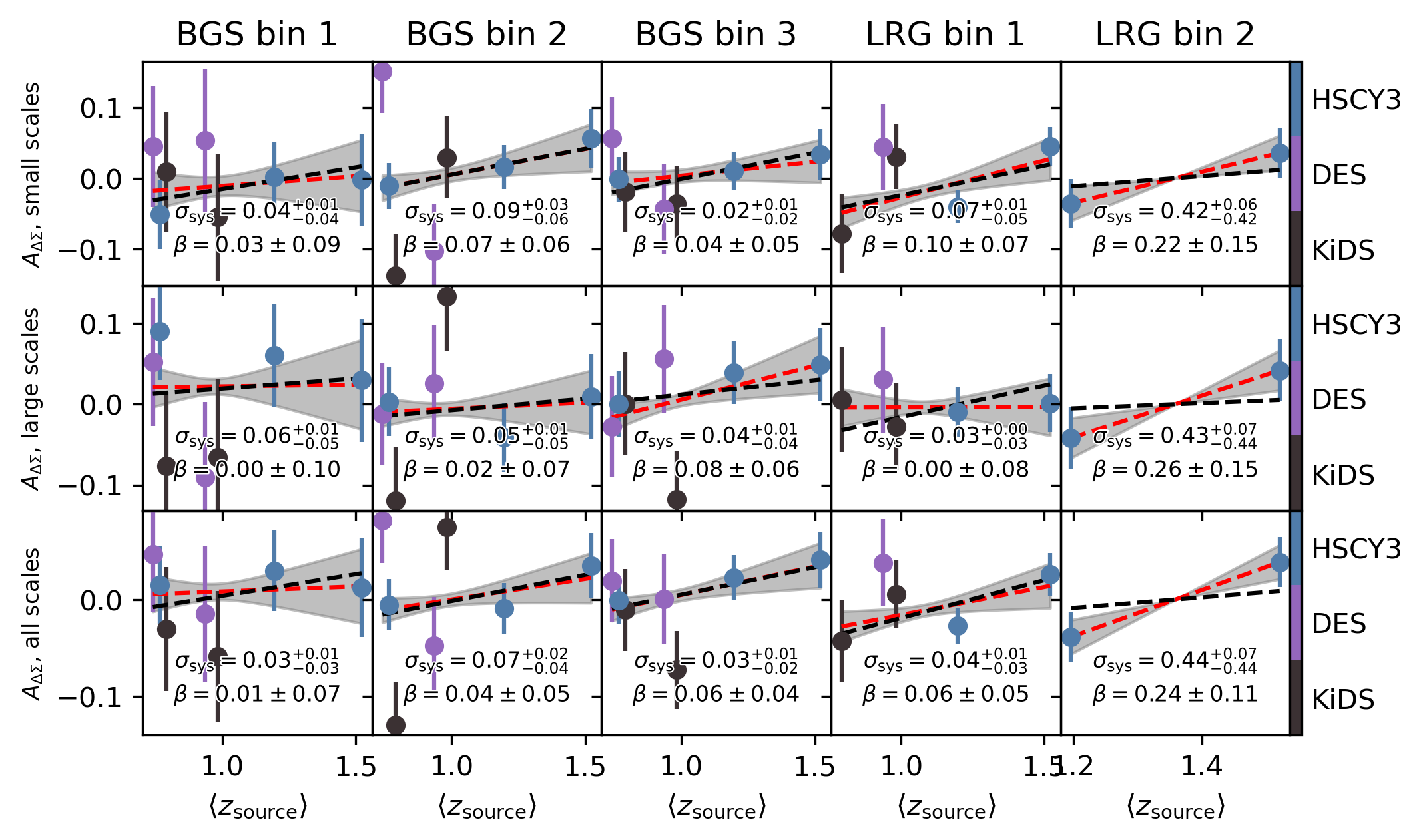}
            \caption{Same as Fig.~\ref{fig:source_redshift_slope}, but the third and fourth \gls{hsc} redshift bin were shifted by the best-fit amounts of \citet{2023PhRvD.108l3518L}. This figure was created post-unblinding.}
            \label{fig:source_redshift_slope_hsc_shifts}
        \end{figure*}

            As outlined in \citet{2023PhRvD.108l3518L}, the redshift distributions of the \gls{hsc} source sample are potentially inaccurate. In particular, a parameter inference with flat and wide priors show a shift of $\Delta z_3=0.115$ and $\Delta z_4=0.192$ to higher redshifts. A similar trend can be spotted in Fig.~\ref{fig:delta_z}, where the third and fourth \gls{hsc} bin prefer shifts to higher redshift. We implement these shifts in our analysis and show the results in Fig.~\ref{fig:source_redshift_slope_hsc_shifts}. As one can see, the trend determined now is completely consistent with the one from the mocks. This means that these shifts in source redshift are a possible explanation of the observed trend. One could even go so far as to say that this work is an additional indication that the lensing data prefer a higher source redshift for the third and fourth \gls{hsc} bins. Subsequent cosmological analyses should keep this issue in mind and allow for priors that can accommodate this shift in redshift.

        \subsubsection{Under-estimation of lens-source overlap due to too narrow photo-z distributions}
            Another potential explanation is that the photometric redshift distributions of the lensing surveys are broader than claimed. This is not entirely unreasonable: As the lensing surveys primarily conduct cosmic shear analyses, they are extremely sensitive to the mean redshift of a tomographic bin, but not to its concrete shape. As a small broadening of photometric redshift distributions can already cause relatively large differences in the overlap, this could cause a similarly increased slope. For example the 4th \gls{kids} source bin and the 2nd \gls{bgs} have very little overlap in Fig.~\ref{fig:all_nofz_plot}, but a small broadening of the \gls{kids} redshift distribution could increase the overlap substantially.
            
            However, due to its high redshift, the first \gls{lrg} bin's source redshift trend is dominated by the \gls{hsc} third and fourth source bin. As can be seen in Fig.~\ref{fig:all_nofz_plot}, a slight increase in the source distributions would only substantially increase the overlap of the 4th \gls{kids} bin with the lens sample. We therefore believe that it is very unlikely that the redshift trend is caused by broader redshift distributions of the photometric source samples.
            
    \subsection{Post-unblinding: potential causes of extra scatter in BGS bins}
        We find signs for excess scatter between lensing amplitudes for the 2nd and 3rd \gls{bgs} bins, as well as for the first \gls{lrg} bin. We note that, when only investigating large scales $\geq 1\,\Mpc/h$, there is no significant evidence for an excess scatter between lensing amplitudes. We also note that this observed scatter is certainly not independent of the detected trends with source redshift. Investigating Fig.~\ref{fig:source_redshift_slope}, we can see that the scatter between measurements is highly consistent with the best-fit linear slope. This means that the scatter, in this case, is likely driven by the source redshift trend. However, for the \gls{bgs} bins we did not detect a significant slope and, again investigating Fig.~\ref{fig:source_redshift_slope}, we do see an excess scatter, driven by low \gls{kids} measurements in both bins and a high \gls{des} measurement in the 2nd \gls{bgs} bin. We note that none of these measurements surpass the lower limit for a residual systematic error at the 2-$\sigma$ level. We note that this scatter is a bit larger, but consistent with independent evaluations of \gls{kids} x \gls{boss} measurements, as evaluated in \citet{Giblin:2021}, where the hypothesis of a single lens amplitude per redshift bin is consistent with the data at the 1.6$\sigma$ level.

        We did not find any evidence for inconsistencies between our measurements and the literature, nor did we find any specific cause for this scatter. We do note, however, that the statistical significance of the scatter strongly depends on our choice of covariance matrix. As detailed in \citetalias{Yuan:2024}, we report a $\sim 10\%$ deviation between the analytical and the simulated covariance matrices at large scales, that is accounted for by artificially boosting the analytical covariance matrix by the measured ratio. If, instead, we introduce an uncertainty of the same magnitude in the covariance matrix and marginalize over this uncertainty, the statistical significance of this scatter disappears.

        We should also note that the presence of additional biases aside from the statistical uncertainties (parametrized by the covariance) is very much expected. We show the residual systematic biases of intrinsic alignments, source magnification, the boost factor, and the reduced shear approximation, estimated from mocks in Fig.~\ref{fig:sigma_sys}. These are completely negligible by design -- we intentionally excluded any lens-source combinations that showed a strong systematic contamination in the mocks. However, we do not include any calibration uncertainties in these best estimates, so uncertainties in photometric redshift, multiplicative shear bias, and other effects which are usually accounted for in a cosmological analysis are not included in this plot. This means that the true systematic error that is usually included in lensing analyses is likely much larger than the one we estimate here. 

    \subsection{Summary and comparison with other works}
    \label{sec:discussion:conclusion}
        We have conducted a comprehensive analysis of the \gls{desiy1} \gls{ggl} measurements. We have not found any evidence for significant unknown systematics in the lensing measurements, but we have found a significant trend of the lensing amplitude with source redshift. We have investigated several potential causes for this trend and found that the most likely explanation is a shift in the mean redshift of the \gls{hsc} source sample, which is also preferred by analyses of the \gls{hsc} collaboration. If these shifts are implemented, all trends with source redshift are consistent with our expectations from the analysis of \citetalias{Lange:2024}. 
        We have also found evidence for excess scatter between lensing amplitudes in the 2nd and 3rd \gls{bgs} bins, as well as in the first \gls{lrg} bin, primarily driven by measurements on small scales. However, we do not record an instance where the statistical significance of this scatter exceeds our lower limit for residual systematics by more than $2\,\sigma$. We have not found any specific cause for this scatter, but we note that the statistical significance of the scatter strongly depends on our choice of covariance matrix, and disappears once we marginalize over our uncertainty in the covariance estimation.
        We have also found no evidence for significant B-modes in our measurements. We have not found any evidence for a trend of the lensing amplitude with potential systematics.

        Comparing our work with the previous \citetalias{Leauthaud:2022}, a few key points are worth mentioning. Our work expands upon the previous analysis in two key ways: Firstly, thanks to previous efforts by \citetalias{Lange:2024} and \citetalias{Yuan:2024}, we were able to make an informed choice about which scales and redshift ranges to use for our analysis, and to estimate the residual contamination of our data by known systematics given our cuts. Secondly, our analysis benefits from a substantially improved signal-to-noise ratio, allowing us to test the homogeneity of the \gls{desiy1} lens sample by splitting the sample into 4 quantiles for each survey, instead of taking the average value of the values of interest for each survey. This allows us to test a larger dynamic range of potential contaminants, and would have enabled us to distinguish between variations caused by different lensing surveys and variations within a single lensing survey, if we had found a contaminant of the \gls{desiy1} sample. This brings us to the first main difference between our findings and those of \citetalias{Leauthaud:2022}: We find no trends of the lensing amplitude with respect to potential contaminants of the \gls{desiy1} sample, whereas \citetalias{Leauthaud:2022} found trends with both stellar density and hemisphere coverage. We further find a substantially lower evidence for excess scatter between lensing amplitudes, which is a testament for how well both calibration efforts for lensing surveys and covariance estimation methods have matured. Like \citetalias{Leauthaud:2022}, we find an increasing trend of the lens amplitude with source redshift for lens bins at higher redshift that is driven by a high value of the \gls{hsc} measurements (compare our App.~\ref{app:hscy1} for an analysis variation using \gls{hscy1} data). Contrary to \citetalias{Leauthaud:2022}, we were able to effectively rule out contaminations by the boost factor or intrinsic alignments thanks to independent investigations of these effects. We suspect that this trend with source redshift could be explained by assuming that the shifts in mean redshift of the \gls{hscy3} sample are indicative of the necessity for a similar shift in the \gls{hscy1} sample. We note that the \citetalias{Leauthaud:2022}-trend is likewise driven by a very low measurement of the CFHTLenS-survey, which we do not analyze here.

        Our measurements are qualitatively very similar to the comparison of lensing signals in \citet{2023MNRAS.518..477A}. However, while they find a consistent lensing signal between the different surveys for all lens bins, we do find potential signs for deviations on small scales. As \citet{2023MNRAS.518..477A} do not compare tomographic measurements, use different lens samples, and a different covariance estimation, it is unclear whether the differences are caused by the analysis choices or simply the increased signal-to-noise of our data.

        Apart from that, our measurements also line up with other, tangentially related works. For example, we measure a trend for a lower lens amplitude in the \gls{kids} survey at the data level, which is reflected both in cosmological analyses where \gls{kids} consistently measures a lower $S_8$ value than its competitors \gls{des} and \gls{hsc} \citep[compare][and references therein]{2022JHEAp..34...49A}, and in direct comparisons of \gls{ggl} amplitudes using \gls{boss} lenses, such as \citet{2023MNRAS.520.5373L,2023MNRAS.518..477A}.

    \section{Conclusion}
    \label{sec:conclusion}
        In this paper, we have conducted a data-driven investigation into potential systematics impacting the \gls{desiy1} \gls{ggl} measurements. We have found evidence for some excess scatter between the measured lensing amplitudes that is driven by small-scale measurements; the excess scatter is not statistically relevant for larger scales $\geq 1\,\Mpc/h$, and vanishes when we marginalize over our uncertainty in the covariance. We have not found any statistically significant trends of lensing amplitudes with respect to splits of the \gls{desiy1} lens sample. We have found a statistically significant trend of the lensing amplitude as a function of source redshift for the first \gls{lrg} bin, and similar trends for other bins at lower levels of significance. We have further shown that both of these findings are independent of what we determine to be significant. We have investigated several potential causes for the source redshift trend and found that the most likely explanation is a shift in the mean redshift of the \gls{hscy3} source sample, which is also preferred by analyses of the \gls{hsc} collaboration. We note that we find a similar trend when investigating the \gls{hscy1} source sample. We attribute the fact that we do not find a source redshift trend (after accounting for a known \gls{hscy3} systematic) to a more stringent source redshift cut, virtually eliminating contributions by intrinsic alignments, and to our use of newer, better-validated data from lensing surveys. We have not found any evidence for trends of the lensing amplitude with potential systematics that might bias the \gls{desiy1} sample. We have also not found any evidence for significant B-modes in our measurements. We further confirm other works that find a trend in lensing amplitudes, where \gls{kids} tend to report a lower lensing signal than \gls{des} and \gls{hsc}. Apart from providing significant and independent evidence of a source redshift shift in \gls{hscy3}, we conclude that the \gls{ggl} measurements are sufficiently validated for subsequent cosmological analyses, if the \gls{hscy3} redshift shifts and the uncertainty in the covariance estimate are taken into account. 
        
    \section{Data Availability}
    \label{sec:discussion:publication}
        We make both our clustering and lensing measurements and covariance estimates public at \url{https://doi.org/10.5281/zenodo.22914838}. In particular, we offer a more conservative choice of source-lens cuts (MC, this work) along with a less conservative choice (LC), as described in Fig.~\ref{fig:all_nofz_plot}, that can be used when intrinsic alignments are forward modeled.

\section*{Acknowledgements}
    This material is based upon work supported by the U.S. Department of Energy (DOE), Office of Science, Office of High-Energy Physics, under Contract No. DE–AC02–05CH11231, and by the National Energy Research Scientific Computing Center, a DOE Office of Science User Facility under the same contract. Additional support for DESI was provided by the U.S. National Science Foundation (NSF), Division of Astronomical Sciences under Contract No. AST-0950945 to the NSF’s National Optical-Infrared Astronomy Research Laboratory; the Science and Technology Facilities Council of the United Kingdom; the Gordon and Betty Moore Foundation; the Heising-Simons Foundation; the French Alternative Energies and Atomic Energy Commission (CEA); the National Council of Humanities, Science and Technology of Mexico (CONAHCYT); the Ministry of Science, Innovation and Universities of Spain (MICIU/AEI/10.13039/501100011033), and by the DESI Member Institutions: \url{https://www.desi.lbl.gov/collaborating-institutions}. Any opinions, findings, and conclusions or recommendations expressed in this material are those of the author(s) and do not necessarily reflect the views of the U. S. National Science Foundation, the U. S. Department of Energy, or any of the listed funding agencies. 

    The authors are honored to be permitted to conduct scientific research on Iolkam Du’ag (Kitt Peak), a mountain with particular significance to the Tohono O’odham Nation.

\bibliographystyle{mnras}
\bibliography{mn-jour,all_refs}
\section*{Affiliations}
\noindent
{\footnotesize $^{1}$ Department of Astronomy and Astrophysics, UCO/Lick Observatory, University of California, 1156 High Street, Santa Cruz, CA 95064, USA \\
$^{2}$ Department of Astronomy and Astrophysics, University of California, Santa Cruz, 1156 High Street, Santa Cruz, CA 95065, USA \\
$^{3}$ Centre for Astrophysics \& Supercomputing, Swinburne University of Technology, P.O. Box 218, Hawthorn, VIC 3122, Australia \\
$^{4}$ Department of Astronomy, Tsinghua University, 30 Shuangqing Road, Haidian District, Beijing, China, 100190 \\
$^{5}$ Department of Physics, American University, 4400 Massachusetts Avenue NW, Washington, DC 20016, USA \\
$^{6}$ Department of Physics \& Astronomy and Pittsburgh Particle Physics, Astrophysics, and Cosmology Center (PITT PACC), University of Pittsburgh, 3941 O'Hara Street, Pittsburgh, PA 15260, USA \\
$^{7}$ Center for Cosmology and AstroParticle Physics, The Ohio State University, 191 West Woodruff Avenue, Columbus, OH 43210, USA \\
$^{8}$ Department of Astronomy, The Ohio State University, 4055 McPherson Laboratory, 140 W 18th Avenue, Columbus, OH 43210, USA \\
$^{9}$ The Ohio State University, Columbus, 43210 OH, USA \\
$^{10}$ Lawrence Berkeley National Laboratory, 1 Cyclotron Road, Berkeley, CA 94720, USA \\
$^{11}$ Department of Physics, Boston University, 590 Commonwealth Avenue, Boston, MA 02215 USA \\
$^{12}$ Dipartimento di Fisica ``Aldo Pontremoli'', Universit\`a degli Studi di Milano, Via Celoria 16, I-20133 Milano, Italy \\
$^{13}$ INAF-Osservatorio Astronomico di Brera, Via Brera 28, 20122 Milano, Italy \\
$^{14}$ Department of Physics \& Astronomy, University College London, Gower Street, London, WC1E 6BT, UK \\
$^{15}$ Institut d'Estudis Espacials de Catalunya (IEEC), c/ Esteve Terradas 1, Edifici RDIT, Campus PMT-UPC, 08860 Castelldefels, Spain \\
$^{16}$ Institute of Space Sciences, ICE-CSIC, Campus UAB, Carrer de Can Magrans s/n, 08913 Bellaterra, Barcelona, Spain \\
$^{17}$ Instituto de F\'{\i}sica, Universidad Nacional Aut\'{o}noma de M\'{e}xico,  Circuito de la Investigaci\'{o}n Cient\'{\i}fica, Ciudad Universitaria, Cd. de M\'{e}xico  C.~P.~04510,  M\'{e}xico \\
$^{18}$ Physics Department, Brookhaven National Laboratory, Upton, NY 11973, USA \\
$^{19}$ NSF NOIRLab, 950 N. Cherry Ave., Tucson, AZ 85719, USA \\
$^{20}$ Department of Astronomy \& Astrophysics, University of Toronto, Toronto, ON M5S 3H4, Canada \\
$^{21}$ University of California, Berkeley, 110 Sproul Hall \#5800 Berkeley, CA 94720, USA \\
$^{22}$ Institut de F\'{i}sica d’Altes Energies (IFAE), The Barcelona Institute of Science and Technology, Edifici Cn, Campus UAB, 08193, Bellaterra (Barcelona), Spain \\
$^{23}$ Departamento de F\'isica, Universidad de los Andes, Cra. 1 No. 18A-10, Edificio Ip, CP 111711, Bogot\'a, Colombia \\
$^{24}$ Observatorio Astron\'omico, Universidad de los Andes, Cra. 1 No. 18A-10, Edificio H, CP 111711 Bogot\'a, Colombia \\
$^{25}$ Center for Astrophysics $|$ Harvard \& Smithsonian, 60 Garden Street, Cambridge, MA 02138, USA \\
$^{26}$ Institute of Cosmology and Gravitation, University of Portsmouth, Dennis Sciama Building, Portsmouth, PO1 3FX, UK \\
$^{27}$ Fermi National Accelerator Laboratory, PO Box 500, Batavia, IL 60510, USA \\
$^{28}$ Institute of Astronomy, University of Cambridge, Madingley Road, Cambridge CB3 0HA, UK \\
$^{29}$ Department of Physics, The Ohio State University, 191 West Woodruff Avenue, Columbus, OH 43210, USA \\
$^{30}$ Department of Physics, University of Michigan, 450 Church Street, Ann Arbor, MI 48109, USA \\
$^{31}$ University of Michigan, 500 S. State Street, Ann Arbor, MI 48109, USA \\
$^{32}$ Department of Physics, The University of Texas at Dallas, 800 W. Campbell Rd., Richardson, TX 75080, USA \\
$^{33}$ CIEMAT, Avenida Complutense 40, E-28040 Madrid, Spain \\
$^{34}$ Aix Marseille Univ, CNRS, CNES, LAM, Marseille, France \\
$^{35}$ Department of Physics and Astronomy, University of California, Irvine, 92697, USA \\
$^{36}$ Department of Physics and Astronomy, University of Waterloo, 200 University Ave W, Waterloo, ON N2L 3G1, Canada \\
$^{37}$ Perimeter Institute for Theoretical Physics, 31 Caroline St. North, Waterloo, ON N2L 2Y5, Canada \\
$^{38}$ Waterloo Centre for Astrophysics, University of Waterloo, 200 University Ave W, Waterloo, ON N2L 3G1, Canada \\
$^{39}$ Sorbonne Universit\'{e}, CNRS/IN2P3, Laboratoire de Physique Nucl\'{e}aire et de Hautes Energies (LPNHE), FR-75005 Paris, France \\
$^{40}$ Departament de F\'{i}sica, Serra H\'{u}nter, Universitat Aut\`{o}noma de Barcelona, 08193 Bellaterra (Barcelona), Spain \\
$^{41}$ Instituci\'{o} Catalana de Recerca i Estudis Avan\c{c}ats, Passeig de Llu\'{\i}s Companys, 23, 08010 Barcelona, Spain \\
$^{42}$ IRFU, CEA, Universit\'{e} Paris-Saclay, F-91191 Gif-sur-Yvette, France \\
$^{43}$ Institute for Astronomy, University of Edinburgh, Royal Observatory, Blackford Hill, Edinburgh EH9 3HJ, UK \\
$^{44}$ Ruhr University Bochum, Faculty of Physics and Astronomy, Astronomical Institute (AIRUB), German Centre for Cosmological Lensing, 44780 Bochum, Germany \\
$^{45}$ Instituto de Astrof\'{i}sica de Andaluc\'{i}a (CSIC), Glorieta de la Astronom\'{i}a, s/n, E-18008 Granada, Spain \\
$^{46}$ Departament de F\'isica, EEBE, Universitat Polit\`ecnica de Catalunya, c/Eduard Maristany 10, 08930 Barcelona, Spain \\
$^{47}$ Department of Physics and Astronomy, Sejong University, 209 Neungdong-ro, Gwangjin-gu, Seoul 05006, Republic of Korea \\
$^{48}$ Queensland University of Technology,  School of Chemistry \& Physics, George St, Brisbane 4001, Australia \\
$^{49}$ Max Planck Institute for Extraterrestrial Physics, Gie\ss enbachstra\ss e 1, 85748 Garching, Germany \\
$^{50}$ SLAC National Accelerator Laboratory, 2575 Sand Hill Road, Menlo Park, CA 94025, USA \\
$^{51}$ National Astronomical Observatories, Chinese Academy of Sciences, A20 Datun Road, Chaoyang District, Beijing, 100101, P.~R.~China \\
}
\clearpage
\begin{appendix}
\setcounter{figure}{0}
\renewcommand{\thefigure}{A.\arabic{figure}}
\section{Secondary figures}
As this manuscript contains many different tests, for which we want to present individual figures to let the reader interpret our results, we place some less critical Figures here to not hamper the flow of the document.

\label{sec:results:homogeneity_tests}
\begin{figure*}
    \centering
    \includegraphics[width=0.9\linewidth]{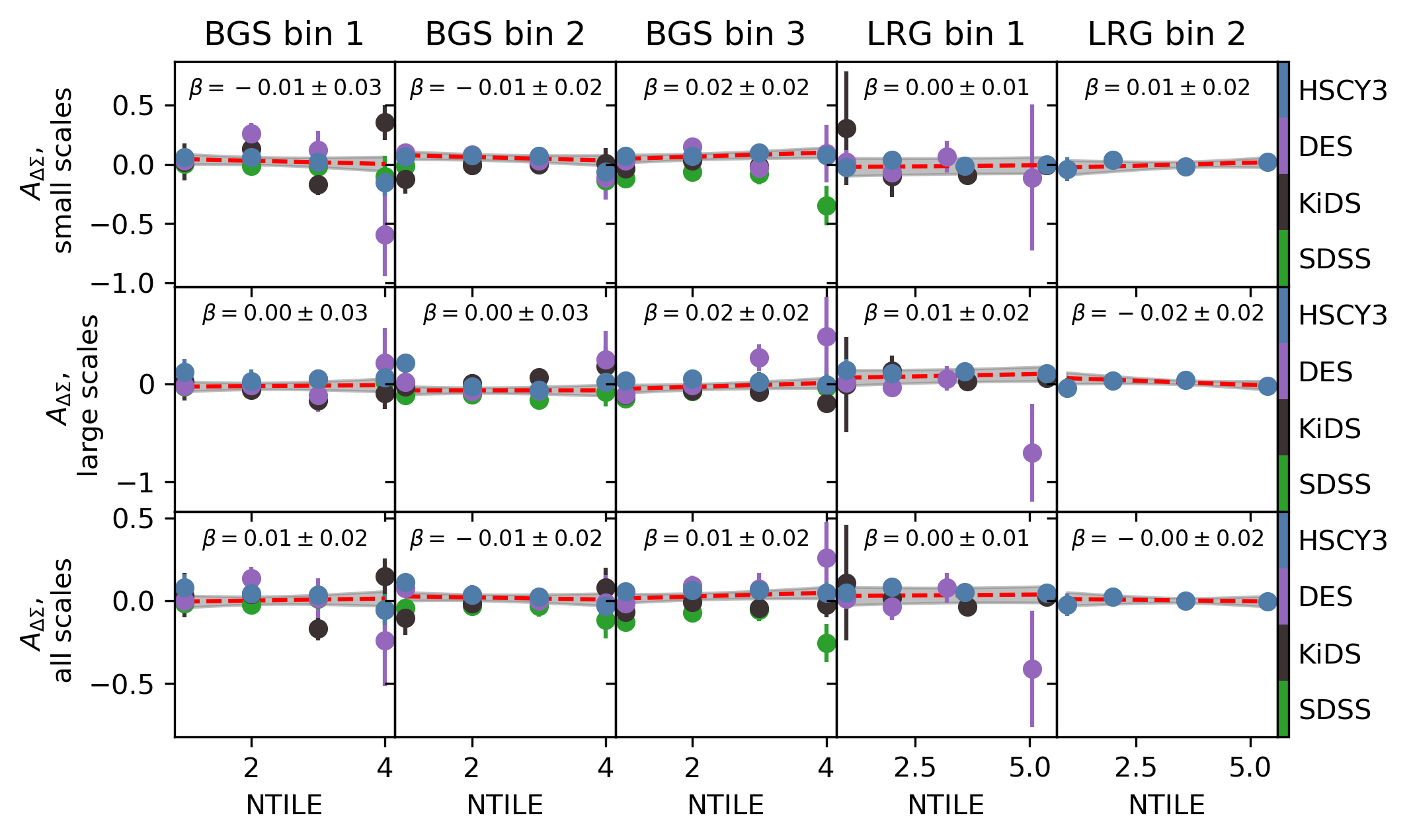}
    \caption{The lensing amplitude $\ADS$ as a function of the parameter NTILE. For the \gls{bgs} galaxies, we split the samples into 4 bins with NTILE=1,2,3,4, respectively. For the \gls{lrg} galaxies, we split into NTILE=1, NTILE=2, NTILE$\,\in \{3,4\}$, NTILE$\,\in \{5,6,7\}$.}
    \label{fig:deltasigma_amplitude_ntile_split}
\end{figure*}

\begin{figure*}
    \centering
    \includegraphics[width=0.9\linewidth]{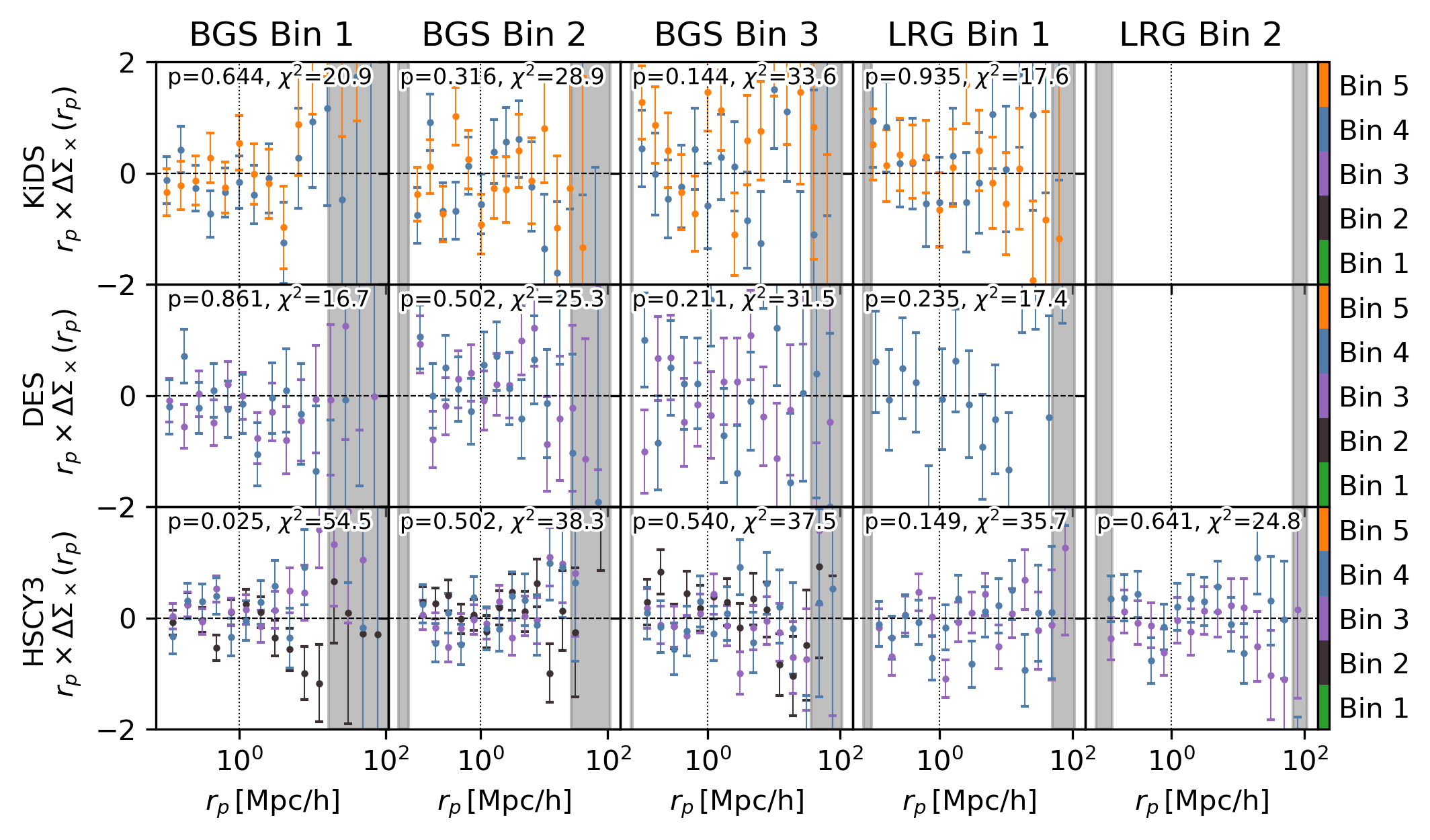}
    \caption{Lensing B-modes for the different source surveys. The grey shaded regions are excluded due to our scale cuts outlined in Sect.~\ref{sec:GGL_measurements:measurement_setup}. The dashed line corresponds to the boundary between small and large scales. We also denote $\chi^2$ and $p$-values for each source survey. We do not detect significant B-modes for any of the bins.}
    \label{fig:deltasigma_bmodes}
\end{figure*}

\begin{figure*}
    \centering
    \includegraphics[width=0.9\linewidth]{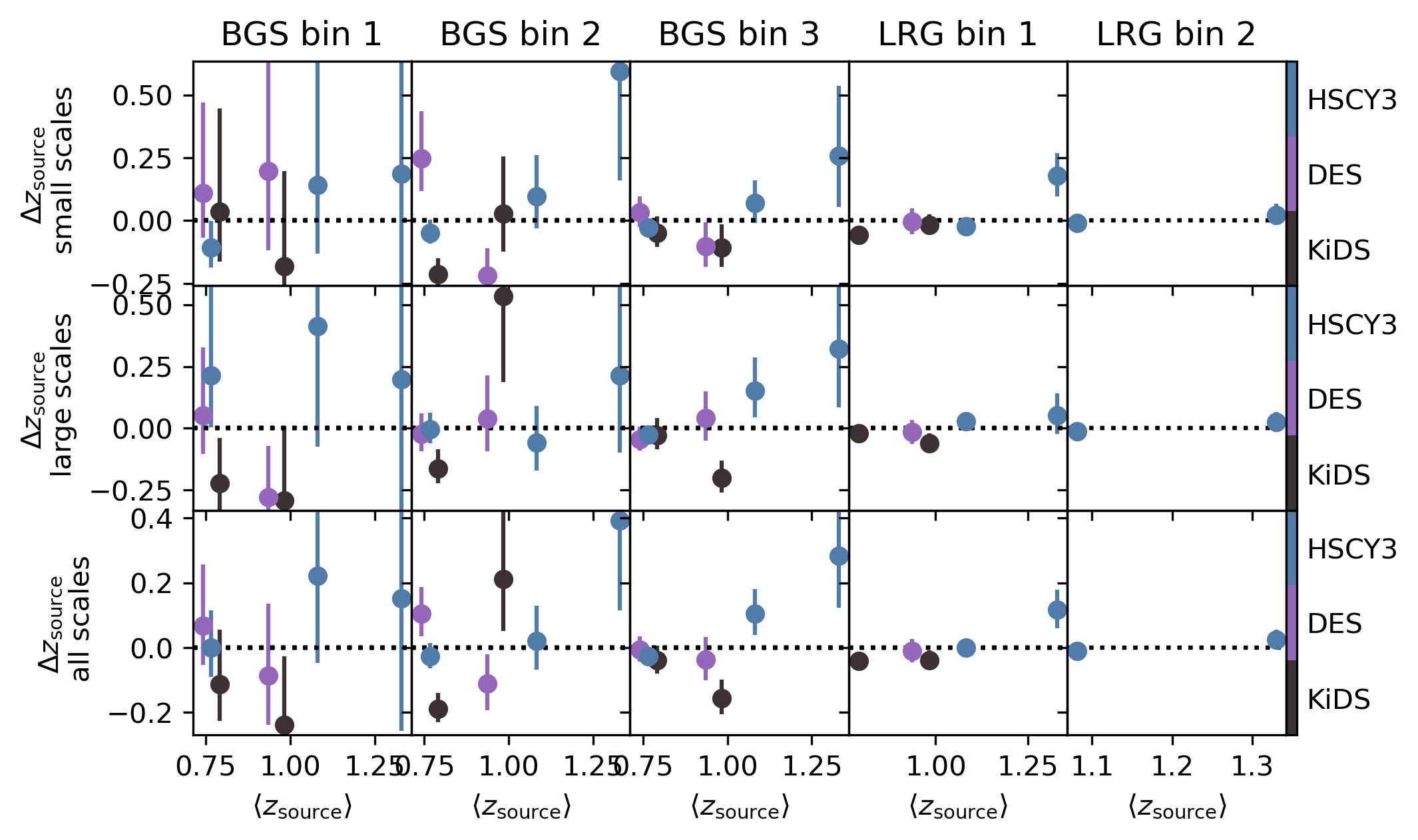}
    \caption{Shifts in mean redshift that would set all lensing amplitudes to the mean lensing amplitude. We calculate the shifts by determining the linear offset of the source redshift distribution that would lead to a $\Sigmacrit$ that yields a lensing amplitude of unity.}
    \label{fig:delta_z}
\end{figure*}

\begin{figure*}
    \centering
    \includegraphics[width=\linewidth]{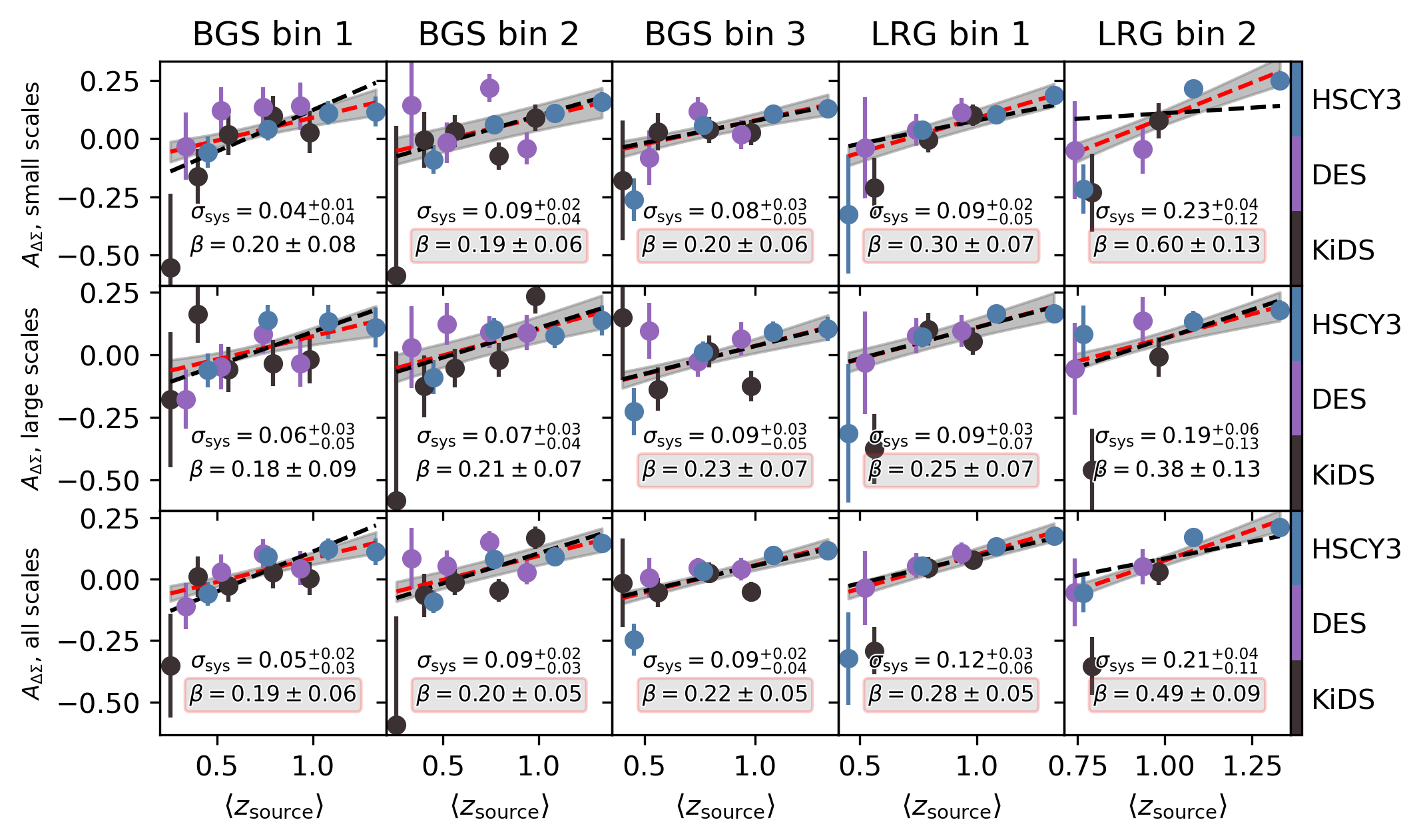}
    \caption{Same as Fig.~\ref{fig:source_redshift_slope}, just including all bins where the sources, on average, are behind the lens. The black dashed line represents the slope expected from systematic effects, as estimated by simulations. This plot was created post-unblinding.}
    \label{fig:source_redshift_slope_allbins}
\end{figure*}

\section{Additional tables for slopes}
\label{app:p-values}
    \begin{table*}
    \centering
        \caption{Measured slopes and uncertainties for the lensing amplitude $A_{\ds}$ as a function of potential systematic biases. Values that qualify as outliers are highlighted in red. Values and errors followed by an asterisk ${}^*$ have been re-scaled by the appropriate order of magnitude to fit the notation pattern.}
        \label{tab:slopes_systematics}
        
    \begin{tabular}{|c|c|c|c|c|}
        \hline
 & & small scales & large scales & all scales \\
        \hline
& & \multicolumn{3}{c|}{\textbf{EBV}} \\
        \multirow{3}{*}{BGS} & Bin 1 & $2.315\pm 1.400$& $-2.083\pm 1.718$& $0.447\pm 1.227$\\
        & Bin 2 & $-0.881\pm 1.018$& $-1.110\pm 1.335$& $-1.032\pm 0.874$\\
        & Bin 3 & $1.029\pm 1.023$& $2.465\pm 1.407$& $1.461\pm 0.928$\\
        \hline
        \multirow{3}{*}{LRG} & Bin 1 & $-1.111\pm 0.896$& $0.645\pm 1.475$& $-0.260\pm 0.899$\\
        & Bin 2 & $0.796\pm 151.365^*$& $3.575\pm 1.826$& $1.758\pm 1.192$\\
        & Bin 3 & -- & -- & -- \\
        \hline
& & \multicolumn{3}{c|}{\textbf{NTILE}} \\
        \multirow{3}{*}{BGS} & Bin 1 & $-0.013\pm 0.027$& $0.466\pm 3.224^*$& $0.599\pm 2.191^*$\\
        & Bin 2 & $-0.014\pm 0.020$& $0.207\pm 25.957^*$& $-0.646\pm 1.663^*$\\
        & Bin 3 & $0.017\pm 0.019$& $0.020\pm 0.024$& $0.011\pm 0.015$\\
        \hline
        \multirow{3}{*}{LRG} & Bin 1 & $0.338\pm 1.270^*$& $0.925\pm 1.761^*$& $0.211\pm 1.199^*$\\
        & Bin 2 & $0.010\pm 0.021$& $-0.017\pm 0.021$& $-0.324\pm 1.540^*$\\
        & Bin 3 & -- & -- & -- \\
        \hline
& & \multicolumn{3}{c|}{\textbf{STARDENS}} \\
        \multirow{3}{*}{BGS} & Bin 1 & $0.438\pm 1.200^*$& $-0.221\pm 0.161^*$& $-0.847\pm 10.304^*$\\
        & Bin 2 & $-0.156\pm 0.111^*$& $-0.237\pm 1.295^*$& $-0.111\pm 0.084^*$\\
        & Bin 3 & $0.965\pm 1.303^*$& $0.347\pm 0.159^*$& $0.140\pm 0.106^*$\\
        \hline
        \multirow{3}{*}{LRG} & Bin 1 & $-0.877\pm 1.263^*$& $0.912\pm 2.081^*$& $0.353\pm 1.232^*$\\
        & Bin 2 & $0.809\pm 2.144^*$& $0.653\pm 2.980^*$& $0.105\pm 0.172^*$\\
        & Bin 3 & -- & -- & -- \\
        \hline
& & \multicolumn{3}{c|}{\textbf{PSFDEPTH\_Z}} \\
        \multirow{3}{*}{BGS} & Bin 1 & $0.305\pm 0.240^*$& $0.588\pm 0.297^*$& $0.290\pm 0.191^*$\\
        & Bin 2 & $0.162\pm 0.181^*$& $0.382\pm 0.208^*$& $0.235\pm 0.141^*$\\
        & Bin 3 & $0.496\pm 1.694^*$& $0.325\pm 0.228^*$& $0.247\pm 0.142^*$\\
        \hline
        \multirow{3}{*}{LRG} & Bin 1 & $-0.116\pm 0.177^*$& $0.187\pm 0.286^*$& $0.891\pm 1.697^*$\\
        & Bin 2 & $-0.238\pm 0.284^*$& $-0.121\pm 0.363^*$& $-0.283\pm 0.219^*$\\
        & Bin 3 & -- & -- & -- \\
        \hline
& & \multicolumn{3}{c|}{\textbf{PSFSIZE\_Z}} \\
        \multirow{3}{*}{BGS} & Bin 1 & $-0.264\pm 0.166$& $-0.287\pm 0.215$& $-0.264\pm 0.142$\\
        & Bin 2 & $-0.118\pm 0.132$& $-0.482\pm 0.194$& $-0.083\pm 0.120$\\
        & Bin 3 & $0.021\pm 0.129$& $-0.201\pm 0.201$& $-0.054\pm 0.120$\\
        \hline
        \multirow{3}{*}{LRG} & Bin 1 & $-0.032\pm 0.124$& $-0.397\pm 0.174$& $-0.098\pm 0.108$\\
        & Bin 2 & $-0.046\pm 0.201$& $-0.242\pm 0.248$& $-0.109\pm 0.154$\\
        & Bin 3 & -- & -- & -- \\
        \hline

    \end{tabular}

    \end{table*}

    \begin{table}
        \centering
        \caption{Measured slopes and uncertainties for the lensing amplitude $A_{\ds}$ as a function of source redshift. Values that qualify as outliers are highlighted in red.}
        \label{tab:slopes_source_redshift}
        
    \begin{tabular}{|c|c|c|c|c|}
        \hline
 & & small scales & large scales & all scales \\
        \hline
        \multirow{3}{*}{BGS} & Bin 1 & $0.076\pm 0.124$& $0.018\pm 0.132$& $0.042\pm 0.092$\\
        & Bin 2 & $0.149\pm 0.082$& $0.103\pm 0.093$& $0.121\pm 0.062$\\
        & Bin 3 & $0.124\pm 0.075$& $0.192\pm 0.084$& $0.152\pm 0.056$\\
        \hline
        \multirow{3}{*}{LRG} & Bin 1 & \textcolor{red}{$0.303\pm 0.093$}& $0.152\pm 0.109$& \textcolor{red}{$0.235\pm 0.071$}\\
        & Bin 2 & $0.153\pm 0.197$& $0.189\pm 0.199$& $0.171\pm 0.141$\\
        & Bin 3 & -- & -- & -- \\
        \hline

    \end{tabular}

    \end{table}

\section{Details on Magnification bias calculations}
\label{app:magnification_bias}
    Estimating the magnification bias for a spectroscopic survey presents a significant challenge. The standard approach of artificially de-magnifying galaxies and counting those that drop out of the selection is complicated by the fact that the selection is determined not only by simple magnitude limits but also by the fiber flux—that is, the amount of flux expected to be observed within a DESI fiber. Magnification increases the solid angle of a galaxy, and the relationship between this effect and the fiber flux is non-trivial, strongly depending on the model used to estimate the galaxy’s flux as a function of distance from its center. \citet{Wenzl:2024} developed a model to address this issue; we have adapted this model to the selection criteria of DESI and applied it in our analysis.

\begin{figure}
    \centering
    \includegraphics[width=0.45\linewidth]{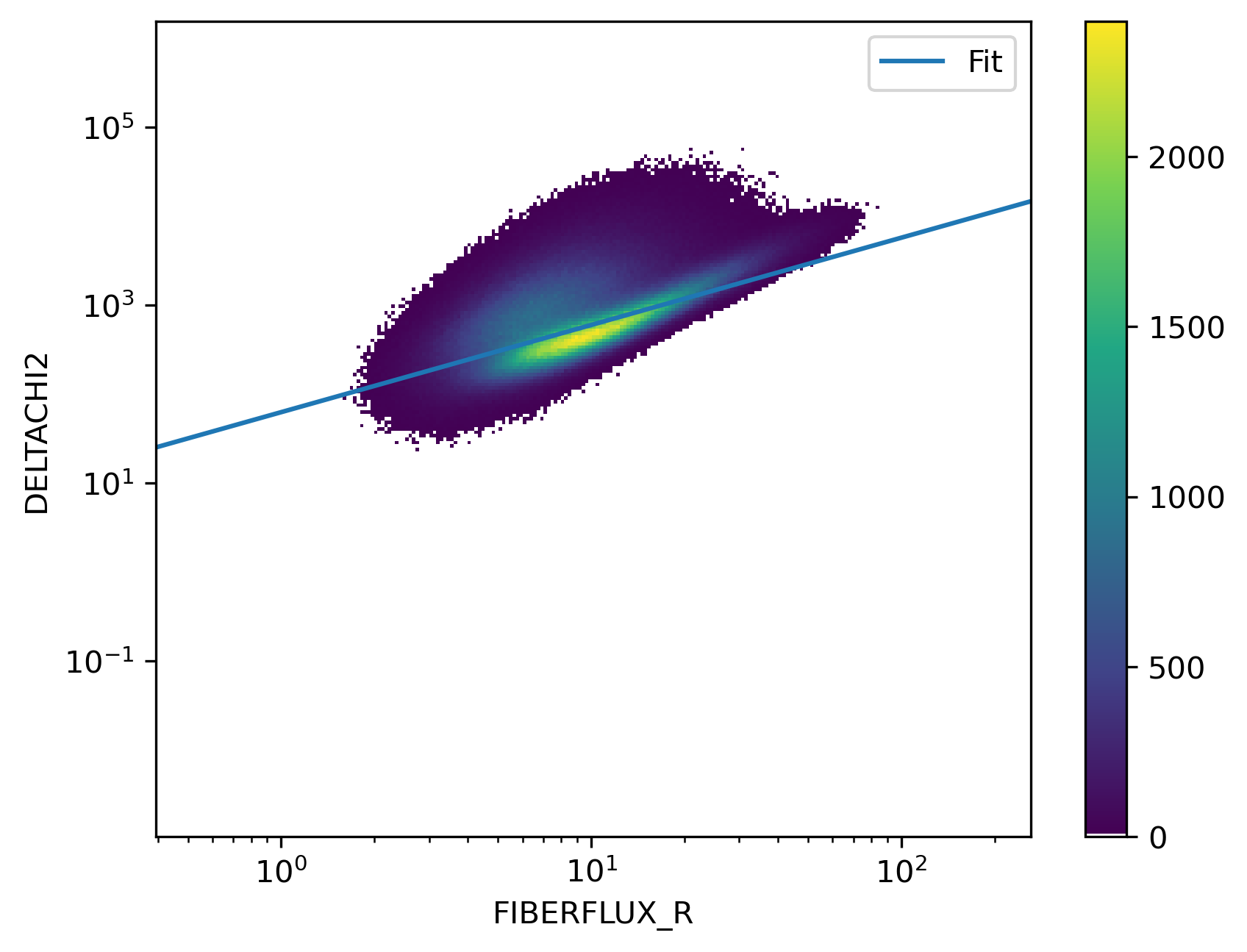}
    \includegraphics[width=0.45\linewidth]{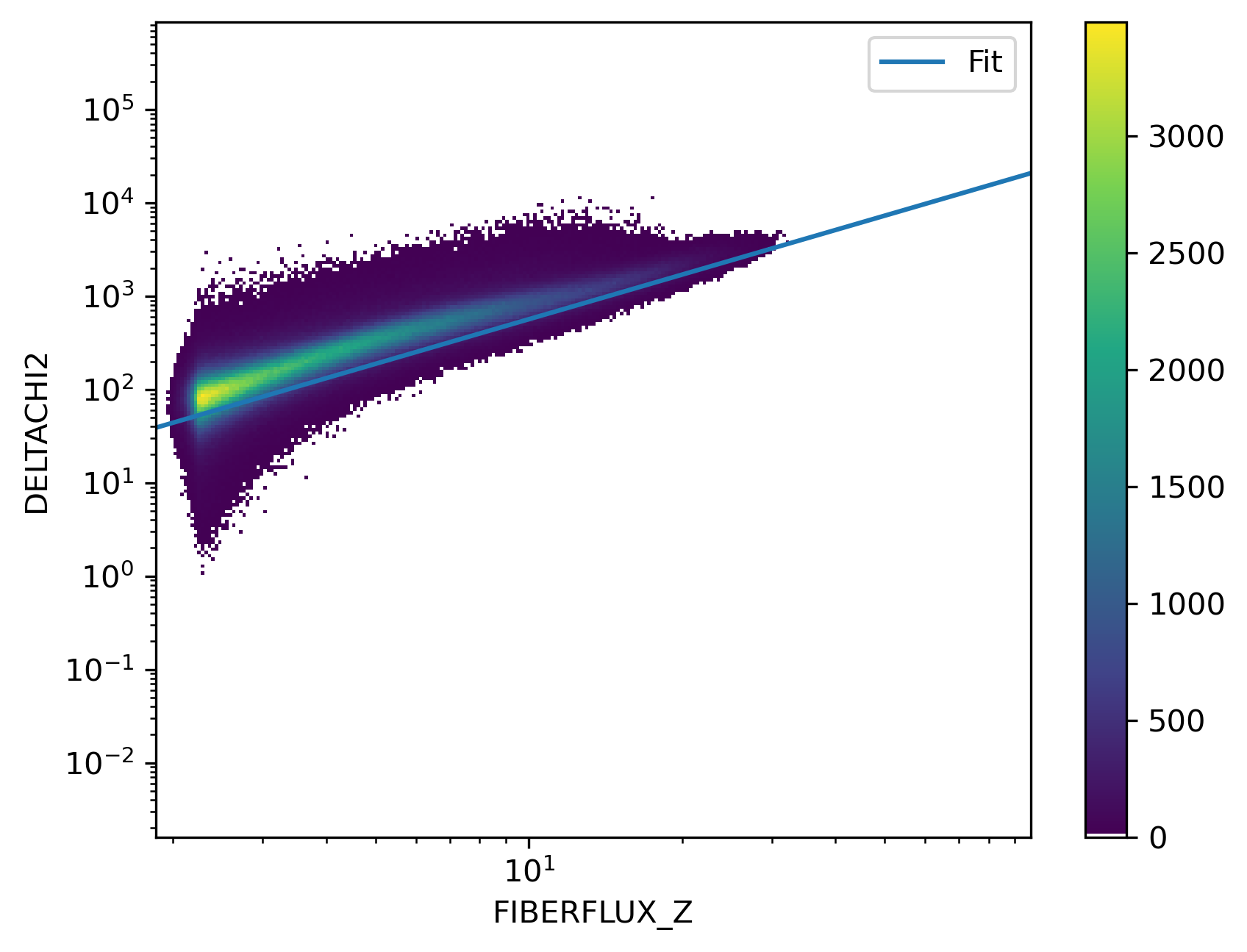}
    \caption{Validation plots for the fit of the \texttt{DELTACHI2} quantity, for which we can not directly model the impact of magnification. We assume that \texttt{DELTACHI2} is solely a function of the fiber flux and model its changes under magnification via the change in fiberflux. Left: BGS, right: LRG}
    \label{fig:magnification_secondary_validation}
    \end{figure}

    An additional secondary selection criterion is the \texttt{DELTACHI2} flag of galaxies. This flag quantifies how much better the best-fitting model fits the data compared to the second-best fit (i.e. \texttt{DELTACHI2}$=\chi^2_\text{2nd best fit}-\chi^2_\text{best fit}$), with a successful observation requiring that \texttt{DELTACHI2} exceeds a certain minimum value. There is no straightforward relationship between this value and magnification effects. Therefore, we approximate it as a power-law function of the fiber flux and fit this model to the available data, as shown in Figure~\ref{fig:magnification_secondary_validation}. While this approximation may be somewhat simplistic, we consider it adequate because this effect is extremely subdominant (see Tables~\ref{tab:magbias_cuts_BGS} and \ref{tab:magbias_cuts_LRG} for comparison).
    
\begin{figure}
        \centering
        \includegraphics[width=0.7\linewidth]{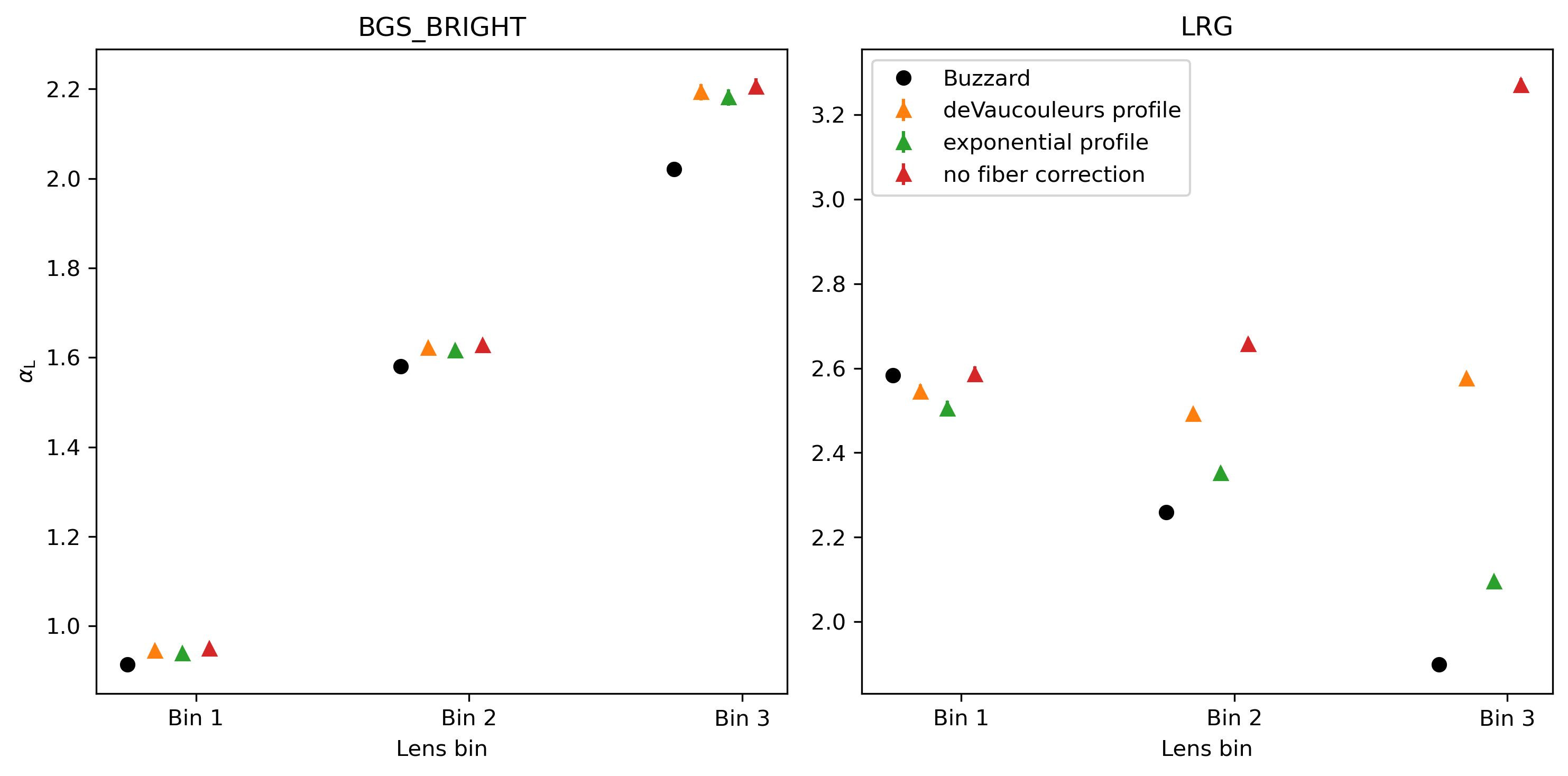}
        \caption{Magnification bias measured with different methods and comparison with simulations}
        \label{fig:magnification_bias}
    \end{figure}

    We present the values of the magnification bias slope, $\alpha_L$, in Figure~\ref{fig:magnification_bias}. One can see that, for the \gls{lrg} sample, the assumption of the light profile yields substantial differences in the measured magnification bias, especially for the last \gls{lrg} bin. Comparing with Tab.~\ref{tab:magbias_cuts_LRG} one can see that this is reasonable: The `faint limit' cut includes a cut on the expected flux within a \gls{desi} fiber, and the model for this effect under magnification strongly depends on the assumed profile shape. We cite the value for the deVaucouleur's profile and estimate the uncertainty as the standard deviation between this, the exponential profile, and the case with no correction for the fiber magnitude. This is certainly an over-estimation of the actual uncertainty, and future work will use the actual tabulated luminosity profiles of the observed galaxies so we do not have to rely on profile shape assumptions. However, since the overall impact of magnification bias is low, this treatment does not introduce any substantial uncertainties in the measurement.
    
    \begin{table}
        \centering
        \begin{tabular}{l|c|c|c|c|c|c|}
Category  & \multicolumn{2}{c|}{Bin 1} & \multicolumn{2}{c|}{Bin 2} & \multicolumn{2}{c|}{Bin 3}\\
& \# cut & $\alpha$ & \# cut & $\alpha$ & \# cut & $\alpha$ \\
\hline
FMC & 2 & 0.00 & 0 & -0.00 & 1 & 0.00 \\
g-r $>$ -1 & 0 & 0.00 & 0 & 0.00 & 0 & 0.00 \\
g-r $<$ 4 & 0 & 0.00 & 0 & 0.00 & 0 & 0.00 \\
r-z $>$ -1 & 0 & 0.00 & 0 & 0.00 & 0 & 0.00 \\
r-z $<$ 4 & 0 & 0.00 & 0 & 0.00 & 0 & 0.00 \\
\textbf{r $<$ 19.5} & 88 & 0.01 & 909 & 0.07 & 8284 & 0.92 \\
r $>$ 12 & 0 & 0.00 & 0 & 0.00 & 0 & 0.00 \\
rfibtotmag $>$ 15 & 0 & 0.00 & 0 & 0.00 & 0 & 0.00 \\
\textbf{absolute magnitude cuts} & 15792 & 0.94 & 20778 & 1.56 & 11911 & 1.32 \\
tsnr cut & 0 & -0.00 & 0 & 0.00 & 0 & 0.00 \\
deltachi2 cut & 185 & 0.01 & 176 & 0.01 & 203 & 0.02 \\
\hline
\# Galaxies in bin  & 841359 & --  & 666330 & --  & 451978 & -- \end{tabular}

        \caption{Contribution to the magnification bias for different cuts done to the \gls{bgs} sample}
        \label{tab:magbias_cuts_BGS}
    \end{table}
    
    \begin{table}
        \centering
        \begin{tabular}{l|c|c|c|c|c|c|}
Category  & \multicolumn{2}{c|}{Bin 1} & \multicolumn{2}{c|}{Bin 2} & \multicolumn{2}{c|}{Bin 3}\\
& \# cut & $\alpha$ & \# cut & $\alpha$ & \# cut & $\alpha$ \\
\hline
zfibertotmag $>$ 17.5 & 0 & 0.00 & 0 & 0.00 & 0 & 0.00 \\
non-stellar cut & 0 & 0.00 & 0 & 0.00 & 0 & 0.00 \\
\textbf{faint limit} & 657 & 0.06 & 7095 & 0.46 & 35108 & 2.04 \\
low-z cuts & 0 & 0.00 & 0 & 0.00 & 0 & 0.00 \\
\textbf{double sliding cuts and high-z extension} & 24546 & 2.42 & 30343 & 1.97 & 8391 & 0.49 \\
absolute magnitude cuts & 0 & 0.00 & 0 & 0.00 & 0 & 0.00 \\
tsnr cut & 0 & 0.00 & 0 & 0.00 & 0 & -0.00 \\
deltachi2 cut & 360 & 0.04 & 601 & 0.04 & 797 & 0.05 \\
\hline
\# Galaxies in bin  & 506911 & --  & 771894 & --  & 859822 & -- \end{tabular}

        \caption{Contribution to the magnification bias for different cuts done to the \gls{lrg} sample}
        \label{tab:magbias_cuts_LRG}
    \end{table}

\section{Analysis of first-year HSC data}
\label{app:hscy1}
    As some projects perform their cosmological parameter inference with \gls{hscy1} data, we repeat the main points of our analysis, in particular the lens homogeneity test and the source redshift test, and replace the third-year \gls{hsc} data with its first-year version, as presented by \citet{Mandelbaum:2018,Hikage:2019}.

    We note that we do not revisit the scale cuts, meaning that our covariance (especially the B-mode covariance) may not be entirely accurate on larger scales. When investigating the redshift distributions, \gls{hscy1} appears to be at slightly larger average redshifts; in particular, their second source redshift bin is included for measurements of the 1st \gls{lrg} bin.

    In a B-mode analysis, we find for the lens bins in ascending order $\chi^2$ values of $43.8, 32.9, 53.1, 55.7,\text{ and } 30.2$, corresponding to $p$-values of $0.174, 0.744, 0.066, 0.077, \text{ and } 0.352$ using the \gls{hscy1} source bins. It can be seen that, in contrast to \gls{hscy3}, no lens bin shows a $p$-value of $p<0.05$. However, at higher redshifts, we find significantly higher $\chi^2$ values for the third \gls{bgs} and the first \gls{lrg} bins, but the significance does not cross the traditional $p<0.05$ threshold.

    \begin{figure*}
        \centering
        \includegraphics[width=0.9\linewidth]{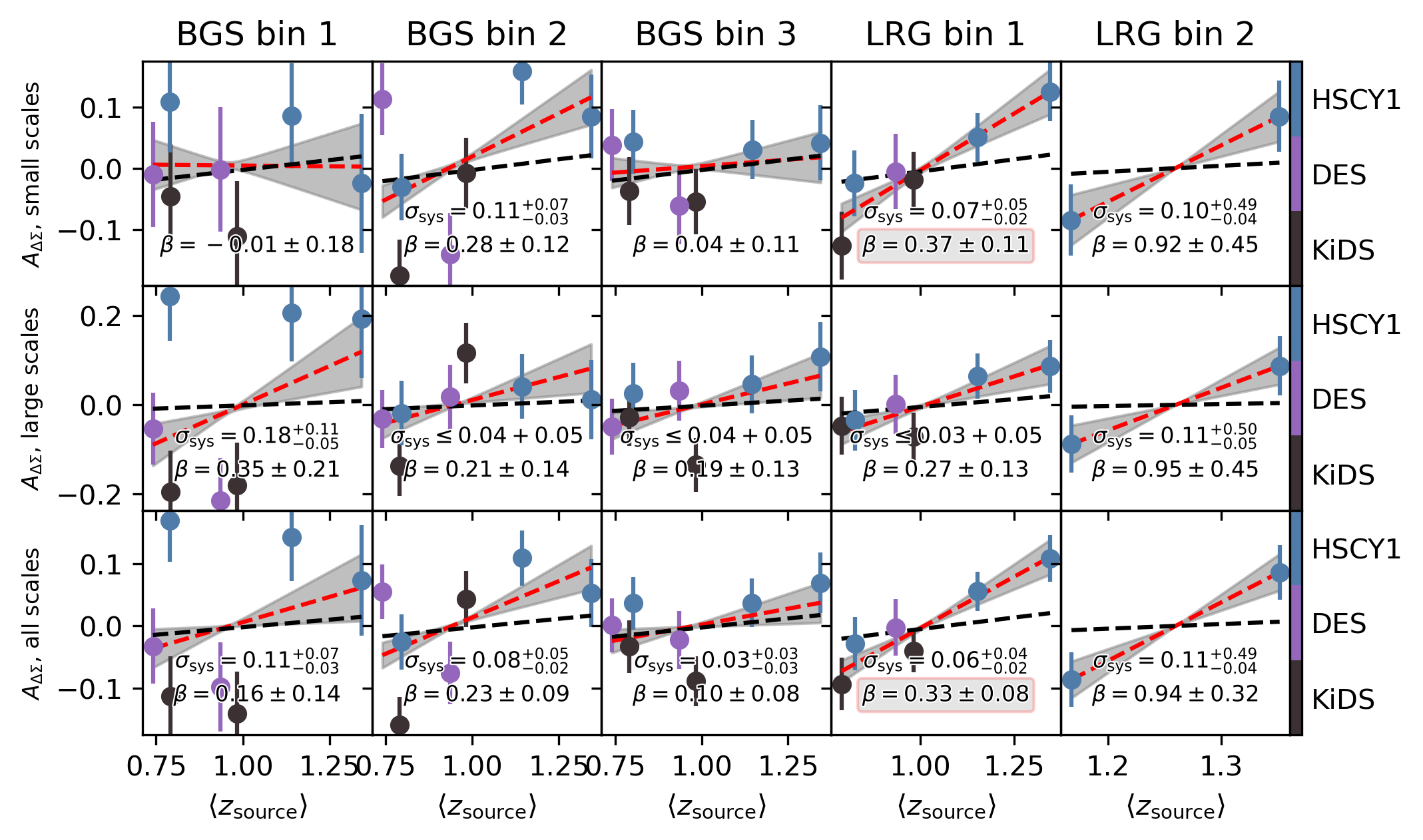}
        \caption{Same as Fig.~\ref{fig:source_redshift_slope}, but replacing the \gls{hscy3} data with \gls{hscy1}.}
        \label{fig:source_redshift_slope_hscy1}
    \end{figure*}

    The lens homogeneity tests do not show any significant trends when the \gls{hscy3} data are replaced with \gls{hscy1}. The source redshift tests also show a similar picture to the fiducial results: Apart from the first \gls{bgs} bin at small scales, all lens bins and all scales prefer a positive slope of the lensing amplitude $\ADS$ as a function of redshift. We show this in Fig.~\ref{fig:source_redshift_slope_hscy1}. We note that \gls{hscy1} adds one additional redshift bin (namely its second source tomographic bin) to the analysis of the first \gls{lrg} bin. We note, however, that this bin is very consistent with the measurements of \gls{kids} and \gls{des} at similar redshifts and is not responsible for the significance of the slope measurement. 
    
    We also note that \gls{hscy1} measures significantly higher lens amplitudes in the first \gls{bgs} bin, particularly at large scales. The first \gls{bgs} bin is most impacted by measurements at large angular scales, so potential issues with the small survey area of \gls{hscy1} could be responsible for this discrepancy. This is also supported by an investigation of the lensing signal around random positions: For both \gls{hscy1} and \gls{hscy3}, the signal around randoms is consistent with zero on small scales. On large scales, however, \gls{hscy1} measures a positive $\ds$ signal around randoms, whereas \gls{hscy3} measures a negative. 

    We conclude with the recommendation to use the third-year \gls{hsc} data for cosmological analyses. Although, in contrast to \gls{hscy3}, \gls{hscy1} did not report issues with B-modes or photometric redshifts, we have reason to believe that the same redshift issues present in the third-year analysis were also present in the first-year data, albeit remaining undetected due to its lower signal-to-noise ratio.

\section{Discussion of matched filter template}
    \label{sec:app:matched_filter_template}
    Here we show that the choice of an inaccurate template datavector $\mathbf{t}'$ yields unbiased comparisons between lensing amplitudes, but achieves a lower signal-to-noise than the optimal template $\mathbf{t}$. We consider a data vector $\mathbf{d}$ modeled as
    \begin{equation}
        \mathbf{d} = A_{\rm true} \mathbf{t} + \mathbf{n},
    \end{equation}
    where $\mathbf{t}$ is the true noiseless signal shape, and $\mathbf{n}$ is noise with covariance $\mathbf{C}$. The optimal linear estimator for the amplitude $A_{\rm true}$ is obtained using weights
    \begin{equation}
        \mathbf{w} = \frac{\mathbf{C}^{-1} \mathbf{t}}{\mathbf{t}^\mathsf{T} \mathbf{C}^{-1} \mathbf{t}}.
    \end{equation}
    Applying these weights, the matched filter amplitude is
    \begin{equation}
        A = \mathbf{w}^\mathsf{T} \mathbf{d} = A_{\rm true} + \mathbf{w}^\mathsf{T} \mathbf{n}.
    \end{equation}
    Since the noise has zero mean, the estimator is unbiased:
    \begin{equation}
        \langle A \rangle = A_{\rm true}.
    \end{equation}
    The variance of $A$ is given by
    \begin{equation}
        \sigma_A^2 = \mathbf{w}^\mathsf{T} \mathbf{C} \mathbf{w} = \frac{1}{\mathbf{t}^\mathsf{T} \mathbf{C}^{-1} \mathbf{t}}.
    \end{equation}
    Thus, the S/N is
    \begin{equation}
        \text{SNR}_t = A_{\rm true} \sqrt{\mathbf{t}^\mathsf{T} \mathbf{C}^{-1} \mathbf{t}}.
    \end{equation}

    \subsection*{Using an Arbitrary Template $\mathbf{t}'$}
    Now we consider an alternative template $\mathbf{t}'$ (which may not match $\mathbf{t}$). The weights for this template are:
    \begin{equation}
        \mathbf{w}' = \frac{\mathbf{C}^{-1} \mathbf{t}'}{\mathbf{t}'{}^\mathsf{T} \mathbf{C}^{-1} \mathbf{t}'}.
    \end{equation}
    Applying these weights to the data, the amplitude estimate is
    \begin{equation}
        A' = \mathbf{w}'{}^\mathsf{T} \mathbf{d} = A_{\rm true} \frac{\mathbf{t}'{}^\mathsf{T} \mathbf{C}^{-1} \mathbf{t}}{\mathbf{t}'{}^\mathsf{T} \mathbf{C}^{-1} \mathbf{t}'} + \mathbf{w}'{}^\mathsf{T} \mathbf{n}.
    \end{equation}
    Defining the bias factor $\alpha$ as
    \begin{equation}
        \alpha \equiv \frac{\mathbf{t}'{}^\mathsf{T} \mathbf{C}^{-1} \mathbf{t}}{\mathbf{t}'{}^\mathsf{T} \mathbf{C}^{-1} \mathbf{t}'},
    \end{equation}
    the expected amplitude is
    \begin{equation}
        \langle A' \rangle = \alpha A_{\rm true}.
    \end{equation}
    While this means that lensing amplitudes are re-scaled by $\alpha$, the comparison between different lensing amplitudes remains unbiased. The variance of $A'$ is
    \begin{equation}
        \sigma_{A'}^2 = \mathbf{w}'{}^\mathsf{T} \mathbf{C} \mathbf{w}' = \frac{1}{\mathbf{t}'{}^\mathsf{T} \mathbf{C}^{-1} \mathbf{t}'}.
    \end{equation}
    Thus, the signal-to-noise ratio using $\mathbf{t}'$ is
    \begin{equation}
        \text{SNR}_{t'} = \frac{\alpha A_{\rm true}}{\sigma_{A'}} = A_{\rm true} \frac{\mathbf{t}'{}^\mathsf{T} \mathbf{C}^{-1} \mathbf{t}}{\sqrt{\mathbf{t}'{}^\mathsf{T} \mathbf{C}^{-1} \mathbf{t}'}}.
    \end{equation}

    \subsection*{Comparison of Signal-to-Noise Ratios}
    We compare the two S/N expressions:
    \begin{align}
        \text{SNR}_t &= A_{\rm true} \sqrt{\mathbf{t}^\mathsf{T} \mathbf{C}^{-1} \mathbf{t}}, \\
        \text{SNR}_{t'} &= A_{\rm true} \frac{\mathbf{t}'{}^\mathsf{T} \mathbf{C}^{-1} \mathbf{t}}{\sqrt{\mathbf{t}'{}^\mathsf{T} \mathbf{C}^{-1} \mathbf{t}'}}.
    \end{align}
    Using the Cauchy-Schwarz inequality\footnote{The inverse covariance matrix is symmetric and positive definite, so the function $(t,t') \mapsto t^\mathsf{T} C^{-1} t'$ is an inner product, and we can apply the Cauchy-Schwarz inequality.}
    \begin{equation}
        (\mathbf{t}'{}^\mathsf{T} \mathbf{C}^{-1} \mathbf{t})^2 \leq (\mathbf{t}'{}^\mathsf{T} \mathbf{C}^{-1} \mathbf{t}') (\mathbf{t}^\mathsf{T} \mathbf{C}^{-1} \mathbf{t}),
    \end{equation}
    we obtain
    \begin{equation}
        \frac{\mathbf{t}'{}^\mathsf{T} \mathbf{C}^{-1} \mathbf{t}}{\sqrt{\mathbf{t}'{}^\mathsf{T} \mathbf{C}^{-1} \mathbf{t}'}} \leq \sqrt{\mathbf{t}^\mathsf{T} \mathbf{C}^{-1} \mathbf{t}}.
    \end{equation}
    Multiplying by $A_{\rm true}$, we conclude that
    \begin{equation}
        \text{SNR}_{t'} \leq \text{SNR}_t.
    \end{equation}

\section{Boost factor}
    \label{app:boost_factor}
\begin{figure}
        \centering
        \includegraphics[width=\linewidth]{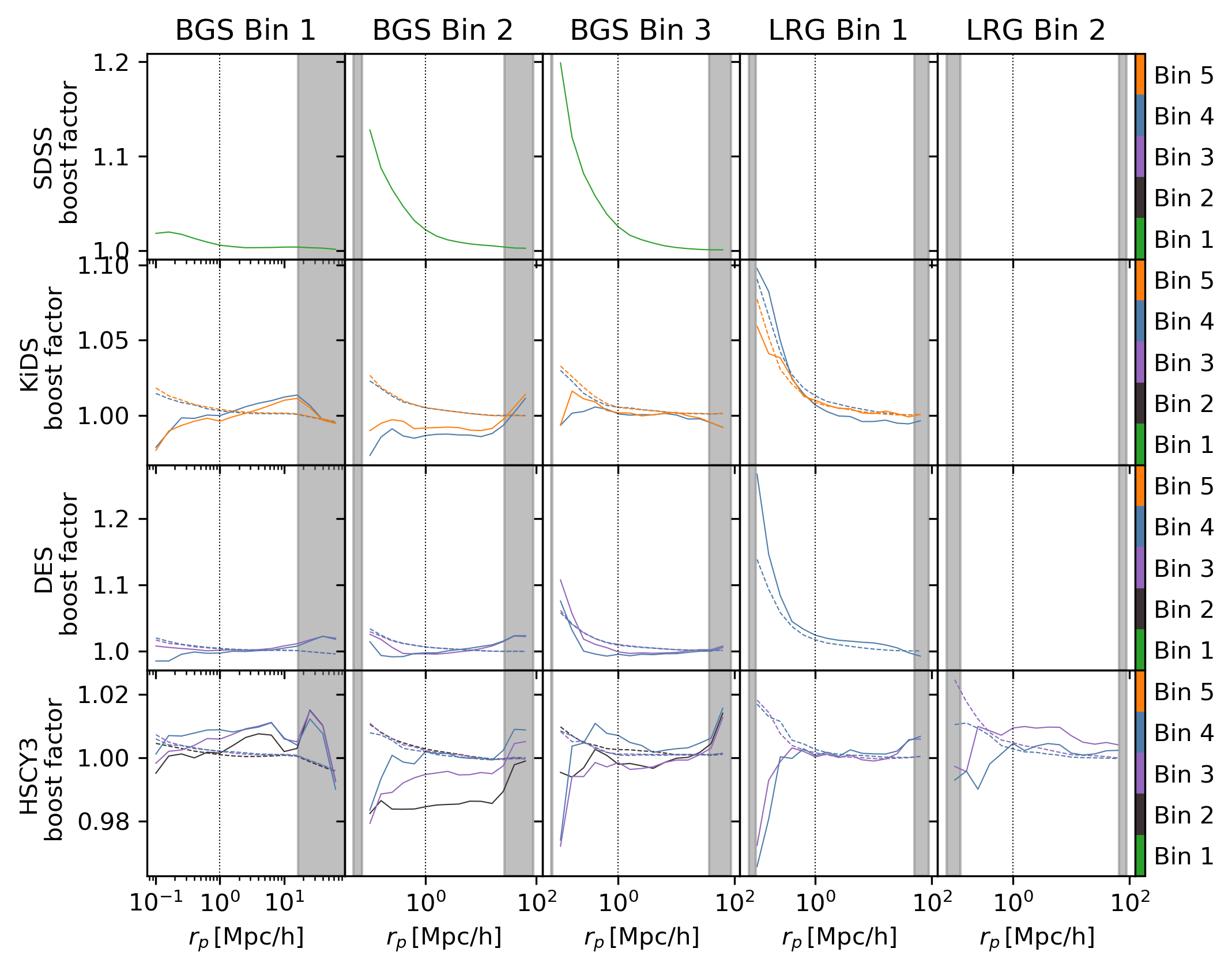}
        \caption{The boost factor for the measurements presented in Fig.~\ref{fig:deltasigma_datavector}. We show the boost factor estimated from the data in solid lines, and compare to the boost factor estimated from the mocks in \citetalias{Lange:2024} in dashed lines. We note that there are no boost factor estimates for \gls{sdss} in the mocks, and the boost factors for \gls{hscy3} in the data are compared to boost factors from \gls{hscy1} in the mocks.}
        \label{fig:boost_factor}
    \end{figure}

    As discussed in Sect.~\ref{sec:lensing_measurements:lensing_estimator:boost_factor}, we do not employ a boost factor correction due to the uncertainty of other contributions to this systematic. We still show the estimated boost factors in Fig.~\ref{fig:boost_factor} and compare them to simulations. We note that the boost factors are generally small due to our efforts to minimize lens-source overlap. In the cases where boost factors are significant, we observe a good agreement between the mock estimates and our measurements. For \gls{hscy3} there are some discrepancies, particularly a boost factor smaller than unity for the lower lens redshift bins, however, the estimates are extremely noisy and the significance of this discrepancy is thus low. We observe that, on scales $\gtrsim 1\,\mathrm{Mpc}/h$, the boost factor is extremely small, both when estimated from the data and from simulations, where discrepancies are usually below one percent, the only notable exception being the 2nd \gls{hscy3} source bin correlated with the 2nd \gls{bgs} bin. For cosmological parameter analyses with reasonable scale cuts, we therefore do not expect substantial boost factor contaminations. If scales $\lesssim 1\,\mathrm{Mpc}/h$ are included in the analysis, an accurate treatment for the boost factor will be required. We note that the boost factors for \gls{hscy3} in the data are compared to boost factors from \gls{hscy1} in the mocks, as no mocks with realistic source populations for \gls{hscy3} were available at the time of writing.

\section{Impact of alternative cosmologies}
    \label{app:alternaive_cosmologies}
    To calculate $\ds$, we have to assume a cosmological model to calculate the angular diameter distance. Throughout this work, we assume the DESI fiducial cosmology, which is the best-fit cosmology of the latest Planck $\Lambda$CDM constraints \citep{2020A&A...641A...6P}. In this Appendix, we briefly revisit the impact of this model choice and show that models with reasonable alternative cosmologies do not significantly impact our results.

    In general, changing the cosmological model impacts the $\ds$ measurements in two ways: First, the comoving angular diameter distance changes, impacting the conversions between angular and comoving distances in a redshift-dependent way. Second, the same change of angular diameter distances modifies the critical surface density $\Sigmacrit$. Technically, differences in $H_0$ also impact distance measurements, but as we report all measurements in Mpc/$h$ units, we are immune to that change. Here, we test our fiducial cosmology against two alternative cosmological models, one where we take the best-fit cosmology of the 9-year WMAP analysis \citep{arXiv:1212.5226}, but artificially impose a low value of $\Omega_\mathrm{m}=0.22$. This is because $\Omega_\mathrm{m}$ (and $\Omega_\Lambda$, which is fixed to $0.78$ as we impose a flat Universe) are the components dominating the expansion history (and thus the distance measures), so we want to force a substantial difference. Our second cosmology retains all cosmological parameters from the Planck analysis, but we impose a dynamic dark energy model with $w_0=-0.48$ and $w_a=-1.34$, corresponding to the best-fit result of the DESI-DR2 BAO analysis \citep{arXiv:2503.14738,2025arXiv250314743D}.

    \begin{figure}
        \centering
        \includegraphics[width=\linewidth]{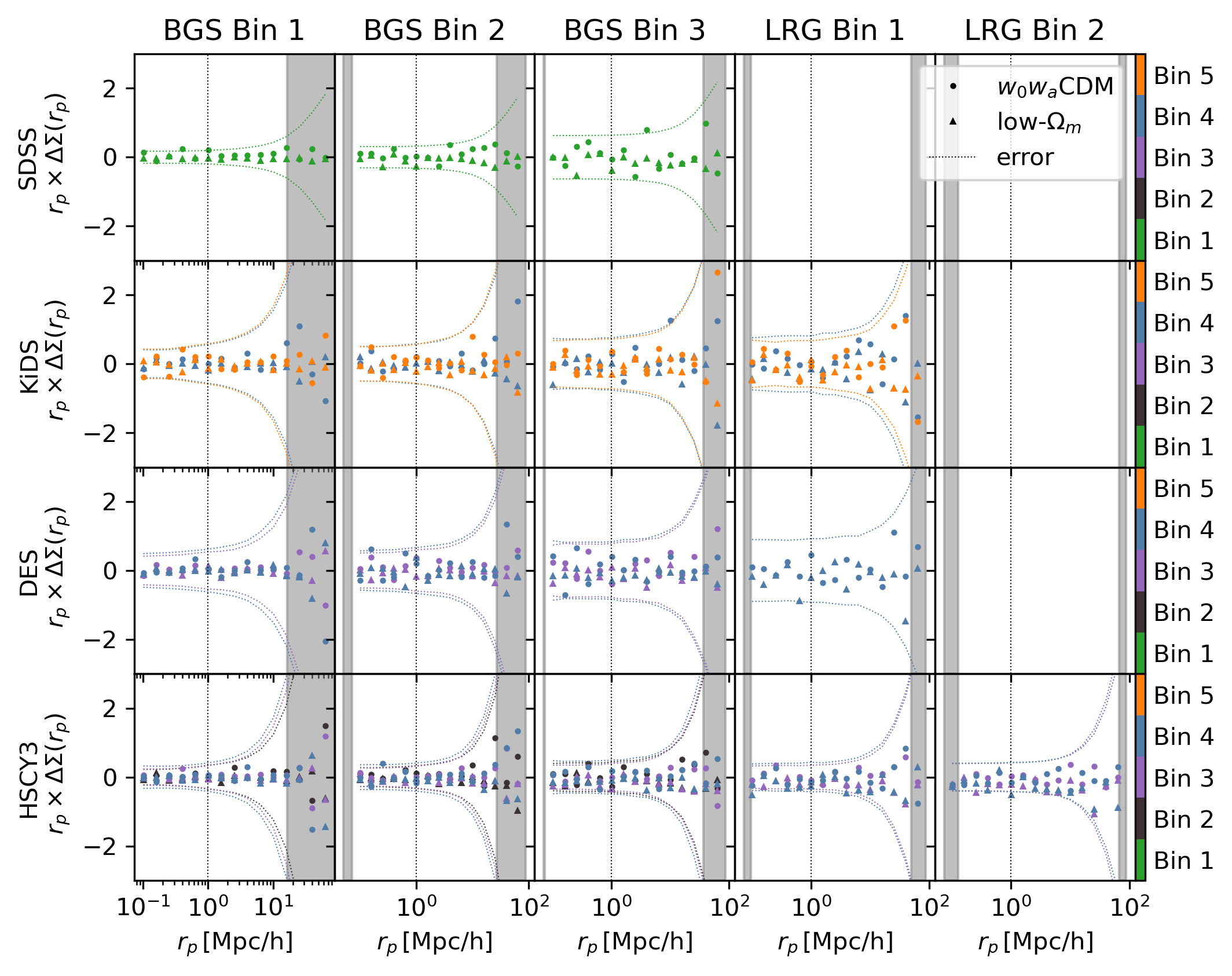}
        \caption{Difference between the $\ds$ measurements of the alternative cosmologies and the fiducial cosmology. The dashed lines indicate the covariance estimates.}
        \label{fig:deltasigma_different_cosmologies}
    \end{figure}
    
    As can be seen in Fig.~\ref{fig:deltasigma_different_cosmologies}, the differences between the two cosmologies and the fiducial cosmology are small compared to the covariance estimates. On small scales, where the covariance is dominated by shape noise, one can observe some scatter, as some source galaxies are attributed to different radial bins. In virtually all cases, the difference is smaller than the covariance estimate and one can observe that there is no bias towards high or low values. At large scales, where the sample variance dominates, the differences become negligible compared to the covariance estimates.
\label{lastpage}


\end{appendix}
\end{document}